\documentclass{emulateapj}
\usepackage{color} 
\usepackage[colorlinks,urlcolor=blue,citecolor=blue,linkcolor=blue]{hyper ref}

\usepackage{epsfig}
\usepackage{subfigure}
\usepackage{graphicx}
\usepackage{amsmath}
\usepackage{natbib}
\newcommand{\pa}{\partial}
\newcommand{\mb}{\boldsymbol}
\newcommand{\mc}{\mathcal}

\shorttitle{MHD-PIC Simulation of the CR Streaming Instability}
\shortauthors{X.-N. Bai et al.}

\begin{document}


\title{Magnetohydrodynamic-Particle-in-Cell Simulations of the Cosmic-Ray Streaming Instability: Linear Growth and Quasi-linear Evolution}


\author{Xue-Ning Bai\altaffilmark{1,2}, Eve C. Ostriker\altaffilmark{3},
Illya Plotnikov\altaffilmark{3,4},
James M. Stone\altaffilmark{3}}
\affil{$^1$Institute for Advanced Study, Tsinghua University, Beijing 100084, China}
\affil{$^2$Tsinghua Center for Astrophysics, Tsinghua University, Beijing 100084, China}
\affil{$^3$Department of Astrophysical Sciences, Peyton Hall, Princeton
University, Princeton, NJ 08544}
\affil{$^4$ Universit\'e de Toulouse, UPS-OMP, IRAP, 9 av. Colonel Roche, BP 44346, F-31028 Toulouse Cedex 4, France }
\email{xbai@tsinghua.edu.cn}




\begin{abstract}
The gyro-resonant cosmic-ray (CR) streaming instability is believed to play a crucial role
in CR transport, leading to growth of Alfv\'en waves at small scales that scatter CRs,
and impacts the interaction of CRs with the ISM on large scales.
However, extreme scale separation  ($\lambda \ll \rm pc$), low cosmic ray number density
($n_{\rm CR}/n_{\rm ISM} \sim 10^{-9}$), and weak CR anisotropy ($\sim v_A/c$) pose
strong challenges for proper numerical studies of this instability  on a microphysical level.
Employing the recently developed magnetohydrodynamic-particle-in-cell (MHD-PIC) method,
which has unique advantages to alleviate these issues, we conduct one-dimensional
simulations that quantitatively demonstrate the growth and saturation of the instability in the
parameter regime consistent with realistic CR streaming in the large-scale ISM.   
Our implementation of the $\delta f$ method dramatically reduces Poisson noise and enables
us to accurately capture wave growth over a broad spectrum, equally shared between left
and right handed Alfv\'en modes. We are also able to 
accurately follow the quasi-linear diffusion of  CRs subsequent to wave growth, which
is achieved by employing phase randomization across periodic boundaries. 
Full isotropization of the CRs in the wave frame requires pitch angles of most CRs to
efficiently cross $90^\circ$, and can be captured in simulations with
relatively high wave amplitude and/or high spatial resolution.
We attribute this crossing to non-linear wave-particle interaction (rather than mirror reflection) by
investigating individual CR trajectories. 
We anticipate our methodology will open up opportunities for future
investigations that incorporate additional physics.
\end{abstract}



\section{Introduction}\label{sec:intro}

Cosmic-rays (CRs) are an essential constituent of the Galaxy. With energy density
comparable to or exceeding other components of the interstellar medium (ISM), CRs likely
play an important role ISM heating, chemistry, and  dynamics 
over timescales of $\sim 10^3 -10^8 {\rm yr}$, as well as galaxy formation and
evolution over Gyr timescales
\citep[see e.g. the reviews of][]{Ferriere01,Zweibel17,Naab17}.  
The dominant CR population is protons, with $\sim$GeV particles representing
the peak of the CR energy distribution; 
the typical 
cosmic ray energy density of $\sim 1 \rm eV\ cm^{-3}$ corresponds to a number density 
$n_{\rm CR} \sim 10^{-9} \rm cm^{-3}$
\citep[e.g.][]{Grenier15}.

CRs are expected to be primarily produced in collisionless shocks, most likely from
young supernova remnants (e.g., \citealp{BaadeZwicky34}), via diffusive shock acceleration
(e.g., \citealp{Krymsky77,Bell78,BlandfordOstriker78,Drury83}).
They are then scattered across the Galaxy,
primarily following magnetic field lines in the turbulent ISM,
and eventually escaping after a few million years \citep{GinzburgSyrovatskii64}.

Being a non-thermal high-energy particle population, the CRs 
only infrequently collide
with the ISM particles (this is greatest at low energies, causing ionization and
heating), but the CRs and bulk ISM gas indirectly interact as they both couple to
the interstellar magnetic field.
Through this indirect coupling, CRs can exert pressure forces on the ISM gas which
are dynamically important. Moreover, being a (trans-)relativistic plasma component,
CRs in the ISM are highly buoyant, with a natural tendency to escape from the Galaxy.
As a result, the CR component can possibly 
contribute to large-scale instabilities and dynamo activity \citep{Parker66,Parker92},
and to driving galactic winds \citep[][see also below]{Ipavich75,Breitschwerdt_etal91,Zirakashvili_etal96,Everett_etal08}.

At microscopic level, the physics of the 
interaction between CRs and background thermal ISM gas
is extremely rich.
Considered as passive test particles, the CRs can
be scattered diffusively by background ISM turbulence at the gyro-radius scale (e.g., 
\citealp{Jokipii66,SchlickeiserMiller98,YanLazarian02}), which is important for understanding
CR transport. For typical $\sim$ GeV CR particles, the gyro-radius for an ISM magnetic field of a few
$\mu G$ is on the order of $10^{12}$cm -- i.e. a micro-pc.
Considering the huge dynamic range from the energy-containing 
scales of ISM turbulence to the relevant wavelength 
(a factor of $10^8$ in length), as well as the anisotropy of the MHD cascade at sub-pc scales \citep{GoldreichSridhar95},
it is considered unlikely for background turbulence to be the
dominant source of CR scattering for GeV particles (e.g., \citealp{Zweibel13}). 
There are even some suggestions from spectral signatures that only higher-energy
($\gtrsim 200 {\rm GeV}$)
particles are affected by background ISM turbulence \citep{Blasi12}.

Anisotropy in the distribution of CRs can destabilize Alfv\'en waves.
Of particular interest
and relevance is the cosmic-ray streaming instability (CRSI,
\citealp{KulsrudPearce69,Wentzel74,Skilling75a,Skilling75b,Skilling75c}).  Instability occurs when the bulk drift
speed of the CRs exceeds the Alfv\'en speed in the background plasma.
The instability drives the growth of Alfv\'en waves at the cost of free energy from CR anisotropy,
and in the absence of wave damping  
the resulting scattering of particles off the waves
will effectively reduce the CR drift speed to Alfv\'en speed. This process limits the bulk flow 
speed of the CRs, and simultaneously 
leads to energy and momentum exchange between the CRs and background gas.
As a consequence, CRSI plays a fundamental role in the interaction between the CRs and ISM gas,
and is the key element behind the picture of CR self-confinement as well as CR-driven winds 
(see, e.g., \citealp{Zweibel13,Zweibel17,Amato18} for reviews).  On much larger scales in galaxy clusters, the CRSI may also be important in regulating heating \citep[e.g.][]{
GuoOh08,Ensslin11,Fujita11,Wiener_etal13,Jacob17,Wiener18a,Wiener18b}.  Here, however, we shall focus on the parameter regime appropriate for cosmic ray interactions with the ISM.  

Studies of CR interactions with the ISM on large scales, whether analytic or numerical, treat the CRs as fluid-like
(e.g., \citealp{Breitschwerdt_etal91,Breitschwerdt93,Enblin_etal07,Sharma10,Pfrommer_etal17,JiangOh18,ThomasPfrommer19}), generally adopting the assumption of streaming, diffusion, or some combination.
This is necessary given the huge scale separation between the macroscopic scale of
astrophysical systems and the tiny gyro-scale of the CRs, but this approach is subject to the
prescriptions adopted to treat the microphysical effects of CR diffusion and streaming.

In recent years, there has been a resurgence of interest in  CR-driven galactic winds.  While most numerical studies to date have adopted diffusion prescriptions 
\citep[e.g.][]{ Jubelgas_etal08,Hanasz_etal13,Booth_etal13,Salem_etal14,SalemBryan14,Simpson16,Pakmor_etal16,Girichidis_etal16,Girichidis18,Jacob18}, there have also been studies that compare wind outcomes when streaming vs. diffusive prescriptions are adopted for the CRs
\citep{Wiener_etal17,Ruszkowski_etal17}.  
The dynamics of winds in which CRs stream  
at the Alfv'en speed have distinctive features, since the effective sound speed of gas {\it increases} as its density decreases, enabling steady-state CR-driven acceleration of cool gas to high velocities in typical galactic potentials \citep{Mao18}. 
Most recently, several studies of wind driving have considered even more sophisticated treatment of the CR fluid, implementing prescriptions for wave damping and for decoupling of CRs from cold ISM components \citep{Recchia_etal16,Holguin18,Farber18}.   
While all of the above studies highlight the important role played by the CRs, it is also
evident that different microphysical prescriptions can yield dramatically different
outcomes for wind properties. Currently, however, microphysical coefficients that are adopted in simulations are motivated on phenomenological grounds or based on classical analytic theory of the CRSI, diffusion, and wave damping.   
Development of a modern theoretical foundation for CR fluid treatments, with coefficients calibrated from numerical studies that directly follow the microphysics, will clearly aid progress in this field.  

At microscopic level,
our current understanding of the CRSI is limited to quasi-linear theory.
We note that on the one hand, the instability can become non-linear when
CR streaming is sufficiently strong, and eventually transitions to the Bell
instability \citep{Bell04} in more extreme environments such as super-Alfv\'enic
shocks in supernova remnants. On the other hand, there are additional
uncertainties related to mechanisms that lead to wave damping (see discussions
in Section \ref{ssec:realistic}), many of which again require numerical studies.

Numerical studies of the CRSI at microphysical level 
{\it per force} must involve
particle-in-cell (PIC) methods, and are inherently challenging 
because of the extreme ratios between the density of thermal and nonthermal particles and the scales relevant for the background thermal plasma compared to CR gyroradii.
Another difficulty is that the weak level of CR anisotropy (of order a few times $v_A/c$)
must be accurately represented, demanding a huge number of computational
particles (see Section \ref{sec:method}). Both of these issues make it very
challenging for conventional particle-in-cell (PIC) methods to study the CRSI, 
especially in the regime relevant for the bulk of the ISM away from CR injection sites
(but see \citealp{HolcombSpitkovsky19} and further discussion 
in Section \ref{ssec:compare}).

Recently, we developed 
a new hybrid MHD-PIC method \citep{Bai_etal15}, 
implemented within the {\it Athena} code package \citep{Stone_etal08}.
By treating the background thermal plasma as a fluid described by magnetohydrodynamics
(MHD), 
we greatly alleviate the issue of scale separation between the thermal gas and CRs. 
Using this method, we 
are able to properly capture the basic properties
of the CRSI, particularly in astrophysically relevant/realistic parameter regimes.
For this purpose, 
in our first study we simplify by ignoring wave damping and concentrating on
one-dimensional simulations that provide the highest possible resolution. 
By employing several novel techniques, we
demonstrate that our method can accurately reproduce the linear growth of CSRI, and
subsequently  follow the quasi-linear diffusion of particles by scattering off of
waves generated by the CRSI. We anticipate this effort will pave the  way for future
studies that incorporate additional physics under a variety of environments.

\begin{figure*}
    \centering
    \includegraphics[width=170mm]{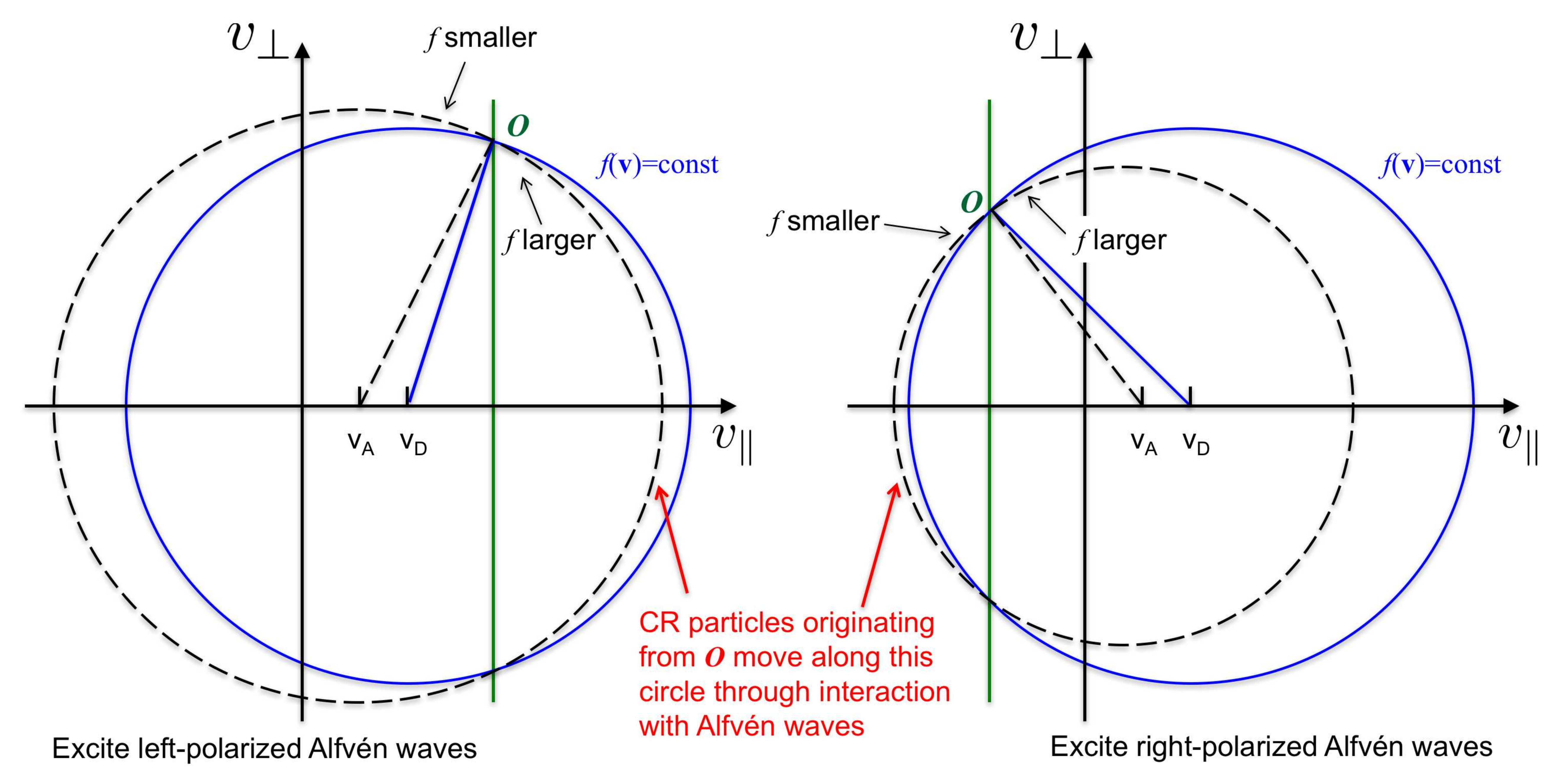}
  \caption{Schematic illustration of the basic physics of the CRSI. When the bulk
  CR drift velocity $v_D$ exceeds the Alfv\'en speed $v_A$, forward-traveling
  CR particles excite left-polarized, forward-propagating Alfv\'en waves, whereas
  backward-traveling CR particles excite right-polarized, forward-propagating
  Alfv\'en waves. We are taking particles located near a position $O$ in velocity space as an example. Also note the sizes of the blue and black circles should be comparable
  but has to be exaggerated in this schematic figure.
  See Section \ref{sec:theory} for details.}\label{fig:schematic}
\end{figure*}

This paper is organized as follows. We describe the basic theory of the CRSI
in Section \ref{sec:theory}. In Section \ref{sec:method}, we present the techniques
and methodology employed in our CRSI simulations using the MHD-PIC approach.
Our simulation setup is described in Section \ref{sec:setup}. In Section \ref{sec:fid},
we present simulation results from two fiducial runs. This is followed by a parameter
study in Section \ref{sec:param}. Our results are discussed in a  broader context in
Section \ref{sec:discussion}, before we conclude in Section \ref{sec:conclude}.
Additional numerical aspects of our simulations are described in the Appendices.

\section[]{Basic Theory}\label{sec:theory}

In this section, we briefly review the theory of the CRSI, which largely follows the
description in \citet{Kulsrud05}. We focus on the growth and saturation of the
instability. While wave damping is another important ingredient for the CRSI, it will
be addressed in our follow-up work and is not considered here.

\subsection[]{Physical Mechanism}\label{ssec:sketch}

Without rigorous mathematical derivation, the basic physics of the CRSI can be
understood schematically as illustrated in Figure \ref{fig:schematic}.
Consider a static background thermal plasma with constant density $\rho_0$ and
constant magnetic field ${\mb B}=B_0\hat{x}$ with $B_0>0$. The Alfv\'en speed is thus
$v_A=B_0/\sqrt{4\pi\rho_0}$. Coexisting with the background plasma is a population
of CRs. Let the initial CR velocity distribution be isotropic in a frame that drifts relative
to the background gas at velocity $v_D$, with the drift velocity along the background field
$\hat{x}$. Without loss of generality, let $v_D>0$ (along the direction of magnetic field).
Let $f_0({\mb v})$ or $f_0({\mb p})$ denote the initial particle velocity or momentum
distribution function, monotonically decreasing with increasing $p$.
Note that individual CR particle speeds greatly exceed $v_D$, hence approximately half
of the particles travel forward/backward with respect to the gas.
For particles with mass $m$, charge $q$ and Lorentz factor $\gamma$, the cyclotron
frequency $\Omega_c$ and gyro frequency $\Omega$ read
\begin{equation}
\Omega_c=\frac{qB_0}{mc}\ , \quad\Omega=\Omega_c/\gamma\ .
\end{equation}

The CR particles can experience gyro-resonance with circularly polarized Alfv\'en waves
when the pattern of particle gyration matches that of the waves, and the frequency of the
waves experienced by the particles matches their gyro-frequency. To satisfy the first
condition, forward traveling CRs can resonate with left-handed, forward-propagating
Alfv\'en waves, whereas backward traveling CRs can resonate with
right-handed, forward-propagating Alfv\'en waves (see Appendix \ref{app:wavedecomp}
for our adopted conventions for left/right handedness). To satisfy the second condition,
the parallel velocity of a particle $v_{\parallel}$ must satisfy
\begin{equation}\label{eq:res}
\omega-kv_{\parallel}=\pm\Omega\approx -kv_{\parallel}\ ,
\end{equation}
where $\omega$ and $k$ are the frequency and wave vector of the Alfv\'en wave
(we only consider wave vector $k$ along the magnetic field), and the $+$/$-$ sign
corresponds to right/left polarized modes, respectively. Note that $\omega=kv_A$,
while typically the CR velocity (and its projection) $v_{\parallel}\gg v_A$. We
can thus omit $\omega$ in the above equation. In other words, the CR
particles travel so rapidly that Alfv\'en waves simply appear as a static pattern.

Now consider CR particles in velocity space decomposed into components parallel
and perpendicular to the background magnetic field $(v_{\parallel}, v_{\perp})$, and
pick a representative particle located at point ``$O$" shown in Figure \ref{fig:schematic}.
To facilitate the illustration, we now assume particles are non-relativistic (but it can be
easily generalized to relativistic CRs), and hence $f_0({\mb v})$ is constant along a
circle centered on $(v_D, 0)$ (the blue solid circle in Figure \ref{fig:schematic}). Clearly,
$f_0$ is larger (smaller) inside (outside) the circle.
For linear Alfv\'en waves, their associated electric field vanishes in the wave frame.
Because magnetic fields do no work on particles, each particle trajectory in
velocity space must follow a circle centered in $(v_A, 0)$ (black dashed
circle in Figure \ref{fig:schematic}). Wave-particle interaction generally leads to a
quasi-linear diffusion (QLD) of particle pitch angle, towards removing the gradient of
$f_0({\mb v})$ across the dashed circle near point $O$. We see that when
$v_D>v_A$, for forward traveling CR particles near $O$ (left panel), the net outcome
should be that more particles with higher $v_{\parallel}$ on the right hand
side of point $O$ diffuse along the dashed circle towards smaller $v_{\parallel}$. 
In net, when $v_D>v_A$ the diffusion (via resonant interactions) of particles irreversibly
gives momentum to the left-handed Alfv\'en waves, feeding their growth.
Similarly, for backward traveling CR particles (right panel), more particles on the right
side of $O$ diffuse along the dashed circle to the left side, with the change in
$v_{\parallel}$ feeding the growth of right-handed Alfv\'en waves.
We can also easily see that if $v_D<v_A$, instead of wave excitation, the Alfv\'en
waves will be damped.

This discussion illustrates how the CRSI is driven by the anisotropy of the CRs.
At the same time, it suggests that CRSI growth is sustained by 
QLD of the CRs 
which produces secular 
and irreversible changes in the background plasma.
This means that for a numerical method to capture linear
growth of CRSI, it must properly capture QLD as well, which is an important fact to bear in
mind when we discuss numerical methods in Section \ref{sec:method}.

\subsection[]{Linear Instability of Waves}\label{ssec:linear}

We now proceed to a more quantitative description of the CRSI.
Its derivation follows from standard procedures, as detailed in \citet{Kulsrud05}.
Here, we only summarize the necessary results needed for setting up and analyzing
the simulations.

For convenience, in the discussions below and also throughout this paper, we
assume that the mass of CR ions and background ions are identical (being $m$).
As will be seen in Section \ref{ssec:formulation}, the value of $m$ has no
physical significance and can always be absorbed into a factor for the
charge-to-mass ratio. We thus drop it and express the momentum of and energy of individual CR
particles as
\begin{equation}
{\mb p}=\gamma{\mb v}\ ,\quad E_p=\gamma c^2\ ,
\end{equation}
where ${\mb v}$ is a particle's
velocity and $\gamma$ is its Lorentz factor. In other words, we effectively take
$m=1$. However, we do retain the symbol $m$ in certain places for better clarity.

As defined earlier, we use $f_0({\mb p})=f_0(p)$ to denote the initial CR distribution
function in the frame moving at $v_D$,
where the distribution is isotropic, with total CR number density given by
$n_{\rm CR}=\int 4\pi p^2f_0(p)dp$.
For Alfv\'en waves with perturbations 
proportional to $\exp{[{\rm i}(kx-\omega t)]}$, the
CR-modified dispersion relation reads
\begin{equation}\label{eq:disp}
\omega^2=k^2v_A^2\mp\frac{n_{\rm CR}}{n_i}\Omega_c(\omega-kv_D)
\bigg[(1-Q_1)\pm{\rm i}Q_2\bigg]\ ,
\end{equation}
where $n_i$ is the density of background ions. The role of the CRs is reflected in
the two dimensionless factors
\begin{equation}
\begin{split}\label{eq:Q1Q2}
Q_1(k)&\equiv\int dp\frac{4\pi p^2f_0(p)}{n_{\rm CR}}\bigg(\frac{p_{\rm res}}{2p}
\log\bigg|\frac{1+p/p_{\rm res}}{1-p/p_{\rm res}}\bigg|\bigg)\ ,\\
Q_2(k)&\equiv\frac{\pi}{2}
\int_{p_{\rm res}}^{\infty}dp\frac{4\pi p^2f_0(p)}{n_{\rm CR}}
\bigg(\frac{p_{\rm res}}{p}\bigg)\ ,
\end{split}
\end{equation}
where the dependence of $Q_1$ and $Q_2$ on $k$ is reflected in the resonant
momentum 
\begin{equation}
p_{\rm res}(k)=\Omega_c/k\ ,
\end{equation}
which is the minimum CR momentum to resonate with an Alfv\'en wave
with wave number $k$ (at zero pitch angle).\footnote{The dispersion relation
shown here is equivalent to Equations (11)-(13) in \citet{Amato09} with slight
differences in sign conventions. The terms $Q_1$ and $Q_2$ arise from integrals
containing $df/dp$ (see also Equation (69) in Chapter 12 of \citet{Kulsrud05}); this
is the mathematical embodiment of the situation depicted in Figure \ref{fig:schematic},
in which a gradient in $f$ leads to wave growth. We have performed integration
by parts to simplify the results, and $df/dp$ no longer appears.}

Physically, the term associated with $1$ in the $(1-Q_1)$ factor is due to the
background CR current along ${\mb B}_0$. The $Q_1$ term is from the
non-resonant response of CR particles to the waves. It is straightforward to
show that $Q_1$ approaches $1$ in long-wavelength limit, and monotonically
decreases to $0$ in short wavelength limit. Thus, the real part of the CR
contribution is bounded. The imaginary part, namely, the $Q_2$ term, results
from the resonant response of CR particles to the waves. Note that at a given
$k$, only particles with $p>p_{\rm res}$ contribute to $Q_2$, as expected.

In the limit $n_{\rm CR}/n_i\ll1$, the real part of the dispersion relation
(\ref{eq:disp}) is clearly dominated by the $k^2v_A^2$ term, and hence the
term proportional to $(1-Q_1)$ can be neglected. In this case, the dispersion
relation largely gives normal Alfv\'en waves, which can slowly grow or damp
owing to the imaginary part of the dispersion relation (i.e., the $Q_2$ term
from resonant response). This is the general CRSI originally derived
\citep{KulsrudPearce69}, where the forward propagating modes, both left and
right handed, grow at the same rates, given by
\begin{equation}\label{eq:growth}
\Gamma(k)=\frac{1}{2}\frac{n_{\rm CR}}{n_i}\Omega_c\bigg(\frac{v_D}{v_A}-1\bigg)Q_2(k)\ ,
\end{equation}
and hence growing modes are essentially linearly polarized.
Clearly, instability occurs only when $v_D>v_A$. Moreover, the maximum growth
rate is of the order $(n_{\rm CR}/n_i)\Omega_c$ at wavelengths that maximizes
$Q_2$ (to reach order unity), which typically correspond to the resonant scale with
lowest energy CRs. Backward propagating modes are damped.

More specifically, the initial distribution function is commonly considered to be
a power law with
\begin{equation}\label{eq:pldist}
f_0(p)\propto p^{-4+\alpha}\ ,\quad p_0\leq p\leq p_{\rm max}
\end{equation}
and $f_0(p)=0$ otherwise. Fermi acceleration gives $\alpha=0$ (e.g.,
\citealp{Bell78,BlandfordOstriker78}). In typical ISM conditions, the Galactic CR
population corresponds to $\alpha\approx0.7$ for CR particle energy beyond
$\sim$GeV, and $n_{\rm CR}/n_i\sim10^{-9}$. The $Q_2$ factor is given by
\begin{equation}
Q_2(k)=\frac{\pi}{2}\bigg(\frac{1+\alpha}{2+\alpha}\bigg)p_{\rm res}
\frac{p_{\rm min}^{-(2+\alpha)}-p_{\rm max}^{-(2+\alpha)}}
{p_0^{-(1+\alpha)}-p_{\rm max}^{-(1+\alpha)}}\ ,
\end{equation}
where $p_{\rm min}={\rm Min}[{\rm Max}(p_{\rm res}, p_0), p_{\rm max}]$.
The maximum growth rate is achieved at $p_{\rm res}=p_0$, or
\begin{equation}
k=k_0\equiv \Omega_c/p_0\ .
\end{equation}
This corresponds to the resonant wavelength for particles with $p=p_0$ (and zero pitch
angle). In the limit $p_{\rm max}\rightarrow\infty$, we have
\begin{equation}
Q_2(k)=\begin{cases}
\dfrac{\pi}{2}\bigg(\dfrac{1+\alpha}{2+\alpha}\bigg)
\bigg(\dfrac{kp_0}{\Omega_c}\bigg)^{1+\alpha}\ , &
k<k_0\ , \\
\dfrac{\pi}{2}\bigg(\dfrac{1+\alpha}{2+\alpha}\bigg)
\bigg(\dfrac{\Omega_c}{kp_0}\bigg)\ , &
k>k_0\ . \\
\end{cases}\label{eq:tstop}
\end{equation}

\begin{figure}
    \centering
    \includegraphics[width=90mm]{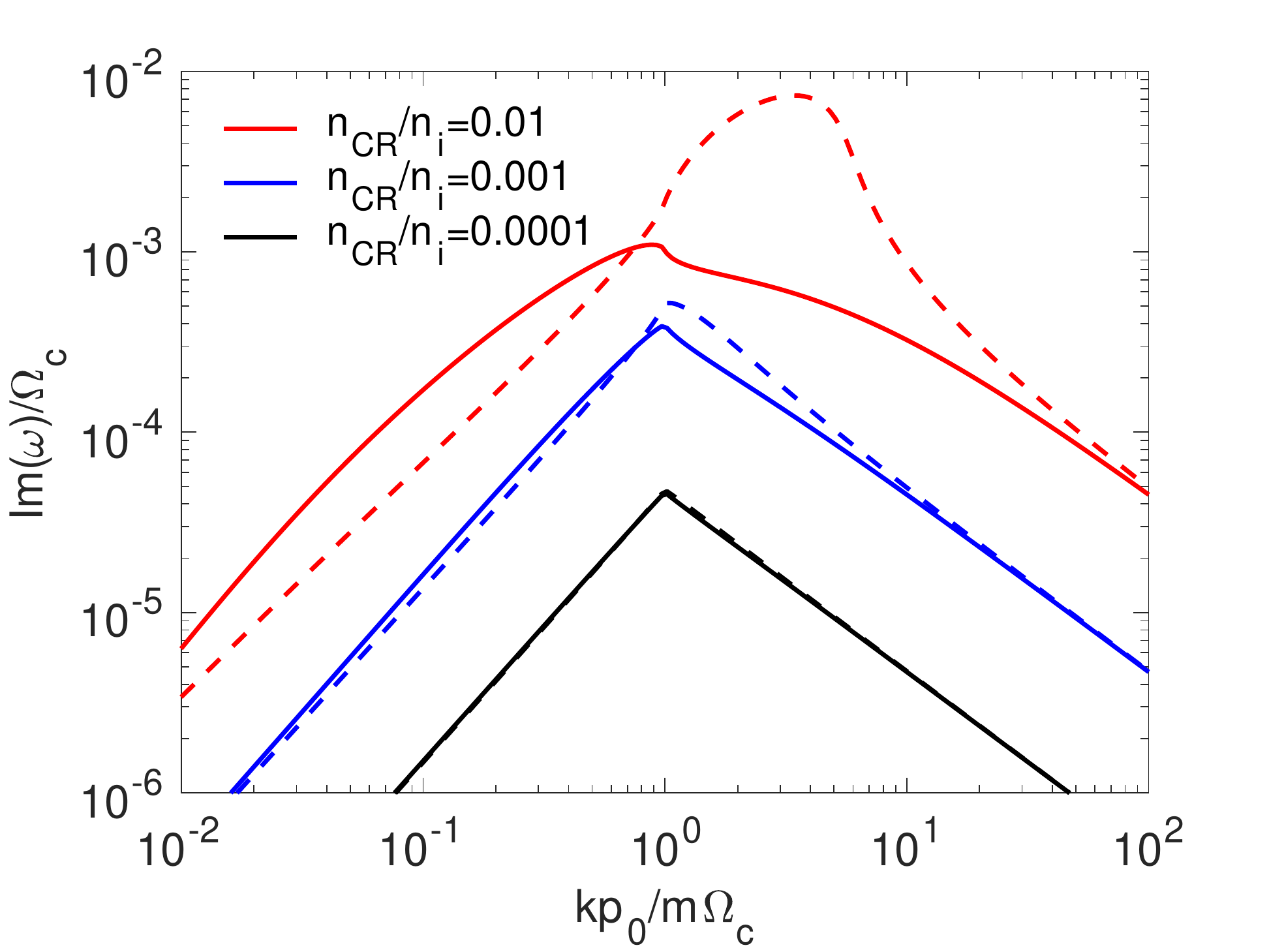}
  \caption{Linear dispersion relation of the CRSI with a truncated power-law
  CR distribution function (\ref{eq:pldist}), with fiducial parameters
  $v_D=2v_A$, $p_0/m=300v_A$, $s=4.5$. 
  Three values of $n_{\rm CR}/n_i$ are adopted, as shown in the red
  ($10^{-2}$), blue ($10^{-3}$) and black ($10^{-4}$) lines. The growth rates for
  left (solid) and right (dashed) handed modes are shown separately. Note the
  gradual development and dominance of the non-resonant Bell mode at
  $n_{\rm CR}/n_i\gtrsim10^{-3}$. We focus on the regime with smaller
  $n_{\rm CR}/n_i$, being more realistic in the bulk of the ISM.}\label{fig:disp}
\end{figure}

Taking into account the term from $(1-Q_1)$ breaks the degeneracy between
left and right handed modes, especially when $n_{\rm CR}/n_i$ increases.
In particular, it makes the right-handed modes grow faster and more dominant
towards short wavelength (see Figure \ref{fig:disp}), and eventually, it smoothly
transitions to the non-resonant Bell mode at $k\gtrsim k_0$ \citep{Bell04}. 
This transition occurs when \citep{Amato09}
\begin{equation}
\frac{n_{\rm CR}}{n_i}\gtrsim\frac{v_A^2}{p_0v_D}\ .
\end{equation}

For simulations in this work, we fiducially adopt $v_D/v_A=2$, $p_0/v_A=300$,
$\alpha=0.5$, and $n_{\rm CR}/n_i=10^{-4}$ to $10^{-3}$. The resulting
dispersion relation is shown in Figure \ref{fig:disp}, which displays all features
discussed above. Note that  
for application to realistic large-scale ISM conditions, a very low value of low
$n_{\rm CR}/n_i\sim 10^{-9}$ is appropriate, such that the left/right handed modes are
degenerate.  However, in practice, the values of $n_{\rm CR}/n_i$ we use in the
simulations have to be much higher 
in order to follow wave growth and particle
quasi-linear diffusion within a reasonable computational time. We see that
for our fiducial parameters, the transition to non-resonant modes
occurs when $n_{\rm CR}/n_i\gtrsim10^{-3}$.

\subsection[]{Quasi-Linear Diffusion of Particles}\label{ssec:qld}

The growth of the CRSI creates a spectrum of Alfv\'en waves. The intensity of the
waves at wave number $k$ is described by $I(k)$, so that
\begin{equation}\label{eq:wavespec}
\int I(k)dk=\bigg\langle\frac{\delta B^2}{B_0^2}\bigg\rangle\ ,
\end{equation}
where angle bracket represents spatial averages.
In Appendix \ref{app:wavedecomp}, we describe the procedures to decompose parallel
propagating Alfv\'en waves into different modes (left/right handed, forward/backward
propagation), from which we obtain $I(k)$ for each mode.

In a spatially uniform media (as in our simulations), quasilinear diffusion (QLD) leads to
pitch angle scattering in the frame of the Alfv\'en waves, with pitch angle $\theta$
more commonly replaced by $\mu\equiv\cos\theta$. We use subscript `$_w$' to
denote quantities measured in this frame. The evolution of $f_w$, the CR distribution
function in the wave frame in a homogeneous system, is described by
(e.g., \citealp{Jokipii66,KulsrudPearce69})
\begin{equation}\label{eq:qld}
\frac{\pa f_w}{\pa t}=
\frac{\pa}{\pa\mu_w}\bigg[\frac{1-\mu_w^2}{2}\nu(\mu_w)\frac{\pa f_w}{\pa\mu_w}\bigg]\ ,
\end{equation}
where
\begin{equation}\label{eq:nu_qld}
\nu(\mu_w)=\pi\Omega k_{\rm res}I(k_{\rm res})\ ,
\end{equation}
and
\begin{equation}\label{eq:kres}
k_{\rm res}\equiv \Omega_c/(p_w\mu_w)
\end{equation}
is the wave number of the resonant waves.\footnote{In calculating $\nu$ from  Equation 
(\ref{eq:nu_qld}), we use $I(|k_{\rm res}|)$ and choose the intensity for left/right handed
waves for positive/negative $\mu_w$. Also note our normalization of $I(k)$ only accounts
for the magnetic energy spectrum (will be doubled for total energy spectrum). Because
of these normalization choices, our expression for $\nu$ does not have a factor $1/4$.}

Our simulations are conducted in the frame where the initial CR distribution is isotropic,
whose relative velocity to the wave frame is $\Delta v\equiv v_D-v_A$. Using subscripts
`$_\parallel$' and `$_\perp$' to denote components parallel and perpendicular to
background field ${\mb B}_0$, and to first order of $\Delta v\ll c$, we have for particle
momenta
\begin{equation}
p_{w,\perp}=p_\perp\ ,\quad p_{w,\parallel}=p_\parallel+\gamma\Delta v\ .
\end{equation}
With conservation of phase-space volume, it follows that the distribution functions
$f=f_w$.
The conversion relations between the two frames read
\begin{equation}
p_w=p\bigg(1+\mu\frac{\Delta v}{v}\bigg)\ ,\quad
\mu_w=\mu+(1-\mu^2)\frac{\Delta v}{v}\ ,
\end{equation}
where $v\sim c$ is the velocity corresponding to momentum $p$.
In the quasi-linear regime, the distribution function only deviates slightly from $f_0$, and
hence we can write $f(p,\mu)=f_0(p)+\delta f(p,\mu)$. Transforming to the wave frame,
and to the leading order of $\delta f/f_0$ and $\Delta v/c$, we obtain
\begin{equation}
\begin{split}\label{eq:frametrans}
&f_w(p_w,\mu_w)=f(p,\mu)\\
\approx &f_0(p_w)+\delta f(p_w,\mu_w)-\mu_w\frac{\pa f_0}{\pa\ln p}\frac{\Delta v}{v_w}\ .
\end{split}
\end{equation}
In the wave frame, we have
\begin{equation}
\frac{\pa f_w}{\pa\mu_w}\approx\frac{\pa\delta f}{\pa\mu_w}-\frac{\pa f_0}{\pa\ln p}\frac{\Delta v}{v_w}\ .\label{eq:dfwdmuw}
\end{equation}
In the fully saturated state, we expect $\pa f_w/\pa\mu_w=0$, and hence the pitch angle
distribution should satisfy
\begin{equation}\label{eq:dfsat}
\frac{\pa\delta f_{\rm sat}}{\pa\mu_w}\approx\frac{\pa f_0}{\pa\ln p}\frac{\Delta v}{v}\ 
\ \Rightarrow\ 
\delta f_{\rm sat}\approx\mu_w\frac{\pa f_0}{\pa\ln p}\frac{\Delta v}{v}\ .
\end{equation}

The amplitude of the Alfv\'en waves in the saturated state can also be estimated.
The 
net momentum density acquired by the CRs after saturation is reached is
\begin{equation}\label{eq:dPCR}
\Delta {\cal P}_{\rm CR}=\int\mu p(\delta f_{\rm sat}) d^3p
=\frac{4\pi}{3}\int\frac{\pa f_0}{\pa\ln p}\frac{\Delta v}{v}p^3dp\ .
\end{equation}
Note that with $\partial f_0/\partial p <0$, this  implies that if initially $\Delta v=v_D -v_A >0$,  in saturation the CR distribution must have acquired a momentum deficit, $\Delta {\cal P}_{\rm CR} <0$.
In the limit that all CRs are relativistic, this yields
\begin{equation}
\Delta {\cal P}_{\rm CR}
\approx - \frac{4}{3}\frac{n_{\rm CR}\Delta v}{c}\langle p\rangle\ ,
\end{equation}
where $\langle p\rangle\equiv \int f_0pd^3p/\int f_0d^3p$. For the non-relativistic CR limit,
the momentum reduction is
$\Delta {\cal P}_{\rm CR}\approx - mn_{\rm CR}\Delta v$.

The deficit in CR momentum must have been transferred to 
forward-propagating Alfv\'en waves. Note that there is an equipartition
in kinetic and magnetic energies in Alfv\'en waves, with effective momentum density in waves
(the Poynting flux divided by $v_A^2$; cf. \citealt{Kulsrud05})
given by
\begin{equation}\label{eq:Pwv}
{\cal P}_{\rm wave}=\rho v_A\bigg\langle\frac{\delta B^2}{B_0^2}\bigg\rangle\ .
\end{equation}
Following the discussion in Kulsrud \citep[see also][]{Wentzel74}, this  momentum density represents the growth of Maxwell and Reynolds stresses in waves needed to transfer the CR momentum to the bulk gas,
and is exhibited as a small fractional change in the momentum of the bulk gas.

Equating $|\Delta{\cal P}_{\rm CR}|$ with ${\cal P}_{\rm wave}$, we can obtain the expected saturation level for magnetic fluctuations.
For ultra-relativistic and non-relativistic cases, we simply have
\begin{equation}
\frac{\delta B^2}{B_0^2}\approx\frac{4}{3}\frac{\langle p\rangle}{mc}
\frac{n_{\rm CR}}{n_i}\frac{\Delta v}{v_A}\ ,\quad
{\rm or}\quad\approx\frac{n_{\rm CR}}{n_i}\frac{\Delta v}{v_A}\ .\label{eq:saturatedE}
\end{equation}
The real situation (as we consider) is of course somewhere in between 
nonrelativistic and ultrarelativistic, but the dependence of wave amplitude on the basic problem parameters is clear.
These relations show that the saturation level of the instability 
is limited by the initial free energy in CR streaming. The above 
upper limit for magnetic field amplitudes would be reduced in the presence of damping.

The timescale for the CRs to relax 
to a state of isotropy in the wave frame 
is characterized by the scattering frequency
$\nu$, which depends on both the pitch angle and the wave amplitudes. This means that
scattering across different pitch angle ranges can take different amounts of time. Still, one may
define a characteristic scattering frequency for the bulk CR population as
\begin{equation}
\nu_{\rm QLD}\sim{\nu(p,\mu)}\sim\Omega\frac{\delta B^2}{B_0^2}\ .
\end{equation}
Note that wave growth is eventually quenched as the gradient in distribution function is smeared out through QLD. This suggests that another way to estimate the wave saturation amplitude is by equating the above $\nu_{\rm QLD}$ with $\Gamma$ from Equation (\ref{eq:growth}), which is indeed similar (to order-of-magnitude) to the more rigorous estimate discussed above.

\subsection[]{Reflection Across 90$^\circ$ Pitch Angle}\label{ssec:90deg0}

In Equation (\ref{eq:kres}), when $\mu_m\rightarrow0$ (i.e., pitch angle approaches
$90^\circ$), $k_{\rm res}\rightarrow\infty$, i.e., the resonant wavelength becomes
infinitely small. There is little energy in the waves generated from the CRSI towards
the shortest wavelength, leading to a bottleneck for particles to be scattered across
this $90^\circ$ pitch angle. This is a well-known problem, and without crossing the
$90^\circ$ pitch angle, the CRSI would saturate prematurely far from fully utilizing the
free-energy in the system.

This problem has been discussed extensively in the literature. Within quasi-linear
theory, relaxing the magnetostatic approximation [Equation (\ref{eq:res})] allows for
additional resonant interactions when both forward and backward propagating waves
are present \citep{Schlickeiser89}, covering a pitch angle of $\Delta\theta\sim 2v_A/c$
around 90$^\circ$.

Magnetic mirroring is more commonly invoked in overcoming the 90$^\circ$ problem
(e.g., \citealp{FeliceKulsrud01}). With magnetic moment $M\equiv p_\perp^2/2B$ being
an adiabatic invariant, a particle will experience a backward mirror force when its guiding
center travels along a positive gradient in total field strength. For particle pitch angle
sufficiently close to 90$^\circ$, this mirror force can lead to reflection, thus directly jumping
over the $90^\circ$ barrier. With total field strength
$B=\sqrt{B_0^2+\delta B^2}\approx B_0(1+\delta B^2/2B_0^2)$, the critical pitch angle
cosine below which mirror reflection can occur is thus given by
\begin{equation}\label{eq:mirrorcond}
\mu_{\rm mir}\approx\frac{1}{\sqrt{2}}\frac{\delta B}{B_0}\ .
\end{equation}
One thus expects mirror reflection to take over when QLD manages to scatter particles
to pitch angle $\mu\sim\mu_{\rm mir}$.

Note that mirror reflection requires slow changes in field strength over many particle
gyro orbits. 
Given that a broad range of wave spectrum is excited by the CRSI, including those
at small scales ($\lesssim k_{\rm res}^{-1}$) for particles of a given momentum, we can also imagine the opposite situation,
where changes in field strength occurs more abruptly. For instance, an abrupt change of
perpendicular field by $\delta B_{dis}$ would change the direction of particle motion by
$\delta B_{dis}/B_0$, and in the mean time, the change the particle pitch angle is also on
the order of $\sim\delta B_{dis}/B_0$. Depending on the gyro-phase, particles have a good
chance to be reflected when $\mu\lesssim\delta B_{dis}/B_0$.

The (extreme) scenario outlined above falls into the category of non-linear wave-particle
interaction: considering particle orbits in perturbed fields (rather than unperturbed fields as
in quasi-linear theory) gives resonance broadening 
(e.g., \citealp{Dupree66,Volk73,Achterberg81}), and alleviates the $90^\circ$ problem.
It is more effective towards larger wave amplitudes, and eventually enters the regime of
strong MHD turbulence \citep{YanLazarian08}.
As we will show, our analysis suggests that
this effect is primarily responsible for overcoming the $90^\circ$ barrier; 
for more discussion see Section \ref{ssec:90deg}.

\section[]{Numerical Method}\label{sec:method}

There are several challenges that confront  numerical study of the CRSI.
First, there is substantial physical scale separation, which is exhibited as follows.
\begin{itemize}
\item The separation between microscopic scales of the background plasma, including
the electron and ion skin depths ($c/\omega_{pe}$ or $c/\omega_{pi}$ that full-PIC or
hybrid-PIC simulations must resolve, where $\omega_{pe}$ and $\omega_{pi}$ are the
electron and ion plasma frequencies), and the resonant scale of the CRSI ($\sim$
gyro-radii of lowest-energy, trans-relativistic CRs). 
The ratio of the two scales is at least of the order $c/v_A$, which is prohibitively large for
conventional PIC simulations if one were to approach realistic conditions.

\item For quasi-linear diffusion of particles, individual CR particles must encounter a sufficient
number of {\it independent} wave packets, so as to experience random-walk behavior in pitch angle evolution.

\item The issue with particles scattering across the $90^\circ$ pitch angle. Particles with pitch
angle near $90^\circ$ have resonant wavelength much shorter than their gyro-radii,
implying that the most unstable wavelength of the CRSI must be very finely resolved.

\end{itemize}

The MHD-PIC approach uniquely alleviates the first scale separation issue by bypassing the microscopic plasma
scales (ion skin depth) of the background thermal plasma. However, the remaining two issues
remain challenging. Clearly, very high resolution is needed to capture scattering across
$90^\circ$ pitch angle. In the mean time, when employing periodic boundary conditions, a
sufficiently long simulation domain that cover many resonant scales is essential so that particles
do not experience the same wave packets upon re-entering the domain. This issue will be
discussed further in Section \ref{ssec:phrand} and Appendix \ref{app:boxsize}.

In addition to the above challenges of physical scale separation, another challenge in modeling the regime of interest for the large-scale ISM is the huge ratio between the number density of CR particles and number density of background ISM particles, $ n_{\rm CR}/n_i \sim 10^{-9}$. Cosmic ray and ISM pressures, however, are generally within an order of magnitude of each other.
For modeling the large-scale ISM with its extremely small $n_{\rm CR}/n_i$, the MHD-PIC method is advantageous in comparison to full- or hybrid-PIC in that there there is no need to represent the background plasma with individual particles.

The third major challenge is that capturing the resonant condition still requires very large number of CR particles. For a
given resonant mode, particles of a given momentum can (if capable) contribute only at
a specific pitch angle. Correspondingly, only a tiny fraction of particles can potentially
be in resonance with a given MHD wave mode. This issue may be seen from another viewpoint.
The level of anisotropy in the CR distribution function is of the order $v_D/c\ll1$. Such weak
level of anisotropy must be accurately represented by the angular distribution of particles,
again demanding for a huge number of particles in each cell.

Even though the MHD-PIC method is able to concentrate all particles in the CRs, 
properly capturing CRSI is still challenging when $v_D/c\ll1$.
To overcome this issue, we employ the $\delta f$ method, which dramatically improves the
signal-to-noise and allows us to successfully simulate the CRSI with substantially reduced
number of particles (see Section \ref{ssec:deltaf} and Appendix \ref{app:noise}).
Below, we describe our methodology in detail.

\subsection[]{Formulation}\label{ssec:formulation}

We solve CR-modified MHD equations in conservative form\footnote{These equations are
simplified from those derived in \citet{Bai_etal15} in that terms related to the CR-modified
Hall effect are dropped. As discussed there, these terms are important only when
$(n_{\rm CR}v_D)/(nv_A)\gtrsim 1$,  corresponding to very strong CR streaming.
In our simulations, this ratio is $\lesssim10^{-3}$, and in reality the ratio is several orders
of magnitude lower.}
\begin{equation}\label{eq:gascont}
\frac{\pa\rho}{\pa t}+\nabla\cdot(\rho\mb{v}_g)=0\ ,
\end{equation}
\begin{equation}
\begin{split}
\frac{\pa(\rho\mb{v}_g)}{\pa t}+\nabla\cdot(&\rho{\mb v}_g^T{\mb v}_g
-{\mb B}^T{\mb B}+P^*{\sf I})\\
=&-(q_{\rm CR}n_{\rm CR}{\boldsymbol{\mc{E}}}+
\frac{{\mb J}_{\rm CR}}{c}\times{\mb B})\ ,
\end{split}
\label{eq:gasmotion}
\end{equation}
\begin{equation}
\frac{\pa E}{\pa t}+\nabla\cdot[(E+P^*){\mb v}_g
-({\mb B}\cdot{\mb v}_g){\mb B}]
=-{\mb J}_{\rm CR}\cdot{\boldsymbol{\mc{E}}}\ ,
\label{eq:engeq}
\end{equation}
where $P^*\equiv P_g+B^2/2$, ${\mc{E}}\equiv-{\mb v}_g\times{\mb B}/c$ is the
electric field, ${\sf I}$ is the identity tensor, and the total energy density of the gas is defined as
\begin{equation}
E=\frac{P_g}{\Gamma-1}+\frac{1}{2}\rho v_g^2+\frac{1}{2}B^2\ .\label{eq:gaseng}
\end{equation}
In the above, $\rho$, ${\mb v}_g$, $P_g$ are gas density, velocity, and pressure,
and $\Gamma$ is the adiabatic index. 
Note we have adopted the units where magnetic permeability is unity so that factors
of $(4\pi)^{-1/2}$ that would otherwise appear with the magnetic field are eliminated.

The CR number density $n_{\rm CR}$ and current
density ${\mb J}_{\rm CR}$ are defined as
\begin{equation}
\begin{split}\label{eq:intfull}
n_{\rm CR}(t, {\mb x})&=\int f(t, {\mb x}, {\mb p})d^3{\mb p}\ ,\\
{\mb J}_{\rm CR}(t, {\mb x})&=q_{\rm CR}\int {\mb v}f(t, {\mb x}, {\mb p})d^3{\mb p}\ ,
\end{split}
\end{equation}
where $q_{\rm CR}$ is individual CR particle charge, and ${\mb p}$, ${\mb v}$ are
the momentum and velocity of individual CR particles, with $f(t, {\mb x}, {\mb p})$ is the local
CR momentum distribution function.

While the MHD formulation is non-relativistic, the CR particles can be relativistic.
We define an artificial speed of light ${\mathbb C}$ for the CR particles, and the overall
formulation is consistent as long as ${\mathbb C}\gg v_A$ (typical MHD velocities).
For an individual particle $j$, we
have ${\mb p}_j=\gamma_j{\mb v}_j$, with
Lorentz factor given by
\begin{equation}\label{eq:Lorentz}
\gamma_j=\frac{\sqrt{{\mathbb C}^2+p_j^2}}{\mathbb C}
=\frac{\mathbb C}{\sqrt{{\mathbb C}^2-v_j^2}}\ .
\end{equation}

The particle equation of motion reads
\begin{equation}
\frac{d\mb{p}_j}{d t}
=\bigg(\frac{q}{mc}\bigg)_j\bigg(c{\boldsymbol{\mc{E}}}
+{\mb v}_j\times{\mb B}\bigg)\ .\label{eq:CRmotion}
\end{equation}
Note that as mentioned earlier, we have dropped the individual particle mass in the
definition of particle momentum. This mass is absorbed to the factor $q/(mc)$, representing
particle charge-to-mass ratio. Also note that the physical speed of
light $c$ has no significance 
in all equations where it appears: inspection of Equations 
(\ref{eq:gasmotion}), (\ref{eq:engeq}), (\ref{eq:intfull}), and (\ref{eq:CRmotion}) shows that $c$
appears only in the combinations $J_{\rm CR}/c$,  $c {\cal E}$, and
$q/mc$, and for numerical purposes can be absorbed into the parameter $(q/mc)$.

In our simulations, ideal MHD equations for the background thermal plasma are solved
using the Athena MHD code \citep{Stone_etal08}, which is a higher-order Godunov code
with constrained transport to enforce the divergence-free condition of the magnetic
field. The corner transport upwind (CTU, \citealp{GardinerStone05,GardinerStone08})
method is adopted for time integration. We use the Roe Riemann solver \citep{Roe81}
and third-order reconstruction in characteristic variables.
CR particles are implemented as Lagrangian particles, and the coupling between CRs
and the background plasma is handled by adding source terms 
in the gas momentum and energy updates, as indicated on the right hand sides of
Equations (\ref{eq:gasmotion}), (\ref{eq:engeq}). The CR particle equation of motion
(\ref{eq:CRmotion}) is solved by the standard Boris integrator \citep{Boris70}.
A standard triangular-shaped cloud (TSC) scheme \citep{BirdsallLangdon05} is
used for interpolating grid quantities to particle locations, and for depositing particle
quantities back to the grid (e.g., to evaluate $n_{\rm CR}$ and ${\mb J}_{\rm CR}$
in Equation (\ref{eq:intfull})).
Details of the CR implementation are described in \citet{Bai_etal15}.

\subsection[]{The $\delta$f Method}\label{ssec:deltaf}

The original CR implementation in \citet{Bai_etal15} interprets individual particles 
as representing the full distribution function $f({\mb x}, {\mb p})$, and evaluates
physical quantities such as $n_{\rm CR}$, ${\mb J}_{\rm CR}$ {\it directly} according
to Equation (\ref{eq:intfull}). We refer to this approach as the ``full-$f$ method'', which can be
subject to large Poisson noise.

In the case where the distribution function is close to some equilibrium distribution
$f_0({\mb x}, {\mb p})$, one can use the fact 
that $f_0$ is already known
analytically, and employ individual particles as Lagrangian markers taken to represent the
difference, $\delta f$, between $f_0$ and the full distribution function $f$. This is known
as the $\delta f$ method (e.g.,
\citealp{ParkerLee93,DimitsLee93,HuKrommes94,DentonKotschenreuther95,Kunz_etal14}). 
The basis of the $\delta f$ method is the Liouville theorem, which requires that the full distribution function $f$ be constant along
particle trajectories in phase space (i.e., characteristics). 

To implement the $\delta f$ method, we first record the initial value of $f$ at $t=0$ for
all particles (which is essentially $f_0$). Then at every time $t$, we assign a weight $w_j$ to
each particle $j$ given by
\begin{equation}\label{eq:dfweight}
w_j\equiv\frac{\delta f(t, {\mb x}_j(t), {\mb p}_j(t))}{f(t, {\mb x}_j(t), {\mb p}_j(t))}
=1-\frac{f_0({\mb x}_j(t), {\mb p}_j(t))}{f(0, {\mb x}_j(0), {\mb p}_j(0))}\ .
\end{equation}
Physical quantities such as the CR number density and current density are obtained by
\begin{equation}
\begin{split}
n_{\rm CR}(t, {\mb x})&=n_{{\rm CR},0}+\int\delta f(t, {\mb x}, {\mb p})d^3{\mb p}\\
&\simeq n_{{\rm CR},0}+\sum_{j=1}^{N_p}w_jS({\mb x}-{\mb x}_j)\ ,
\end{split}
\end{equation}
\begin{equation}
\begin{split}
{\mb J}_{\rm CR}(t, {\mb x})
&={\mb J}_{{\rm CR},0}+q_{\rm CR}\int {\mb v}\delta f(t, {\mb x}, {\mb p})d^3{\mb p}\\
&\simeq{\mb J}_{{\rm CR},0}+q_{\rm CR}\sum_{j=1}^{N_p}w_j{\mb v}_jS({\mb x}-{\mb x}_j)\ ,
\end{split}
\end{equation}
where $S({\mb r})$ is the shaping function used in particle interpolation (i.e., the TSC
scheme), and $n_{{\rm CR}, 0}$ and ${\mb J}_{{\rm CR}, 0}$ are obtained analytically
from $f_0$. By contrast, full-$f$ method corresponds to setting $w_j=1$, $n_{{\rm CR}, 0}=0$
and ${\mb J}_{{\rm CR}, 0}=0$.
In this way, the $\delta f$ method dramatically reduces the Poisson noise of
the background particle distribution, allowing the signal from the $\delta f$ part to be
substantially boosted.

In practice, we have found that the $\delta f$ method is essential. Without
employing it (i.e., using the full-$f$ method), and using our fiducial simulation parameters,
we barely observe the development of the CRSI even using $\sim10^4$ particles per cell,
and the system is almost entirely dominated by Poisson noise (see  Appendix
\ref{app:fullf} for more information).

We note that with the $\delta f$ method, exact conservation of the total
(gas and CR) momentum and energy, achieved in the full-$f$ method, is
lost. This is inevitable, since the formulation of the $\delta f$ method is intrinsically
non-conservative. However, the benefit from low noise is of overwhelming importance
for the CRSI problem, and in practice we find that the error in total momentum and energy
is negligible throughout all our simulations.

\begin{table*}
\caption{List of main simulation runs}\label{tab:params}
\begin{center}
\begin{tabular}{c|cc|cccc|c}\hline\hline
 Run & $v_D/v_A$ & $n_{\rm CR}/n_i$ & Domain size & Domain size & resolution &  $N_p$ per bin &  Runtime  \\
   & & & $L_x$ ($d_i$) & $L_x/\lambda_m$ &($d_i$ per cell) &  (per cell) & ($\Omega_c^{-1}$) \\\hline
Fid & 2.0 & $1.0\times10^{-4}$ & $9.6\times10^4$ & 51 & $10$ & $256$ & $10^6$ \\
M3 & 2.0 & $1.0\times10^{-3}$ & $9.6\times10^4$ & 51 & $10$ & $256$ & $10^6$ \\
M5 & 2.0 & $1.0\times10^{-5}$ & $3.2\times10^4$ & 17 & $10$ & $256$ & $1.5\times10^6$ \\
vD4 & 4.0 & $1.0\times10^{-4}$ & $9.6\times10^4$ & 51 & $10$ & $256$ & $10^6$ \\
vD8 & 8.0 & $1.0\times10^{-4}$ & $9.6\times10^4$ & 51 & $10$ & $256$ & $10^6$ \\
Fid-Hires & 2.0 & $1.0\times10^{-4}$ & $6.4\times10^4$ & 34 & $4$ & $256$ & $10^6$ \\
Fid-Short & 2.0 & $1.0\times10^{-4}$ & $3.2\times10^4$ & 17 & $10$ & $256$ & $10^6$ \\
Fid-Np64 & 2.0 & $1.0\times10^{-4}$ & $9.6\times10^4$ & 51 & $10$ & $64$ & $10^6$ \\
Fid-Np1024 & 2.0 & $1.0\times10^{-4}$ & $9.6\times10^4$ & 51 & $10$ & $1024$ & $2\times10^5$ \\
\hline\hline
\end{tabular}
\end{center}
Fixed parameters: ${\mathbb C}/v_A=300$, $p_0/mv_A=300$, $\kappa=1.25$,
and initial wave amplitude $A=10^{-4}$. 
Note the most unstable wavelength is 
$\lambda_m=2\pi p_0/(m\Omega_c)\approx 1885 d_i$ for all models.
\end{table*}

\begin{figure}
    \centering
    \subfigure{
    \includegraphics[width=90mm]{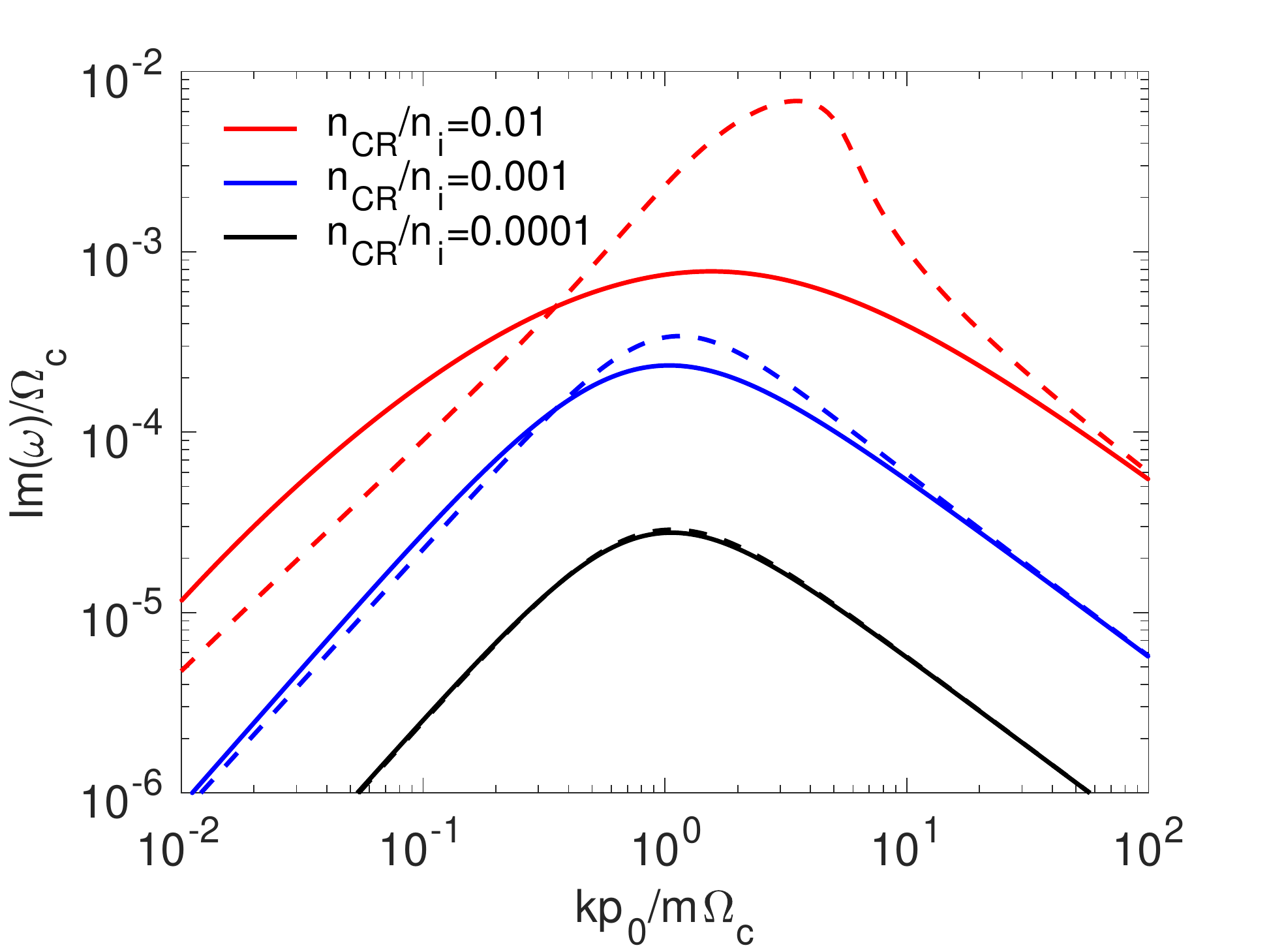}}
    \subfigure{
    \includegraphics[width=90mm]{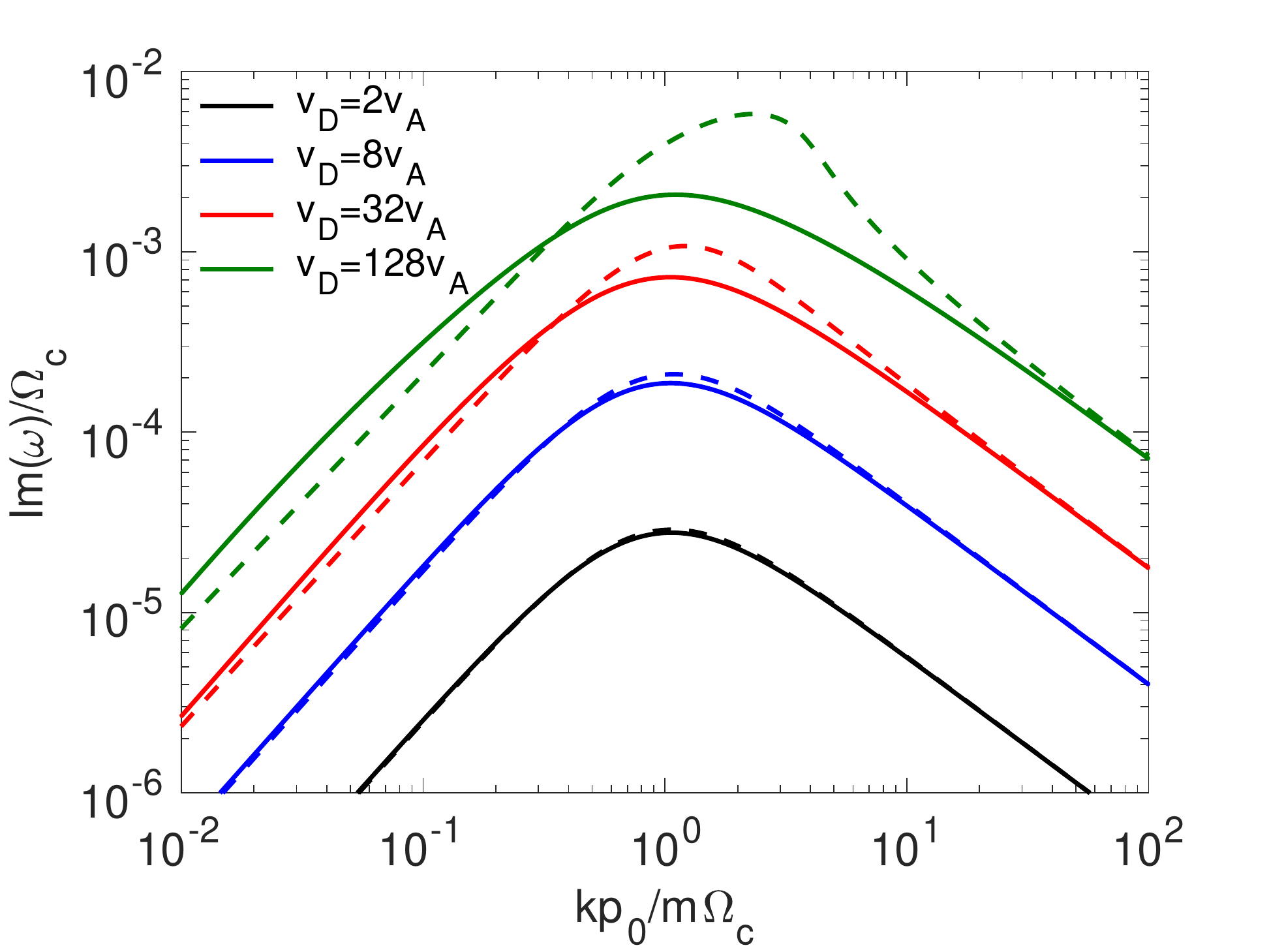}}
  \caption{
  Linear dispersion relation of the CRSI for a $\kappa$ distribution
  (\ref{eq:kappadist}) with $\kappa=1.25$, with fiducial parameters
  $n_{\rm CR}/n_i=10^{-4}$, $v_D=2v_A$, $p_0/m=300v_A$. 
  The top panel varies $n_{\rm CR}/n_i$, while the bottom panel
  varies $v_D/v_A$, as indicated in the legends. The growth rates for
  left (solid) and right (dashed) handed modes are shown separately.
  }\label{fig:disp_kappa}
\end{figure}

\subsection[]{The $\kappa$ distribution}

While the CRSI is commonly analyzed with $f_0$ being a simple truncated power law
(\ref{eq:pldist}), this is incompatible with the $\delta f$ method. The reason is that
the $\delta f$ method
requires $f_0$ to be finite at all $p$ since it serves as the normalization factor in the
weighting function, Equation (\ref{eq:dfweight}).
To avoid divergence, we modify $f_0$ to be a $\kappa$ distribution (e.g., see
\citealp{SummersThorne91} and references therein for motivation)
\begin{equation}
f_0(p)=\frac{n_{\rm CR}}{(\pi\kappa p_0^2)^{3/2}}\frac{\Gamma(\kappa+1)}{\Gamma(\kappa-\frac{1}{2})}
\bigg[1+\frac{1}{\kappa}\bigg(\frac{p}{p_0}\bigg)^2\bigg]^{-(\kappa+1)}\ ,\label{eq:kappadist}
\end{equation}
which has the property that $f(p)$ is a constant when $p\ll p_0$,
and $f(p)\propto p^{-2(\kappa+1)}$ as $p\gg p_0$.

The dispersion relation remains the same except with factors of $Q_1(k)$ and $Q_2(k)$ in
Equation (\ref{eq:Q1Q2}) replaced by
\begin{equation}
\begin{split}
Q_1
=&\frac{\Gamma(\kappa+1)}{s_0^3(\pi\kappa)^{3/2}\Gamma(\kappa-\frac{1}{2})}\int_0^{\infty}
\frac{2\pi s\cdot ds}{[1+(s/s_0)^2/\kappa]^{\kappa+1}}\log\bigg|\frac{1+s}{1-s}\bigg|\ ,
\end{split}
\end{equation}
\begin{equation}
Q_2
=\frac{\sqrt{\pi}}{\kappa^{3/2}}\frac{\Gamma(\kappa+1)}{\Gamma(\kappa-\frac{1}{2})}
\frac{1}{s_0[1+1/(\kappa s_0^2)]^\kappa}\ ,\label{eq:Q2kappa}
\end{equation}
where the dependence on $k$ is expressed in 
$s_0\equiv p_0/p_{\rm res}=kp_0/m\Omega_c$.
The $Q_1$ integral can be evaluated numerically, and in the limit $n_{\rm CR}/n_i\ll1$,
the growth rate can be obtained analytically by substituting (\ref{eq:Q2kappa}) into
(\ref{eq:growth}).

In Figure \ref{fig:disp_kappa} (top panel), we show the linear growth rate in the same way as in Figure
\ref{fig:disp}, but for $f_0$ being a $\kappa$ distribution. 
We see that in the long wavelength limit ($p_{\rm res}\gg p_0$ or $s_0\ll1$), we have
$Q_2\propto s_0^{2\kappa-1}$. In the short wavelength limit, we have
$Q_2\propto s_0^{-1}$. The results are consistent with a power law distribution
function in the same limits, with $\alpha=2(\kappa+1)$. The growth rate varies smoothly
in between the two limits, with peak growth rate only slightly reduced compared to the
truncated power-law case. It is this analytical growth rate that we aim to test with further study.

We further show in the bottom panel of Figure \ref{fig:disp_kappa} the analytical growth rate
for different drift velocities at fixed $n_{\rm CR}/n_i=10^{-4}$. Increasing the
drift velocity has a similar effect as increasing $n_{\rm CR}/n_i$, which increases the overall
growth rate (roughly linearly). It also makes the right-handed mode stand out, which becomes
more notable for $v_D\gtrsim10v_A$. In this paper, we restrict our explorations to $v_D\lesssim8v_A$.

\subsection[]{Simulation Box Size and Phase Randomization}\label{ssec:phrand}

Even equipped with the $\delta f$ methods and a $\kappa$ distribution function, 
it is difficult to accurately recover the linear growth of the CRSI with a numerical simulation.
Over a large suite of tests with wide range of numerical setups, we have found that sustained growth of the
CRSI at the rate predicted by theory 
is only achieved when the simulation box size is extremely long.
Otherwise, the desired growth rates are found only at very early stages, and they
become erroneous shortly afterwards (we demonstrate this in Appendix \ref{app:boxsize}).

Further adjusting the numerical speed of light ${\mathbb C}$ reveals that 
in order to properly recover growth rates,
the required simulation box length $L$ should be such that the time for the
CR particles to traverse the simulation domain, $t_{\rm cross}=L/(\mu{\mathbb C})$, exceeds the growth timescale for CRSI, $1/\Gamma_{\rm max}$.  
 Given that
$\Gamma_{\rm max}\sim (n_{\rm CR}/n_i)\Omega_c$, and using Equation \ref{eq:kres}, 
this may be further translated to
\begin{equation}\label{eq:Lmin}
L>L_{\rm min}\equiv\frac{{\mathbb C}\mu}{\Gamma_{\rm max}}
\sim \frac{n_i}{n_{\rm CR}}\frac{k_{\rm res}^{-1}}{\gamma}.
\end{equation}

This phenomenon is related to the validity of the random phase approximation for QLD
discussed at the beginning of Section \ref{sec:method}, which feeds instability growth.
For CRSI, CR particles
should constantly experience different wave packets throughout the growth process.
With wave growth, we may regard the wave packets to be altered over the timescale of
$\sim\Gamma_{\rm max}^{-1}$. With the fastest particles traveling at $\mathbb C$ along
the background field, a simulation box longer than $L_{\rm min}$ guarantees all particles
experience ``new" wave packets after traversing the entire box and reenter from the
other side.

However, for reasonable simulation values of $n_i/n_{CR}$, the required $L_{\rm min}$ could easily reach a few
hundreds or thousands times the resonant/most-unstable wavelengths 
(using $n_i/n_{CR}\sim 10^9$ as in the bulk ISM, the number could be billions).
The condition of $L\gtrsim L_{\rm min}$ thus would impose a prohibitive requirement for
simulating the CRSI, even in 1D.

We resolve this issue by realizing that the effect of long simulation box can be
equivalently achieved by directly randomizing particle phases when a particle crosses
the periodic boundary: both approaches allow particles to effectively see different wave
packets at all times to conform to the random phase approximation.
In our simulations, upon boundary crossing, we fix a particle's total velocity, as well as
its parallel velocity along background magnetic field, whereas we randomize its gyro-phase.
By this simple approach, we completely eliminate the requirement of $L>L_{\rm min}$.
Also, under the $\delta f$ method, the phase randomization procedure does not
introduce more errors to momentum/energy conservation.

\section[]{Simulation Setup and Diagnostics}\label{sec:setup}


We set up the simulations in the frame of the ``unperturbed'' CRs, where their background
distribution function is isotropic $f_0({\mb p})=f_0(p)$. This is also the $f_0$ for the
$\delta f$ method, with background CR current ${\mb J}_{{\rm CR}, 0}=0$.
We set background density to $\rho_0=1$ and pressure $P_0=0.6$, with an ideal gas
equation of state with adiabatic index $\Gamma=5/3$ (so that sound speed
$c_s=\sqrt{\Gamma P_0/\rho_0}=1$). While thermodynamics has no effect to the
system, we adopt adiabatic rather than isothermal equation of state to better characterize
energy conservation of the system (see Section \ref{sssec:satleveel}).
In 1D (along the $\hat{x}$ axis), the background magnetic field is set to
$B_0=1$ along $\hat{x}$, so that the Alfv\'en speed $v_A=B_0/\sqrt{\rho_0}=1$.

For all simulations, we initialize the background gas to a velocity ${\mb v}_0=-v_D\hat{x}$.  
Since we are working in the frame where the CRs are initially isotropic, this means that the
CR distribution streams to the right (along the direction of the mean magnetic field) with
speed $v_D$ relative to the gas. 
On top of the background state, we initialize the system with a spectrum of Alfv\'en
waves propagating along background ${\mb B}_0$. For each wavenumber $|k|$,
there are four modes corresponding to forward and backward propagation, and left
and right polarization. The wavenumber coverage ranges from $k=2\pi/L_x$ to
$k=2\pi/(2\Delta x)$ (except that the longest wave is initiated with 0 amplitude),
where $L_x=N_x\Delta x$ is the simulation domain size, and
$N_x$ and $\Delta x$ are the number of grid cells and cell size in $\hat{x}$.
Correspondingly, a total of $4(N_x/2-1)$ modes are initialized. We choose the wave
amplitudes to be such that
\begin{equation}\label{eq:initI}
I(k)=A^2/k\ ,
\end{equation}
so that wave energy is equally distributed in logarithmic $k$-space, and the total
wave energy integrated over one dex in $k$ is a fraction $2A^2\ln{10}$ of the
background field energy 
(cf. Equation \ref{eq:wavespec};  the factor of two accounts for velocity perturbations).
Each mode is initialized with a random phase.

We choose code units such that individual CR particles  have cyclotron frequency
$\Omega_c=1$ in the background field $B_0$. This corresponds to the factor $q/mc=1$ in
Equation (\ref{eq:CRmotion}). Thus, our code unit of length is $d_i = v_A/\Omega_c=1$.
We note that  while irrelevant to MHD, $d_i$ is equivalent to the ion skin depth
$c/\omega_{pi}$ for the background plasma. Our MHD-PIC framework typically applies to
scales greater than $d_i$.

The CR particles are injected with $f_0(p)$ being a $\kappa$ distribution.
We choose $\kappa=1.25$ throughout this work, corresponding to $f_0(p)\propto p^{-4.5}$,
at large $p$ or $\alpha=0.5$ in Equation (\ref{eq:pldist}).
In practice, we divide the momentum space into 8 bins ranging from $0.01p_0$ to
$100p_0$, with half a dex per bin. We inject equal number ($N_p$) of CR particles
per cell per bin so as to guarantee sufficient number of particles in each bin
(otherwise there would be too few particles towards lower and higher energies).
Within each bin in each cell, particles are sampled according to $f_0(p)$ in that
momentum range, and are symmetrically distributed in pitch angle and phase.
More specifically, once a set of random momentum $p$, pitch
angle $\theta$ and gyro-phase $\phi$ is generated, four particles are injected at
cell center whose momenta are
$(p\cos\theta, p\sin\theta\cos\phi, p\sin\theta\sin\phi)$ and its permutations with
$\theta\rightarrow\pi-\theta$ and $\phi\rightarrow\pi+\phi$.

We note that particles with $p<0.01p_0$ and $p>100p_0$ are not included in the
simulation whereas we still use a background distribution function $f_0$ that covers
the entire momentum space. This will introduce some inconsistencies. However,
since these momenta are sufficiently far from the peak of $f_0$, waves that
they drive are typically beyond what can be accommodated in our simulation box
(and moreover the simulation time is typically not long enough for these modes to grow).

\subsection{Choice of Parameters}

The interstellar medium in our Galaxy spans a wide range of conditions in various
phases. In a averaged sense, $n_{\rm CR}/n_i$ is on the order of $\sim10^{-9}$,
and $v_A/c\sim10^{-5}-10^{-4}$, and the bulk of the energy  is in GeV CR particles that are mildly
relativistic. While the parameters that we adopt are by no means fully realistic 
(this would be unachievable), we aim to
achieve as much scale separation as possible to mimic conditions in the large-scale ISM.

We choose the artificial speed of light ${\mathbb C}/v_A=300$, and
choose the characteristic momentum 
in the CR distribution to be 
$p_0/v_A=300={\mathbb C}/v_A$.  
With this choice, Equation (\ref{eq:Lorentz}) yields $\gamma_0 = \sqrt{2}$ so as to mimic the fact that the dominant CR population
in the ISM is trans-relativistic $\sim$GeV particles.

The fact that ${\mathbb C}=300$ is larger than $v_A=1$ by more than two orders of
magnitude allows for sufficient separation, making the transformation between
different reference frames well in the non-relativistic limit (e.g., a drift velocity
as large as $\sim10v_A$ only amounts to about $3\%$ of ${\mathbb C}$).
With our choices, the wavelength for most unstable mode is given by
$\lambda_m\approx 2\pi/k_0=2\pi p_0/\Omega_c\approx1885$ in code units.

Based on Equation \ref{eq:growth} for the growth rate of CRSI, the two main physical parameters are the CR number density ratio $n_{\rm CR}/n_i$,
and the CR drift velocity $v_D/v_A$. As already seen in Figure \ref{fig:disp_kappa}, given
our choices of ${\mathbb C}$ and $p_0$, the density ratio $n_{\rm CR}/n_i$ has to
be well below $10^{-3}$ to make left and right handed modes grow at about equal
rates (as in standard CRSI). We thus choose $n_{\rm CR}/n_i=10^{-4}$ as fiducial,
and vary it within $[10^{-5}, 10^{-3}]$. Note that further smaller values would make the
waves grow too little within a reasonable computational time.
We take $v_D=2v_A$ for our fiducial model, and also consider higher drift velocities up to
$8v_A$.\footnote{In terms of energy density, we have $E_{\rm CR}/E_B\sim45$ under the fiducial
parameters,
which is much larger than the order-unity value that applies through most of the ISM.
We note, however, that while the total energy density of CRs matters for large-scale ISM
dynamics, 
it is not directly relevant for the CRSI because the free energy of the CRSI comes from
CR anisotropy (i.e., streaming) rather than total CR kinetic energy. 
Nonetheless, with  $E_{\rm CR}/E_B \sim (n_{\rm CR}/n_i)({\mathbb C}/v_A)^2$, our run
with $n_{\rm CR}/n_i=10^{-5}$ better approaches the realistic limit. One can also
reduce ${\mathbb C}/v_A$ to 100 or smaller to reduce this ratio, although
this would limit the range of $v_D$ that can be more reliably covered by simulations.}

The remainder are numerical parameters. Fiducially, we choose simulation box size
$L/d_i=9.6\times10^4$, resolved by $9600$ cells ($\Delta x/d_i=10$).
The fiducial box 
is about $\sim50$ times the most unstable wavelength, which is more than
sufficient for particles with $p\sim p_0$, but it better accommodates more energetic
particles with $p\gtrsim10p_0$. Note that without implementing phase randomization,
from Equation (\ref{eq:Lmin}) the box size $L_{\rm min}$ would have to be 
$\sim 300$ times longer than our fiducial choice.
We also conduct a simulation with shorter box size for comparison.

The level of noise in the simulation depends on the number of particles. Fiducially, we
choose $N_p=256$ particles per momentum bin per cell (in total $2048$ per cell),
but we also consider a case with fewer particles.  The initial wave amplitude is fiducially
chosen to be $A=10^{-4}$.
Note that $A$ should not be too small, otherwise the initial evolution would be noise
dominated. It should neither be too large, which would trigger artificial QLD of particles
from the beginning. A more detailed discussion about noise in our simulations is
given in Appendix \ref{app:noise}.

In Table \ref{tab:params}, we list all simulation runs presented in this paper. We will
primarily focus on fiducial run Fid and M3 (higher $n_{\rm CR}/n_i$). For other simulations,
we generally vary just one parameter from run Fid at a time, and examine the role of
both physical and numerical parameters. Note that in the simulations, almost all
computational cost is spent on the CR parts given the large number of particles per
cell. For our fiducial run, a typical timestep is $6\times10^{-2}\Omega_c^{-1}$, and
the entire run to $t=10^6\Omega_c^{-1}$ costs about $4\times10^4$ CPU hours on
intel Xeon E5-2690 v4 CPUs.

\begin{figure}
    \centering
    \includegraphics[width=90mm]{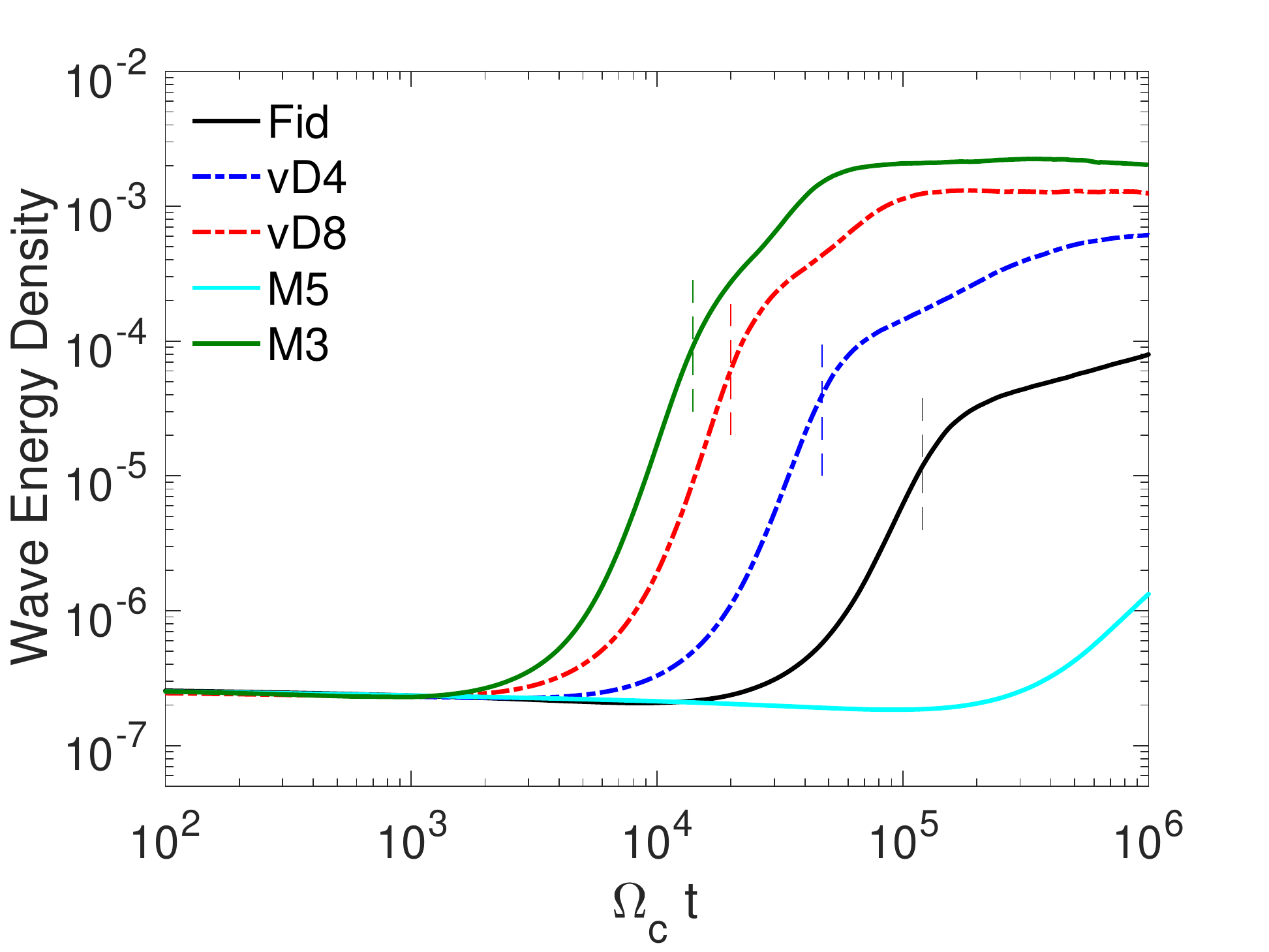}
  \caption{Time evolution of total wave energy in runs Fid (black), M3 (green), M5 (cyan),
  as well as runs vD4 (blue dotted), vD8 (red dotted). All in code units. Vertical dashed lines
  mark the end of the linear growth phase.}\label{fig:Ehst}
\end{figure}

\begin{figure*}
    \centering
    \subfigure{
    \includegraphics[width=87mm]{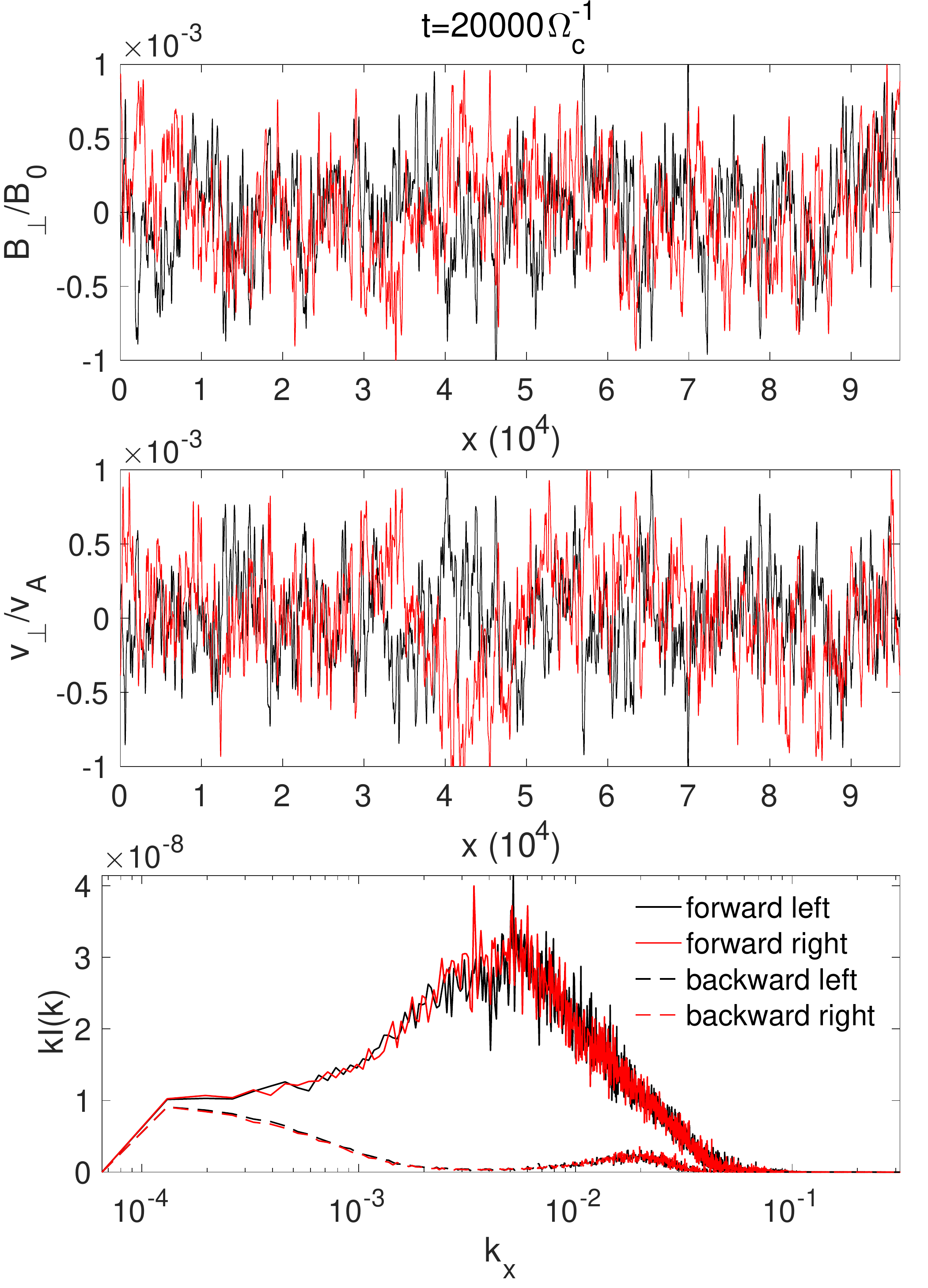}}
    \subfigure{
    \includegraphics[width=87mm]{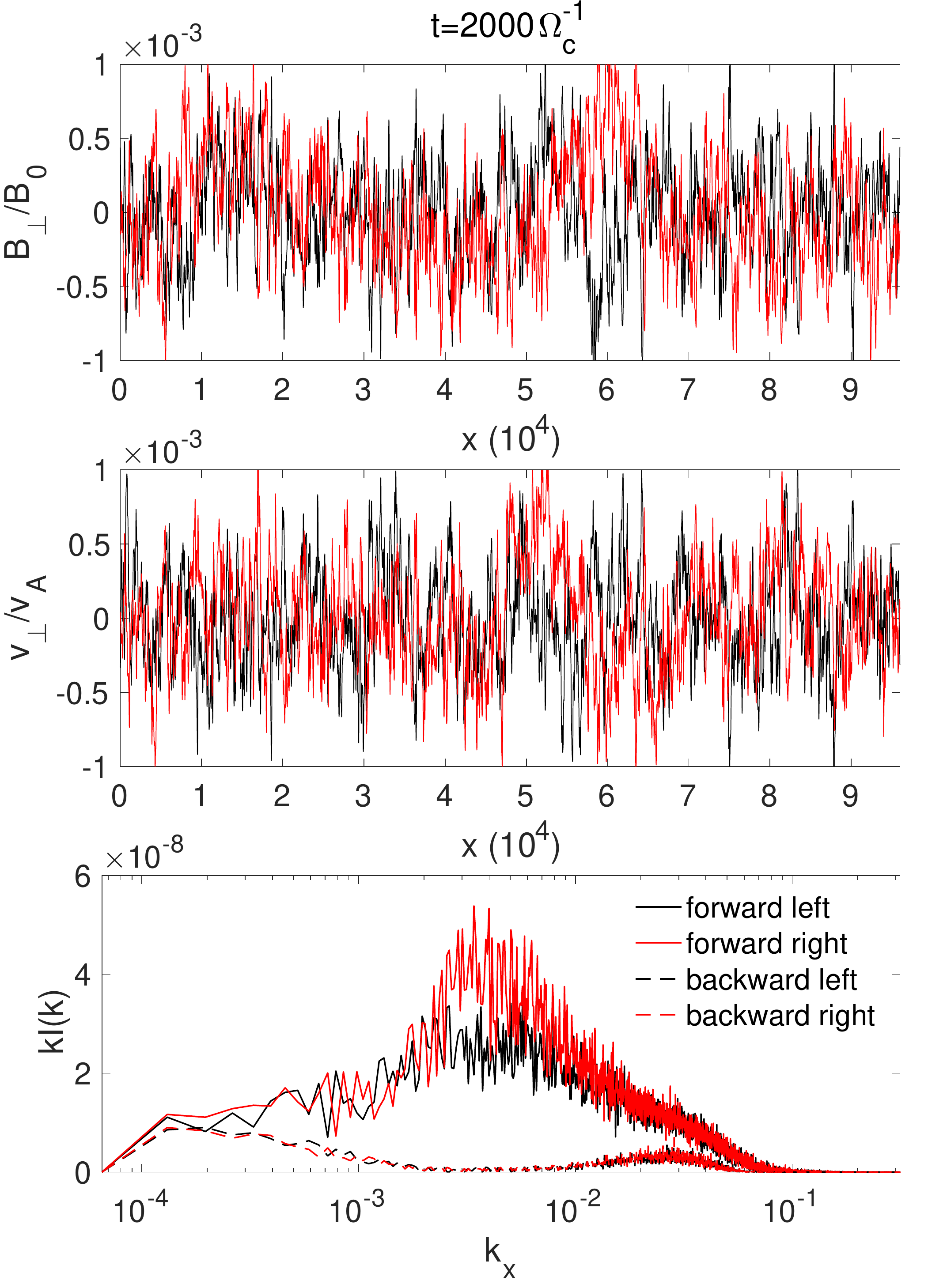}}
  \caption{Snapshot at the linear growth phase in run Fid (left) and M3
  (right). Top and middle panels show the profiles of the perpendicular
  components of magnetic field and velocity (black and red for the $y$ and
  $z$ components), normalized to $B_0$ and $v_A$. The bottom panel
  shows the wave spectra (dimensionless, in $kI(k)$) for forward (solid)
  and backward (dashed) propagating modes. Left/right handed branches
  are marked in black/red. }\label{fig:FidM3Wave}
\end{figure*}

\subsection[]{Diagnostics}

We first measure the linear growth rate of the CRSI. Over constant time intervals,
we decompose the perpendicular magnetic field and velocity profiles into four
Alfv\'en modes and obtain $I(k, t)$ for each mode (see Appendix \ref{app:wavedecomp}).
A linear fit is then performed to $\ln[I(k,t)]$, with the slope being $2\Gamma(k)$.
This is done within early phases of evolution before QLD starts to modify the CR distribution
function. 

Another diagnostic is to directly measure the CR distribution function.
Thanks to the $\delta f$ method, this again can be done with great precision.
In the simulation frame, we simply compute $f(p,\mu)=f_0(p)+\delta f(p,\mu)$,
where $f_0$ is analytic and $\delta f$ can be obtained in the same way binning particles
onto a $p-\mu$ grid using the weighting scheme in Equation (\ref{eq:dfweight}). 

We can further combine the above information aiming to test QLD directly.
This involves transforming $f(p,\mu)$ into the wave frame using Equation (\ref{eq:frametrans}),
and then solving Equation (\ref{eq:qld}), with $I(k)$ feeding into the calculation.
The results can be compared with the measured CR distribution function.

Finally, we have tracked a (small) subset of particles in certain runs over certain
time intervals; we combine this information  with the field data to investigate how particles overcome
the $90^\circ$ barrier.

\section[]{Simulation Results: Run Fid and M3}\label{sec:fid}

In Figure \ref{fig:Ehst}, we show the evolutionary history of total wave energy
for the first six runs listed in Table \ref{tab:params}. 
In this section, we first focus on the two main simulation runs: Fid and M3,
and will also briefly discuss run M5.
With wave amplitude $A=10^{-4}$, total wave energy integrated over the wave
spectrum in $k-$ space, and summed over all four modes, is about
$3.2\times10^{-7}$ (compared to 0.5 for background magnetic field energy
density). Wave energy slowly decreases at first, which is due to 
numerical damping of high-$k$ modes at grid scale, as well as damping of
backward propagating modes. Shortly afterwards, exponential growth of
the CRSI 
becomes clear,
leading to rapid increase of wave energy. This
continues to time $t\sim10^5$ for run Fid and $t\sim10^4$ for run M3,
after which wave growth slows down as QLD starts to substantially modify the particle
distribution function. Subsequent growth is slower, and is characterized by a
combination of saturation of the fastest growing modes excited by low-energy
particles, and the linear growth of longer wavelength modes resonant with
(less abundant) higher-energy particles. Eventually, for run M3, the wave energy
plateaus, marking system saturation. The wave energy density in the
saturated state for run M3 is comparable to the estimates in Equation
(\ref{eq:saturatedE}). However, for run Fid, in which the saturated energy
density should simply be one order of magnitude smaller, this is not yet reached.

Clearly, the wave energy evolution is much better discussed on the
$k$-by-$k$ basis, and should be accompanied by the evolution of particle
distribution function.These will be addressed in the following subsections,
for individual evolutionary phases.

\begin{figure}
    \centering
    \includegraphics[width=90mm]{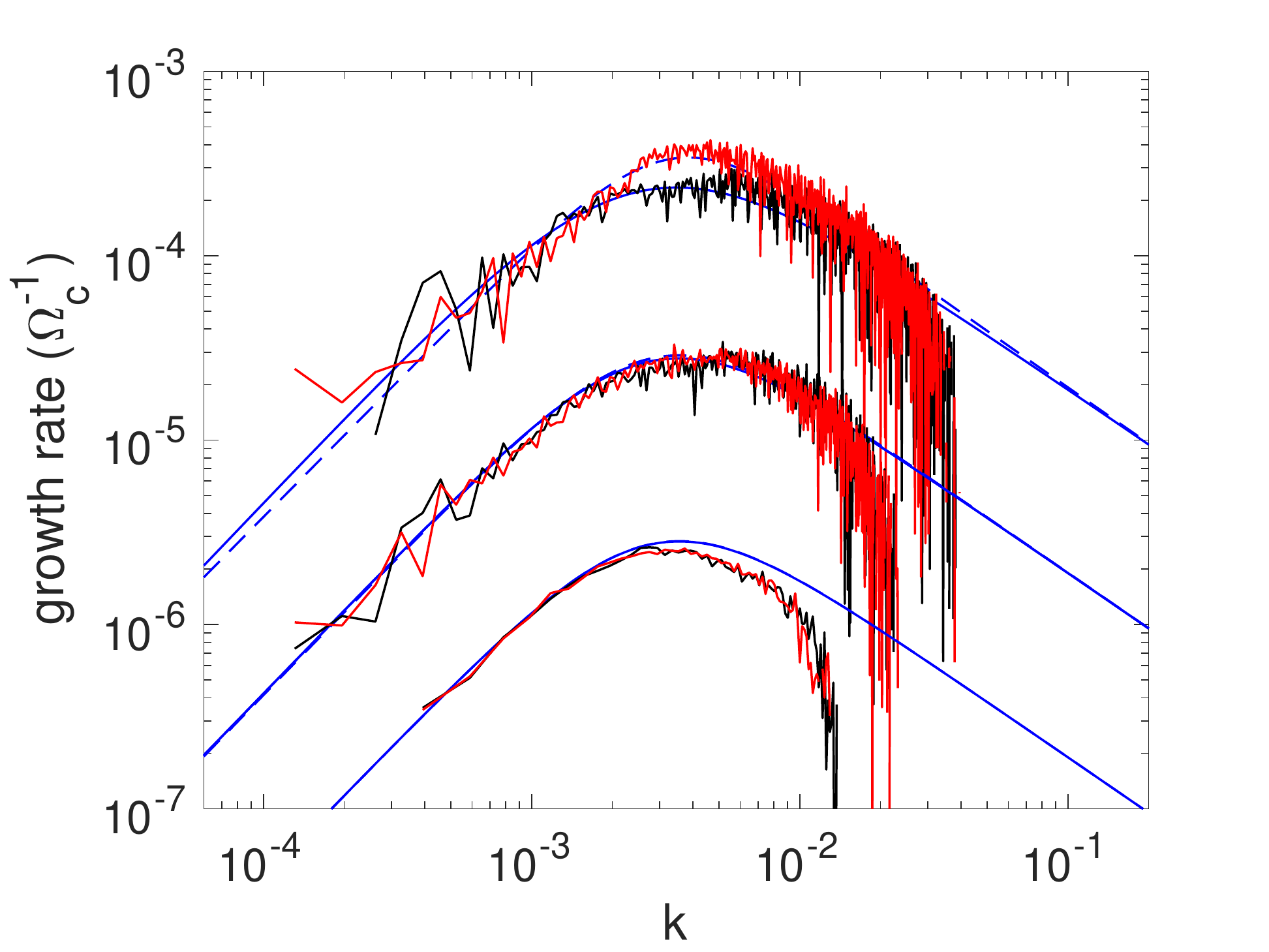}
  \caption{The linear growth rate of the CRSI measured in run M3 (top),
  Fid (middle) and M5 (bottom), where black/red curves mark the left/right
  handed polarization. Blue solid/dashed lines mark the analytical growth
  rate expected from a $\kappa$-distribution for left/right handed modes.}\label{fig:growth}
\end{figure}

\begin{figure*}
    \centering
    \includegraphics[width=180mm]{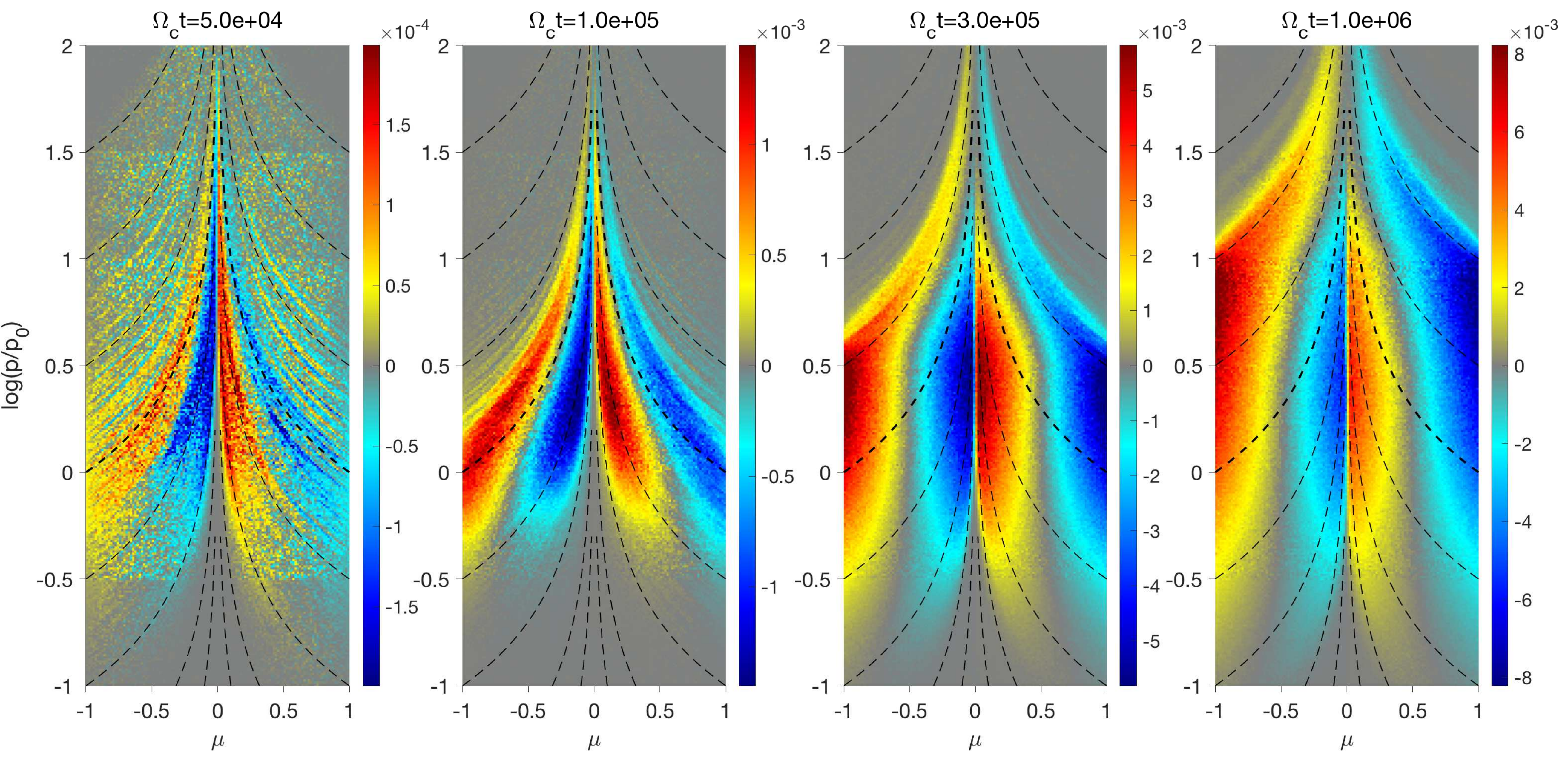}
  \caption{The 2D distribution function $\delta f/f_0$ in the lab (simulation) frame at
  four snapshots in run Fid. The dashed lines are contours in momentum space that
  are resonant with the same wave ($p\mu=$const), with the thick line marking 
  $p\mu=p_0$. Note that color ranges are different among different panels.}\label{fig:dfof_fid}
\end{figure*}

\begin{figure}
    \centering
    \includegraphics[width=85mm]{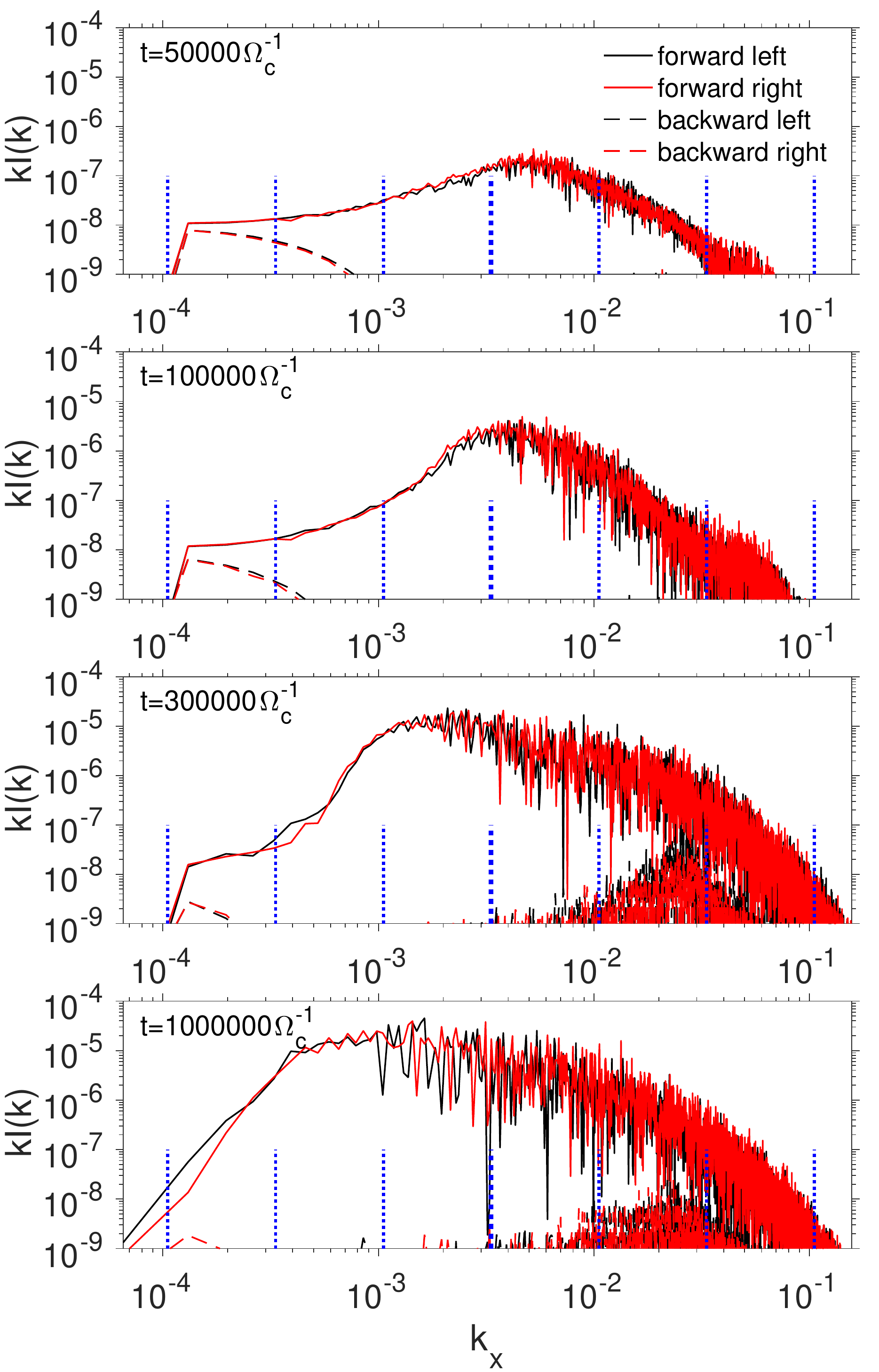}
  \caption{Wave spectrum at different times during QLD in run Fid. The vertical dotted
  lines mark the wave numbers that particles located at the black dashed lines in
  Fig. \ref{fig:dfof_fid} can resonate with.}\label{fig:spectseries_fid}
\end{figure}

\subsection[]{Linear Wave Growth}\label{ssec:lin}

In Figure \ref{fig:FidM3Wave}, we show the Alfv\'en wave spectrum (in wave
intensity $kI(k)$ for the four modes (forward/backward propagation, left/right
polarization) at some early times during the linear phases for the two runs.
Initial wave intensity of all modes has $kI(k)=10^{-8}$ being constant, with
$\delta v/v_A\sim\delta B/B_0$ at the level of a few times $10^{-4}$.
Modes with $k\gtrsim$ a few times $10^{-2}$ are numerically damped.
With very low numerical noise, linear growth of all other forward-propagating
modes is observed from the beginning, whereas backward propagating modes
always damp. The fastest growing mode, resonant with particles near $p=p_0$,
is found to be near $k=k_m\approx3.3\times10^{-3}=\Omega_c/p_0$ as expected.

More quantitatively, we measure the linear growth rate of forward-propagating modes
on the $k$-by-$k$ bases. This is done within $t=3\times10^4\Omega_c^{-1}$ for
run Fid and $t=10^4\Omega_c^{-1}$ for run M3. We also consider run M5, and
measure its growth rate within $t=5\times10^5\Omega_c^{-1}$. The results are
shown in Figure \ref{fig:growth}, and are compared with analytical growth rates.

Overall, even though the modes are very densely populated, the numerical growth rate
matches analytical results remarkably well. For runs Fid and M5, both left and right
handed modes grow at about the same rate, whereas for run M3, right-handed
mode grows slightly faster, all being consistent with theoretical expectations.
We stress that phase randomization is essential in achieving these results,
as is further discussed in Appendix \ref{app:boxsize}.

The growth of the CRSI is better captured towards low $k$, whereas for high $k$,
instability growth competes with wave damping by numerical dissipation.
As a result, the measured growth rates cut off at some $k\sim k_{\rm cut}$, which
decreases from run M3 to run M5 as the instability growth rate decreases by about two
orders of magnitude. In fact, because of the small CRSI growth rate, $k_{\rm cut}$
corresponds to about $20$, $30$, and $45$ grid cells for runs M3, Fid, and M5.
This again indicates much higher numerical resolution is needed to study the 
CRSI 
compared with typical pure MHD instabilities.

Note that for run M3, although the difference in growth rate between left and
right handed modes is small, the resulting difference in wave amplitudes can
be significant after a few e-folding times. This is already evident in the right
bottom panel of Figure \ref{fig:FidM3Wave}: within one e-folding time,
the intensity of the right handed mode is already about twice that of the left
handed mode. This will affect subsequent QLD of particles to be discussed later.

As a side note, we have also conducted simulations with drift speed $v_D$
between $0$ and $v_A$. We find that all wave modes are damped, and
damping is the fastest for at $k\sim k_m$. In the case of $v_D=0$, 
forward/backward propagating modes damp at the same rate, whereas for
$0<v_D<v_A$, damping of forward propagating modes are slower than that
for backward propagating modes. They all agree well with the dispersion relations.

\subsection[]{Quasi-Linear Evolution for run Fid}\label{ssec:qlinFid}

Past the linear stage, QLD starts to modify the particle distribution functions, and
drives the particle distribution towards isotropy in the wave frame. In Figure
\ref{fig:dfof_fid}, we show the time evolution of the particle distribution function
($\delta f/f_0$) via four successive snapshots of the our fiducial run. For each snapshot,
the color scale represents occupation at a given $p$ and $\mu$. We have also
marked with dashed lines the loci in phase space resonating with the
same wavelength
\begin{equation}
p\mu=\Omega_c/k={\rm\ const}\ .
\end{equation}
From Figure \ref{fig:dfof_fid}, it is clear that over time a deficit of particles develops near $\mu \sim 1$ while an excess develops near $\mu \sim -1$.  This change in $\delta f$ reflects the tendency of QLD to scatter CRs towards a new distribution that is isotropic in a frame that is moving to the left with respect to the initial CR distribution.  Figure \ref{fig:dfof_fid} also shows that an excess and deficit of particles develops respectively to the right and left of $\mu=0$; we discuss this further below.

In Figure \ref{fig:spectseries_fid} we show the wave spectrum for the same
snapshots as Figure \ref{fig:dfof_fid}. In these plots, we mark with dashed vertical lines the $k$ values for
the corresponding dashed lines in Figure \ref{fig:dfof_fid}.
Over time, the spectrum of forward-propagating waves grows at all but the lowest and highest $k$, with comparable amplitudes for left- and right- polarizations.

\subsubsection[]{Overall Evolution}\label{sssec:qlinFidql}

At $t\sim3$ to $10\times10^4\Omega_c^{-1}$, QLD has started to make appreciable
changes to the particle distribution function. The initial QLD is the fastest around
particles in resonance with the fastest growing mode at
$k\sim k_m=3.3\times10^{-3}$, as can be clearly identified in the second
panel of Figure \ref{fig:dfof_fid}.\footnote{The initial wave spectrum also
leads to some QLD that slightly affects the appearance of the distribution function (owing
to contributions from high-$k$ modes) at early times $t\lesssim5\times10^4\Omega_c^{-1}$.}
As discussed in Section \ref{ssec:sketch}, the way QLD acts to isotropize the CR
distribution is by scattering forward-traveling CRs to from higher to lower (positive) parallel
velocities, and by scattering backward-traveling CRs from lower to higher (negative)
parallel velocities. This is exactly what we see in Figure \ref{fig:dfof_fid}, with excess
and deficits on the left and right sides of the thick black dashed line.

With some CRs resonant with these waves giving up their free energy through
QLD, the growth of the most unstable mode slows down. 
More slowly growing modes at longer wavelengths have not yet grown to sufficient
amplitudes to yield appreciable QLD to particles in resonance with them at this time,
thus still grow at the same rate. As a result, the peak of the wave spectrum shifts towards
longer wavelength, as seen from the time sequence in Figure \ref{fig:spectseries_fid}.

Such wave growth then leads to further QLD for more energetic particles, as can be
seen in the later two panels of Figure \ref{fig:dfof_fid}. By the end of our simulation
at $t=10^6\Omega_c^{-1}$, all particles with $p\lesssim10 p_0$ have undergone
substantial QLD, whereas for particles with $p>10p_0$, only the ones with smaller
pitch angle have undergone some QLD. 
While the trend should continue over time, we are also limited by the simulation box
size, with corresponds to 50 times the most unstable wavelength. Modes available for scattering
particles with $p>10p_0$ are fairly limited and discretized, precluding following QLD
accurately (the discreteness can already be seen in the last panel of Figure \ref{fig:dfof_fid}).
We thus do not run this simulation for longer.

Across $\mu=0$, there is substantial excess (deficit) of particles with $\mu>0$ ($\mu<0$)
for $p_0\lesssim p\lesssim 10p_0$, which clearly illustrates the $90^\circ$ pitch angle
problem. Evidently, redistribution of particle pitch angle occurs only on either side of
$\mu=0$ but particles do not scatter across $\mu=0$. 
In part, the failure to scatter across $\mu=0$ owes to lack of wave power at large
$k_x$; particles with small $|\mu|$ would only be able to resonate and scatter off of
short-wavelength waves.  In our simulation, $k_x > 0.1$ waves have very low amplitude
due to damping (see Figure \ref{fig:spectseries_fid}; waves at $k_x >  \pi N_x/L_x$ are
not resolved at all). In addition, for run Fid the amplitude $\delta B/B_0$ is
insufficient to trigger reflection near $m=0$ for the bulk CR particles.  We further
discuss the circumstances that lead to ``pre-mature'' saturation in the next subsection.

Particles with $p\lesssim p_0$ have also undergone QLD, which is caused by
shorter-wavelength modes with $k>k_m$. These modes initially grow slower than the
fastest growing mode at $k=k_m$, but they catch up towards later time, and
eventually saturate at similar amplitudes, as can be seen from Figure
\ref{fig:spectseries_fid}. Correspondingly, the evolution in the distribution function
also propagates towards lower $p$ side, as seen in Figures \ref{fig:dfof_fid}.
However, this is eventually limited by numerical dissipation. Given our resolution, we
already see from Figure (\ref{fig:growth}) that modes with $k\gtrsim 6k_m$ grow much
slower than theoretical expectation due to numerical dissipation. Considering scattering
across the entire pitch angle range, and given that even particles with $p\sim p_0$
are subject to pre-mature saturation, we do not discuss particles with $p<p_0$ in general.

We have also examined run M5. The overall evolution is largely similar to run Fid:
the evolutionary stage at time $t$ in run M5 is similar to that at time $t/10$ in run Fid,
which is also visible from Figure \ref{fig:Ehst}. This is reasonable, given the exactly
factor of 10 difference in wave growth rates. QLD starts to modify the CR distribution
function after $t\sim3\times10^5\Omega_c^{-1}$, and by the end of the simulation at $t=1.5\times10^6\Omega_c^{-1}$, QLD is already close to being completed for particles
with $p\sim1-3p_0$, except that they are stuck at $\mu=0$. Given even smaller
wave amplitudes in run M5 than in run Fid, we do not expect further evolution would help
with this $90^\circ$ problem.

\begin{figure*}
    \centering
    \subfigure{
    \includegraphics[width=185mm]{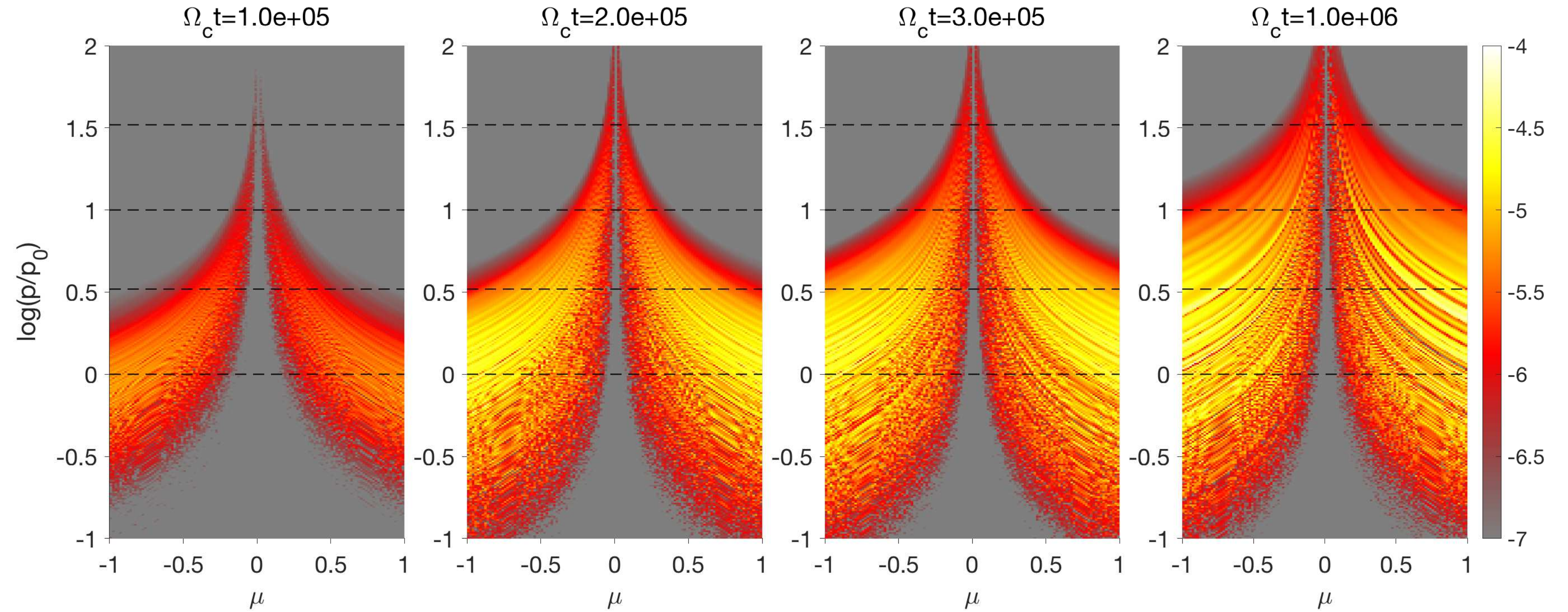}}
    \subfigure{
    \includegraphics[width=185mm]{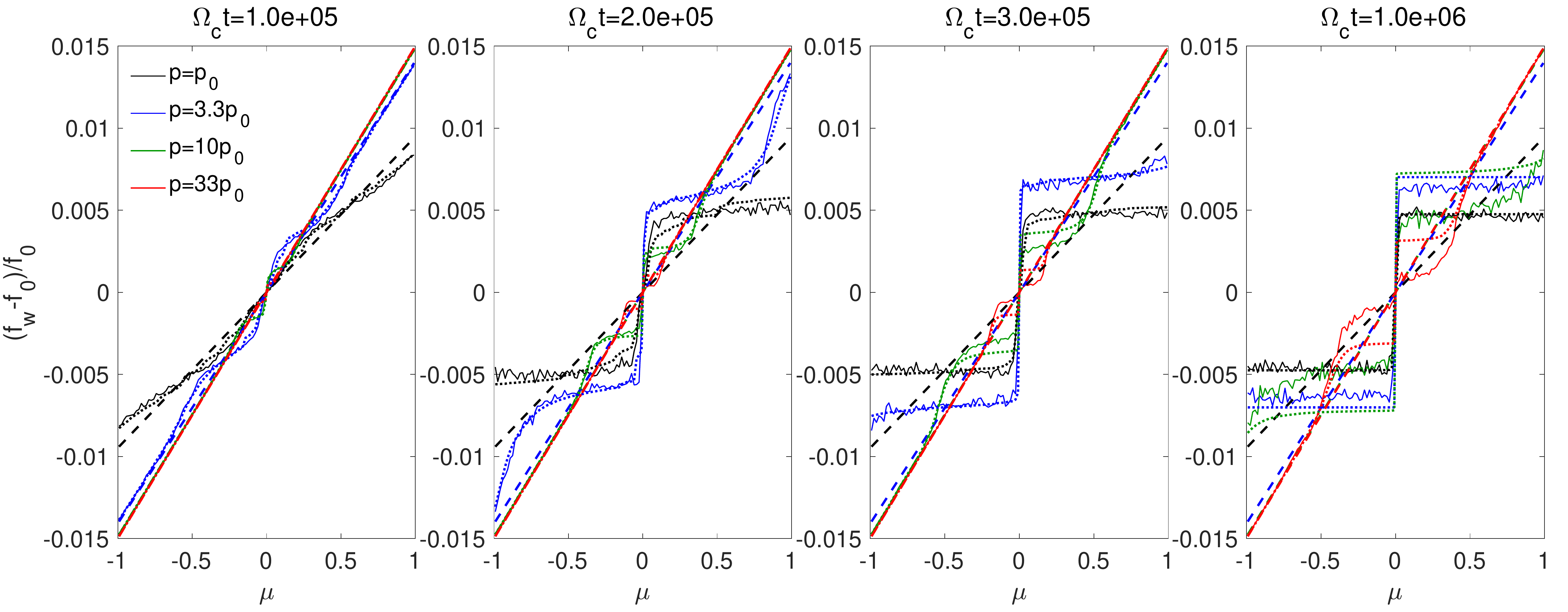}}
  \caption{Top: four sequential snapshots of the scattering frequency $\nu$ of QLD (see
  Equation \ref{eq:nu_qld}) in momentum space in run Fid. Shown in color is
  $\log_{10}(\nu/\Omega_c)$. Bottom: particle distribution function in the wave frame
  $f_w$ at the same four snapshots as in the top panel. Shown are $(f_w-f_0)/f_0$ for four
  representative momenta as a function of pitch angle cosine $\mu$, indicated by different
  colors (see legend). These momenta are also marked as black dashed lines in the top panel.
  Solid lines are measured from the simulations, thick dashed lines mark the initial distribution,
  while the dotted lines are reconstructed distribution functions by solving the QLD Equation
  (\ref{eq:qld}) in time.}\label{fig:qld_fid}
\end{figure*}

\subsubsection[]{Quantitative Analysis}\label{sssec:qlinFidqt}

To see how well our simulation results compare with the quasilinear theory, we
transform the distribution function to the wave frame using Equation (\ref{eq:frametrans}).
For each particle bin $(p_w, \mu_w)$ at a given snapshot, we can find its resonant
wave number $k_{\rm res}$ and its corresponding $I(k_{\rm res})$, and compute the
scattering frequency $\nu$ from Equation (\ref{eq:nu_qld}). Then, we solve the QLD
Equation (\ref{eq:qld}) over time numerically, with $I(k)$ being updated over a time
interval of $\Delta t=500\Omega_c^{-1}$ (which is the frequency of simulation data output).

In Figure \ref{fig:qld_fid}, we show four snapshots in this process. Note that for better
comparison, we have chosen snapshots 1, 3 and 4 to correspond to snapshots 2, 3, 4 in
Figures \ref{fig:dfof_fid} and \ref{fig:spectseries_fid}. The scattering frequency gradually
increases to the level of $\nu_{\rm max}\sim10^{-4}\Omega_c$ as the waves grow, and
its appearance in $p-\mu$ space again shows the characteristic features corresponding
to $p\mu=$const from the resonance condition (modulo that $\nu$ decreases towards
larger $p$ as particles become relativistic). 
With short wavelength modes being damped by numerical dissipation, the scattering
frequency for small values of $|p\mu|$ has a cutoff, leaving a gap around $\mu=0$
that is broader towards smaller $p$. This causes the 90$^\circ$ barrier for QLD, where
no particles can cross $\mu=0$, as is the case when solving Equation (\ref{eq:qld}).

In the bottom panel, we show the particle distribution function in the wave frame.
Initially, $f_w-f_0$ is simply given by Equation (\ref{eq:frametrans}) taking $\delta f=0$,
which is linear in $\mu_w$.
Note that for particles with different momenta, both
$\pa\ln f_0/\pa\ln p$ and $v_w$ can be different, yielding different slopes. With the
$\kappa$ distribution, the slope is smaller for smaller $p$, and approach a constant
$2(\kappa+1)\Delta v/c$ for relativistic particles (thus the red and green dotted lines
overlap). Subsequent evolution has already been discussed in Section \ref{sssec:qlinFidql}
in the simulation frame, which is straightforward to be translated to the wave frame.
Several important results and features are worth noticing.

First, at early time ($t\lesssim2\times10^5\Omega_c^{-1}$), evolution of the numerical distributions follow the QLD prediction 
remarkably well. This gives us confidence that our simulations properly capture QLD,
thanks to the phase randomization technique that we employed.

Second, QLD drives the {\it local} slope of $f_w-f_0$ towards zero 
on either side of $\mu=0$ [see Equations (\ref{eq:dfwdmuw}), (\ref{eq:dfsat})];
piecewise flattening is
achieved at later stages of evolution, reached by lower energy particles first. 
This marks the saturation of the CRSI in the {\it local} momentum space.

Third, particles undergoing QLD tend to get stuck at $\mu=0$. 
Eventually, 
the CRSI saturates because $f_w=$const on both
sides of $\mu=0$ for each $p$, although there is a discontinuous jump at $\mu=0$.
This is the hallmark of pre-mature saturation of QLD.

Fourth, 
solving the QLD equation allows us to well reproduce the saturated state
for lower-energy particles
($p\lesssim3p_0$), but this is not true for higher energy particles ($p\gtrsim10p_0$). 
Given that the only missing ingredient in the QLD equation (\ref{eq:qld}) is the process that governs the crossing of $90^\circ$ pitch angle, this indicates
that the $\mu=0$ crossing does happen (slowly) for higher-energy particles in our simulations.

Given the above results, we now check the requirements for QLD to 
scatter particles to the
critical pitch angle such that the condition for mirror reflection
(\ref{eq:mirrorcond}) can be satisfied. Seen from Figure \ref{fig:Ehst}, we
have the rms $\delta B/B_0\sim9\times10^{-3}$, giving $\mu_{\rm mir}\sim6\times10^{-3}$.
Within our simulation time of $10^6\Omega_c^{-1}$, from Equation (\ref{eq:qld}) we need $kI(k)\gtrsim10^{-6}$
for effective QLD. Reading from Figure \ref{fig:spectseries_fid}, this gives a maximum
wave number $k_{\rm cut}\sim8k_m$. Based on resonance conditions, the
minimum momentum for which QLD can scatter the particle to reach
$\mu=\mu_{\rm mir}$ is then 
\begin{equation}\label{eq:pmin}
p_{\rm min}\sim\frac{k_m}{\mu_{\rm mir}k_{\rm cut}}p_0\sim20p_0\ .
\end{equation}
In other words, waves in our simulation only have sufficient power to scatter
particles with $p\gtrsim p_{\rm min}$ to reach pitch angle $\mu_{\rm mir}$, potentially
allowing the mirror effect to yield reflection (and deviate from pure QLD result).
However,
from the bottom right panel of Figure \ref{fig:qld_fid} representing the
last snapshot of the simulation, we see that even for $p$ as small as $\sim3p_0$, deviation from QLD
can be identified, though just by a very small amount. Reading off the figure, 
deviation from QLD appear to begin for $p_{\rm min}$ between $3p_0$ and $10p_0$, which is in tension with the expectation  from
mirror reflection (Equation \ref{eq:pmin}). Further discussion of this issue is deferred to Section \ref{ssec:90deg},
where a more detailed study based on run M3 is presented.  

\begin{figure*}
    \centering
    \subfigure{
    \includegraphics[width=87mm]{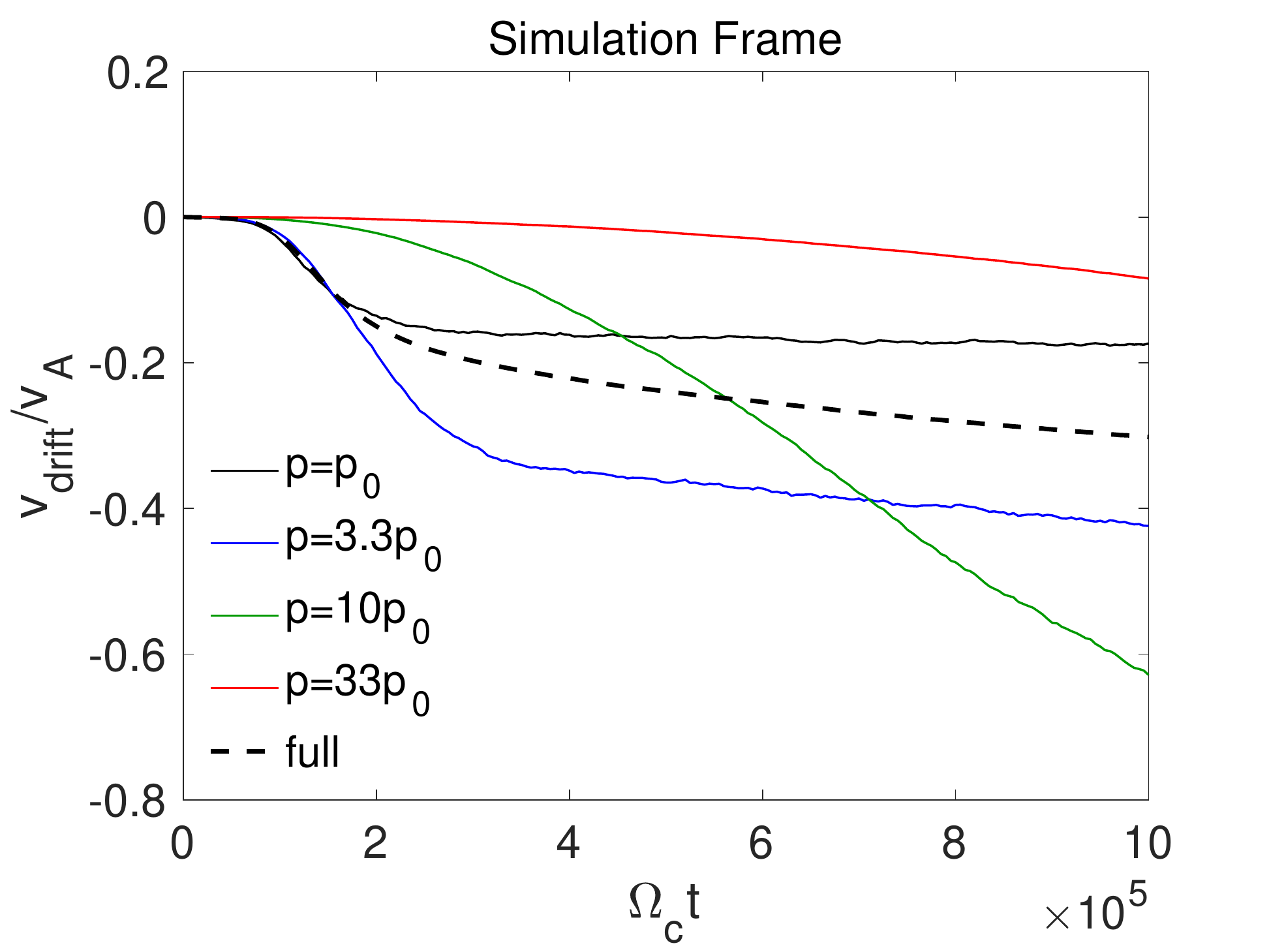}}
    \subfigure{
    \includegraphics[width=87mm]{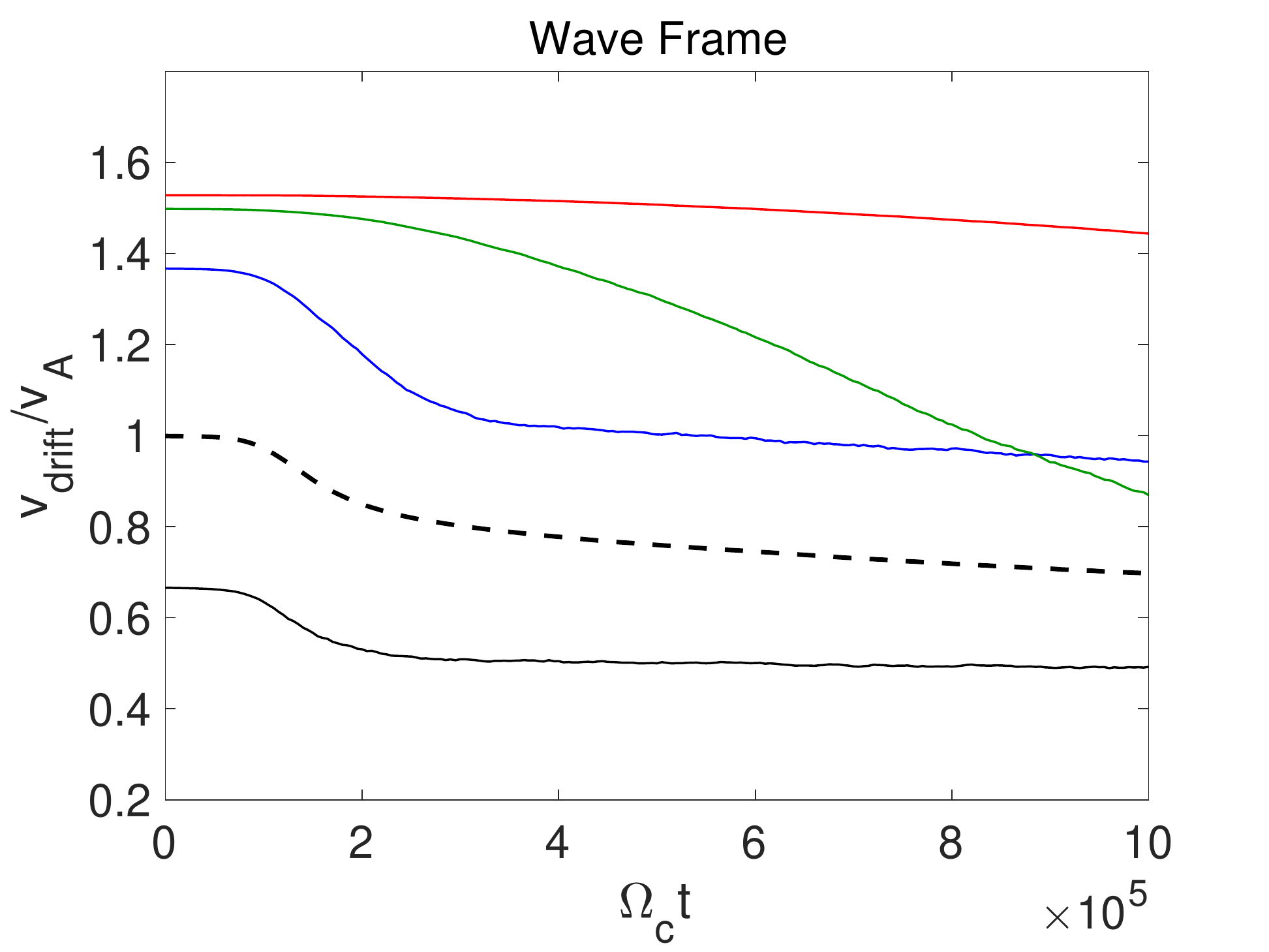}}
  \caption{Time evolution of mean particle drift velocity in different momentum bins, together with the
  full drift velocity (see legend), for run Fid.  We show velocities measured in the simulation frame (left) and wave
  frame (right). Thick dashed line represent the mean drift velocity averaged over the entire
  particle population.}\label{fig:vd_fid}
\end{figure*}

\subsubsection[]{Evolution of CR Drift Velocity}

While it is generally convenient to use the reduction of CR bulk drift velocity to describe how much
CR particles have undergone scattering through the CRSI, this drift velocity is clearly energy-dependent.
We can evaluate the energy-dependent drift velocity $v_d(p)$ (to distinguish from the initial CR bulk
drift velocity $v_D$), in the simulation frame, as 
\begin{equation}
v_d(p)=\frac{\int f(p,\mu)v(p)\mu d\mu}{\int f(p,\mu)d\mu}\ ,
\end{equation}
where $v(p)$ is the velocity that corresponds to momentum $p$, and $f=f_0+\delta f$. Values of $v_d(p)$ can be simply
obtained integrating over a horizontal line in Figure \ref{fig:dfof_fid}. 
We can further integrate the distribution function over $p$, to obtain the full drift velocity $v_{d,{\rm full}}$.

In the left panel of Figure \ref{fig:vd_fid},
we show the results for four different representative particle momenta from $p=p_0$ to $p=33p_0$.
Based on the previous discussion, the results are straightforward to interpret. The initial drift velocity in the simulation
frame is by definition 0. Low-energy CRs, which are responsible for driving faster wave growth, undergo more rapid initial
QLD and a corresponding reduction their drift speed; higher-energy CRs take much longer time to catch up. However, the CRSI and drift reduction of low-energy CRs saturates pre-maturely, leading to only modest reduction of $v_d$, whereas
higher-energy CRs ($p\gtrsim10p_0$) can overcome the $\mu=0$ barrier and are on the way to achieve complete isotropy. The overall average drift $v_{d,{\rm full}}$ decreases over time. Contribution to the overall reduction is mostly from lower-energy particles with momenta between $\sim p_0$ and $\sim3p_0$ at early
times, and from higher-energy particles with $p\sim10p_0$ at late times.

The drift velocity is frame-dependent. Similar to the above procedure, we can measure the drift velocity in the wave
frame $v_{d,w}(p)$, with the results are shown in the right panel of Figure \ref{fig:vd_fid}. While the velocity
difference between the two frames is just $v_A$, initial particle drift speeds within individual momentum bins do
not equal $v_A$. This is because after a frame transformation, particles with a given momentum $p$
come from a range of momenta near $p$ from the original frame, making the results depend on $d\ln f_0/d\ln p$
(see Equation (\ref{eq:frametrans})). Only the full drift speed $v_{d,w,{\rm full}}$ precisely equals $v_A$, as
expected. On the other hand, 
in a fully saturated state, we would expect  $v_{d,w}(p)=0$ for all $p$. 
However, we see clearly from the Figure that low-energy particles are stuck
at nonzero $v_{d,w}(p)$ whereas higher-energy particles require a longer time to
achieve saturation.

\begin{figure*}
    \centering
    \includegraphics[width=172mm]{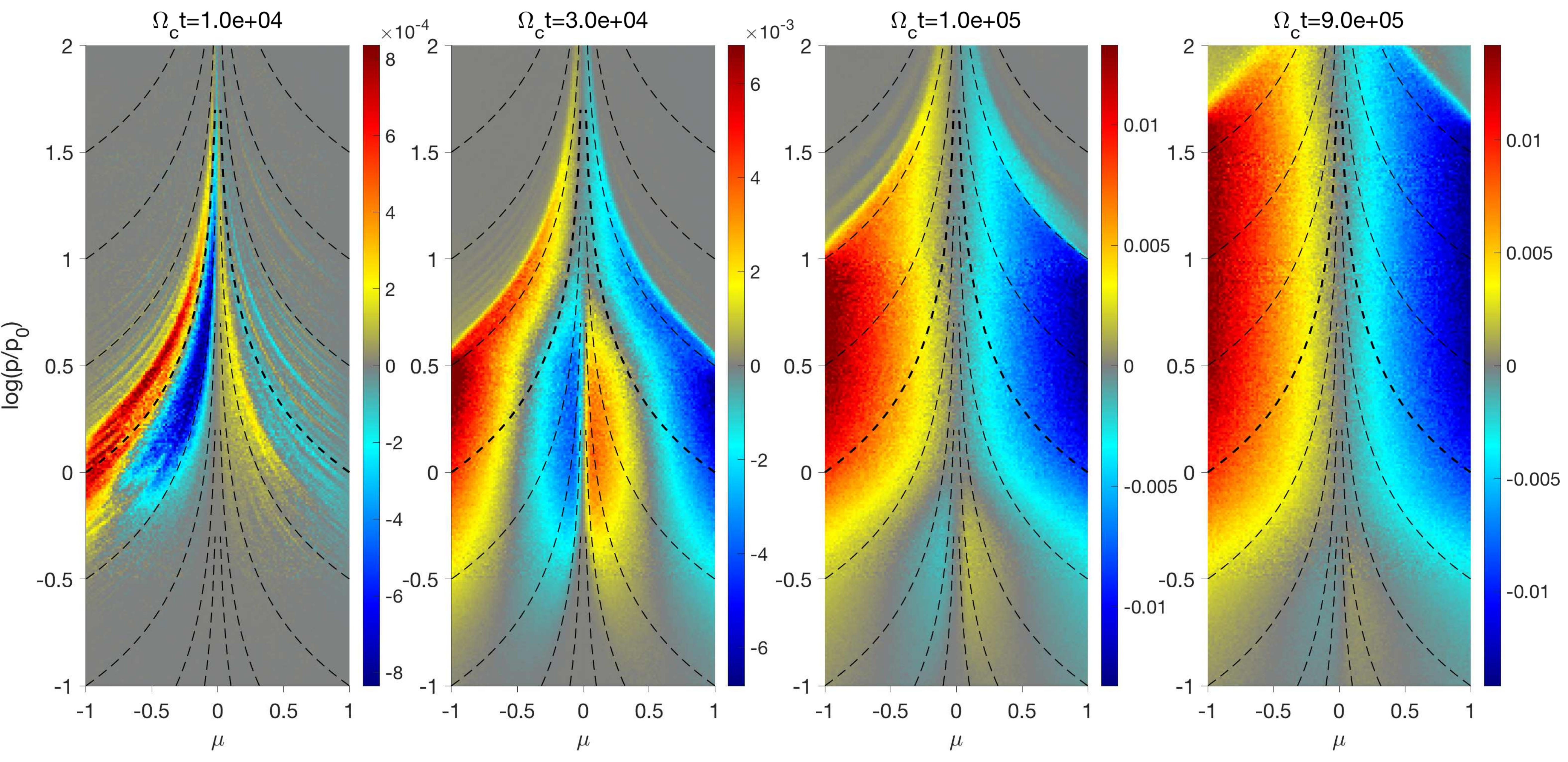}
  \caption{Same as Figure \ref{fig:dfof_fid}, but for run M3, showing the 2D distribution function
  of $\delta f/f_0$ in the simulation frame at four snapshots.}\label{fig:dfof_m3}
\end{figure*}

\begin{figure}
    \centering
    \includegraphics[width=84mm]{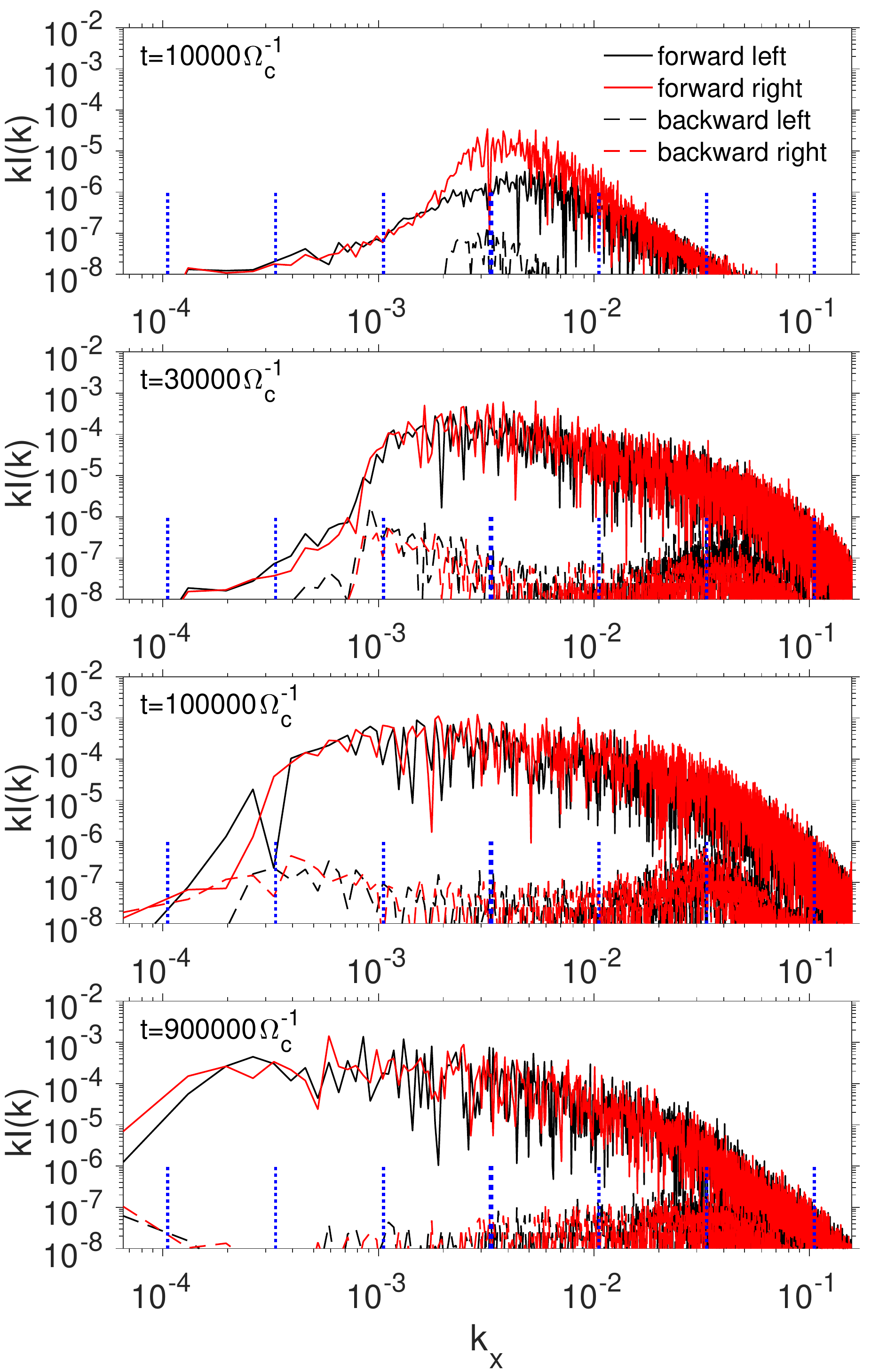}
  \caption{Same as Figure \ref{fig:spectseries_fid}, but for run M3, showing the wave spectrum
  at the same four snapshots as in Figure \ref{fig:dfof_m3}.}\label{fig:spectseries_m3}
\end{figure}

\subsection[]{Quasi-Linear Evolution for run M3}\label{ssec:qlinM3}

With $n_{\rm CR}/n_i$ a factor of 10 larger, run M3 evolves much more rapidly than run Fid. We
choose four representative snapshots at $\Omega_ct=10^4$, $3\times10^4$, $10^5$
and $9\times10^5$ and analyze the results.

\subsubsection[]{Overall Evolution}

In Figure \ref{fig:dfof_m3}, we show the particle distribution function at these snapshots,
while Figure \ref{fig:spectseries_m3} shows the corresponding wave intensity spectra.
As discussed earlier, right-handed modes grow faster near $k\sim k_m$. This continues
to $t\gtrsim10^4\Omega_c^{-1}$. Since these modes are resonant with backward-traveling
CRs, we see that at early time, QLD proceeds much more rapidly for $\mu<0$ particles
resonant with these waves, leading to an initial asymmetry in the particle distribution function.

With asymmetric wave growth and QLD, backward traveling CRs have used up more
free-energy, thus slowing down the growth of right-handed waves. The left-handed modes,
on the other hand, continue to grow normally. This allows left-handed modes to eventually
catch up with the right-handed modes, and we see that by the time
$t=3\times10^4\Omega_c^{-1}$, left and right handed modes at all wavelengths 
have similar amplitudes.  
At the same time, the distribution functions for forward and backward traveling CRs also
tends to become more symmetric about $\mu=0$. However, the $\mu=0$ barrier remains
at this stage for low-energy particles.

In between $\Omega_ct=3\times10^4$ and $10^5$, we see that crossing of the $\mu=0$
barrier occurs, paving the way for full saturation. The release of more free energy has led to
some further wave growth. This process continues until the end of our simulation, with
QLD continuing to extend to higher and higher energies.

\begin{figure*}
    \centering
    \subfigure{
    \includegraphics[width=85mm]{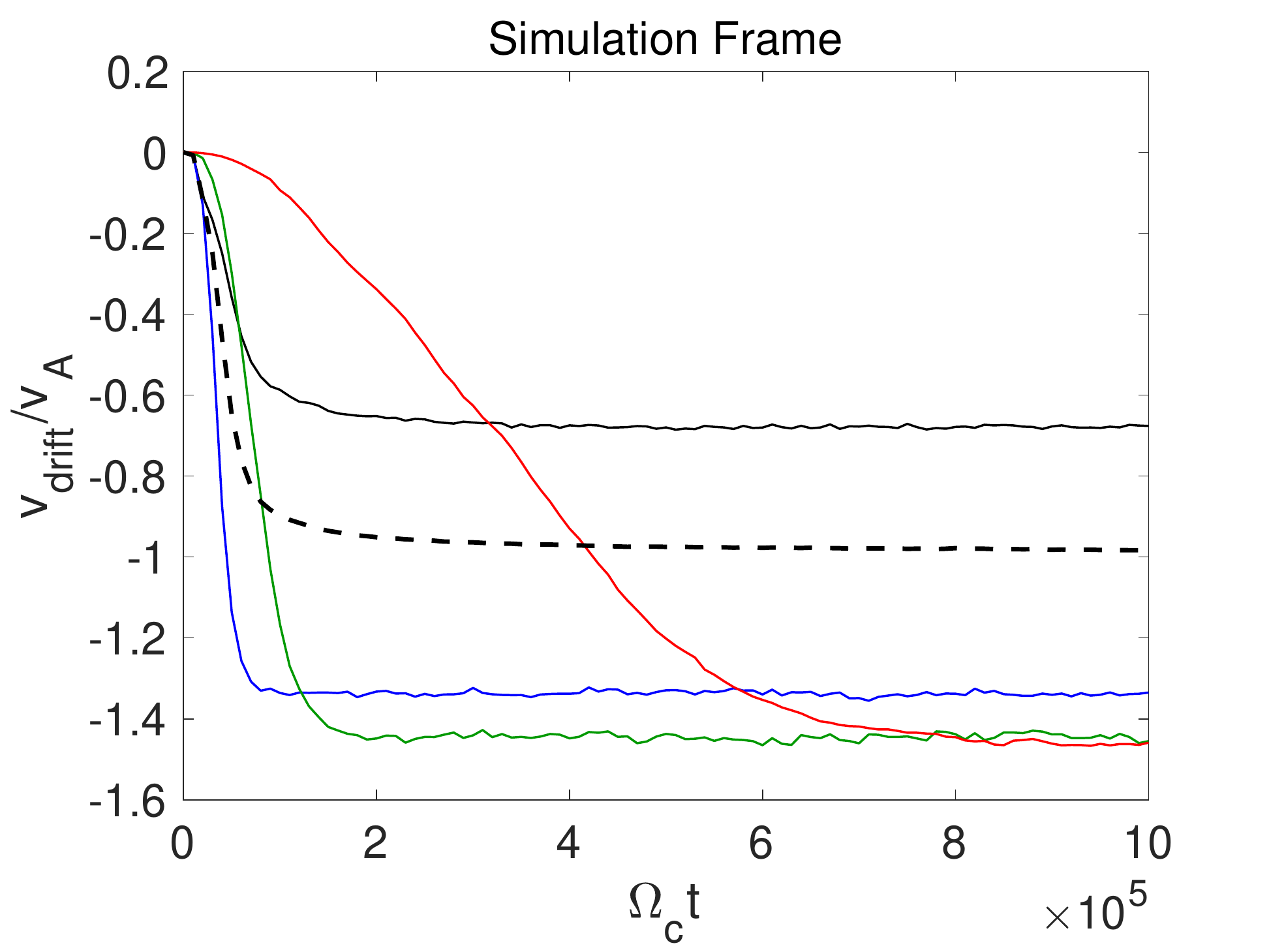}}
    \subfigure{
    \includegraphics[width=85mm]{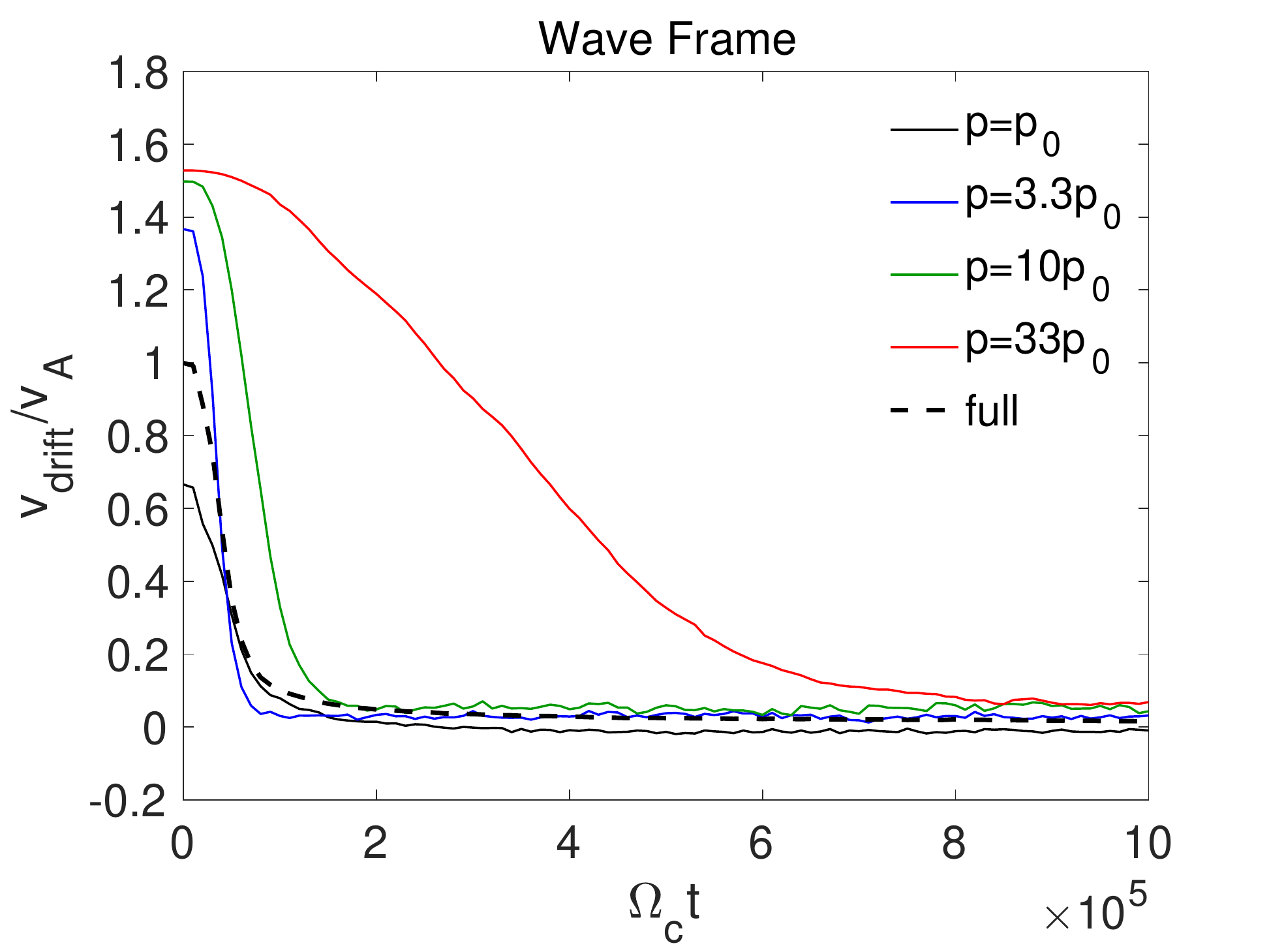}}
  \caption{
Same as Figure \ref{fig:vd_fid}, but for run M3, showing evolution of mean particle drift velocity in different momentum bins, together with the
  full drift velocity (see legend).  
}\label{fig:vd_m3}
\end{figure*}

\begin{figure*}
    \centering
    \includegraphics[width=183mm]{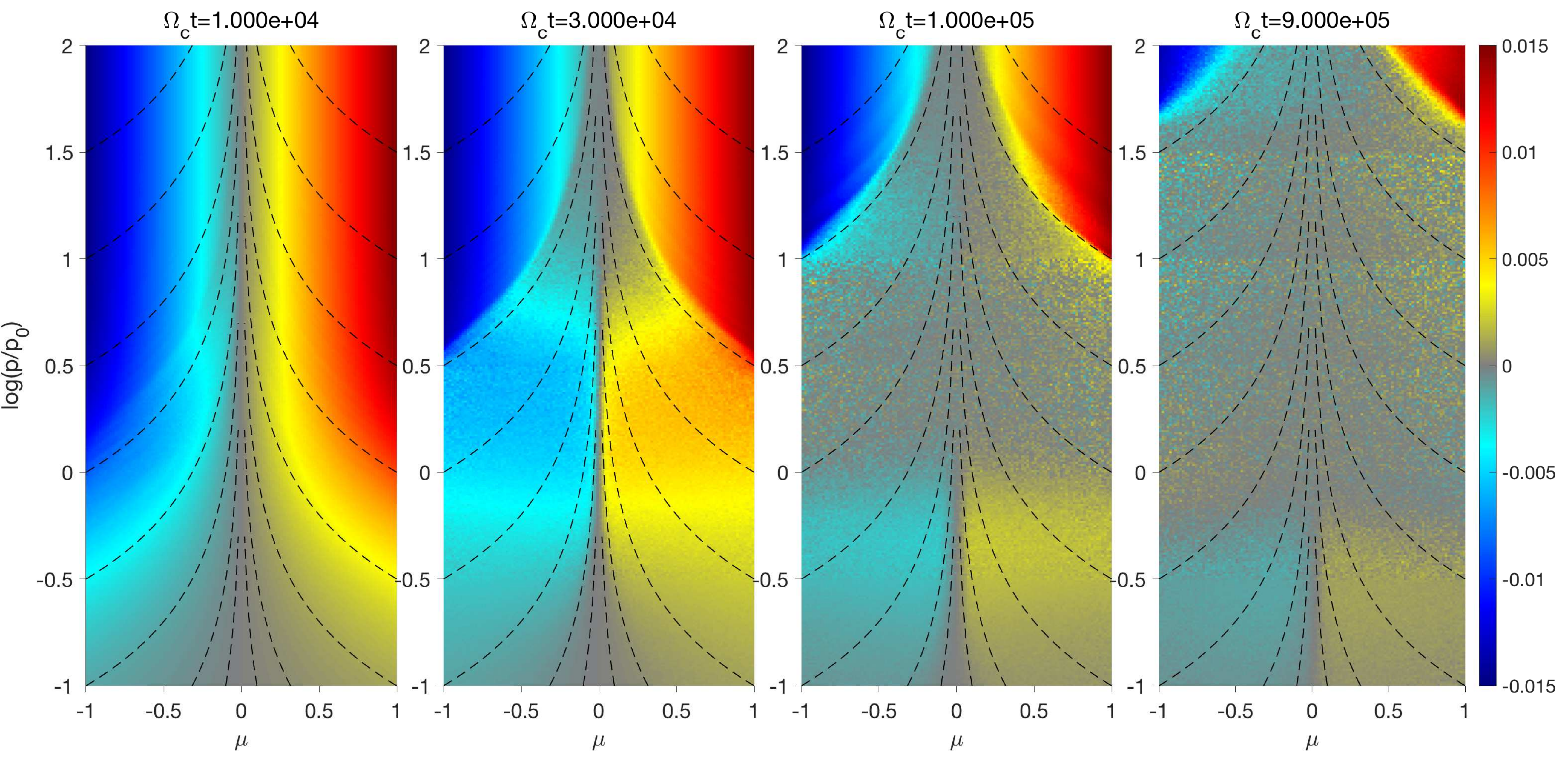}
  \caption{The 2D distribution function $\delta f_w/f_0$ in the wave frame for 
  four snapshots from run M3. The dashed lines are contours in momentum space with
  the same resonance condition.}\label{fig:dfwof_m3}
\end{figure*}

\begin{figure*}
    \centering
    \subfigure{
    \includegraphics[width=185mm]{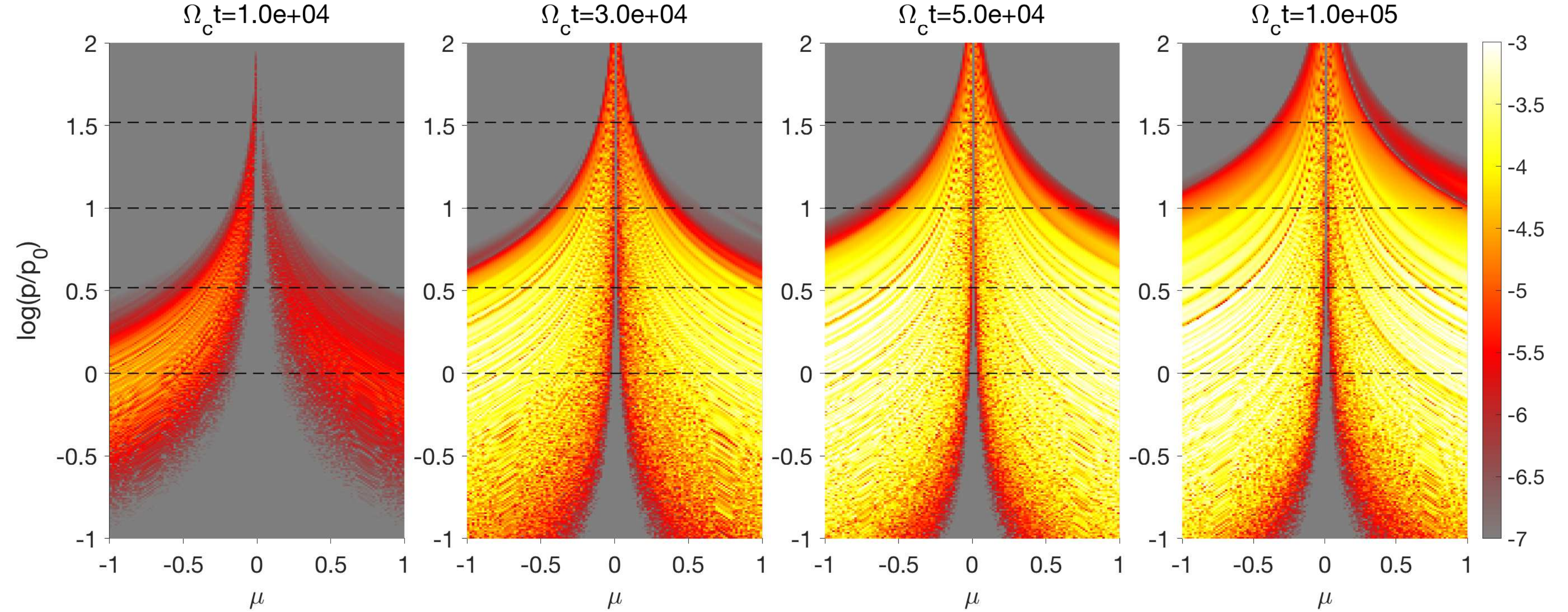}}
    \subfigure{
    \includegraphics[width=185mm]{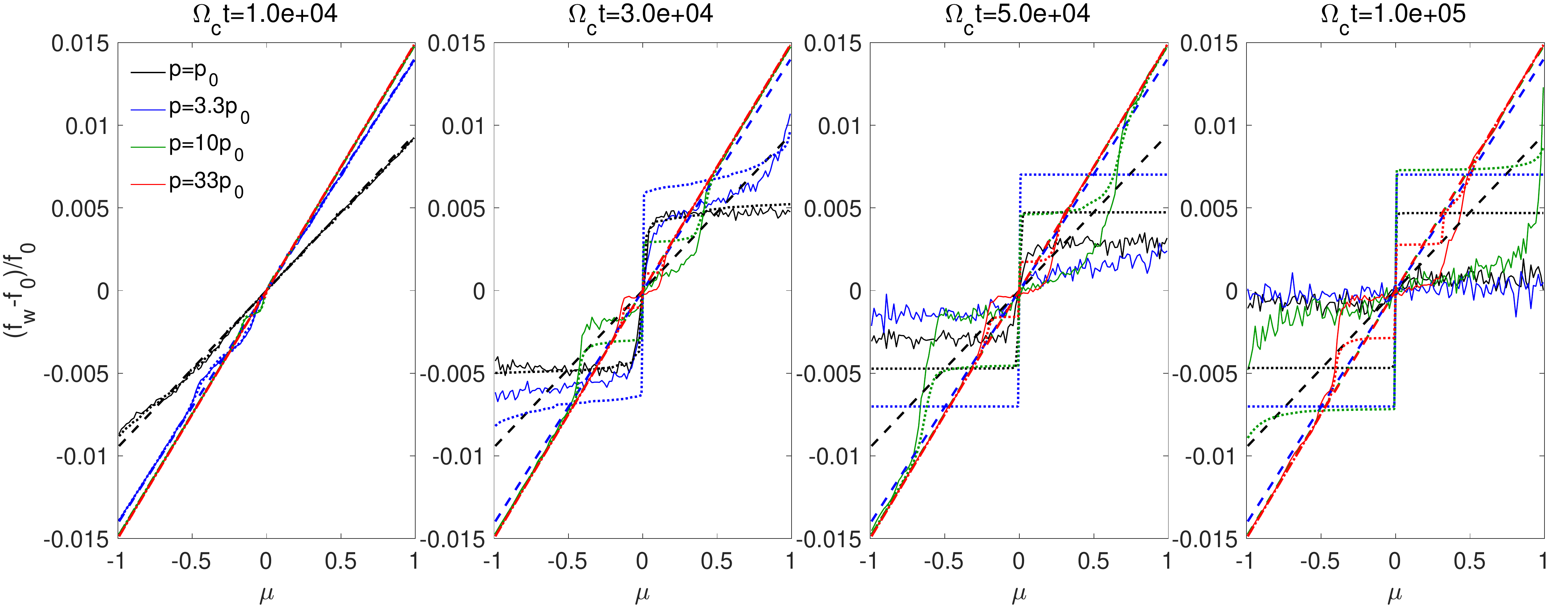}}
  \caption{Same as Figure \ref{fig:qld_fid}, but for run M3 at four different snapshots.}\label{fig:qld_m3}
\end{figure*}

\subsubsection[]{More Detailed Analysis Towards Saturation}\label{sssec:satM3}

As before, for run M3 we compute the CR drift velocity in four different momentum bins for both
the simulation frame and the wave frame, and show the results in Figure \ref{fig:vd_m3}.
We see that in the simulation frame, particles of all four momentum bins eventually flatten
at different levels of $v_d(p)$.  In the wave frame, they all reach $v_d(p)$ equal to zero\footnote{The reason that the
saturated drift velocity in the wave frame is not exactly zero in Figure \ref{fig:vd_m3} is purely
numerical, depending on the resolution employed in binning particle momentum during
simulation output (due to the sensitive dependence of $f_0$ on $p$). Finer binning makes the
results closer to zero but with more noise, and vice versa.}, indicating that they are
fully isotropized and have  achieved complete saturation.

This saturation is more evident from Figure \ref{fig:dfwof_m3}, showing a time sequence of
the full distribution function in the wave frame, with the same snapshots as in previous Figures
\ref{fig:dfof_m3} and \ref{fig:spectseries_m3}. We can clearly identify the isotropization of
particles in the later two snapshots, extending from the range of $p_0\lesssim p\lesssim10p_0$
at $t=10^5\Omega_c^{-1}$ to covering most of the particle population at the end of the simulation. Because of this, and with most of the particle population having momentum close to $p_0$, we see that the full drift speed $v_{d,{\rm full}}$
averaged over all particle population shown is reduced almost exactly
by $v_A$ (Figure \ref{fig:vd_m3}).

For run M3, we have also computed the scattering frequency according to Equation
(\ref{eq:nu_qld}) and solve the QLD equation (\ref{eq:qld}) based on output data, and
show the results at four different snapshots in Figure \ref{fig:qld_m3}. The asymmetry
in scattering of forward/backward traveling CRs at early times ($t\sim10^4\Omega_c^{-1}$)
can be clearly identified in the leftmost upper panel. At later times, we can see by comparing with Figure \ref{fig:dfof_m3}
that the extent to which QLD has proceeded in momentum space is strongly related to the region
where scattering frequency is high. 

The bottom panels of Figure \ref{fig:qld_m3} give more
quantitative information about how quasi-linear evolution has progressed over time.
For lower-energy particles (with $p\sim p_0-3p_0$), evolution proceeds rapidly and can initially
be well reproduced by solving the QLD equation. During subsequent evolution, particle
distribution functions in fixed momentum bins are largely flat (due to rapid QLD) on both sides with a jump at $\mu=0$.
The jump gradually decreases, as particles get scattered across $\mu=0$, which is the
rate-limiting process for particle isotropization.
For higher-energy particles ($p\gtrsim10p_0$), on the other hand, crossing of $\mu=0$ is
more rapid (leading to substantial deviation from directly solving the QLD equation). The
rate-limiting process for isotropization then becomes QLD itself, especially at large $\mu$.

A key result from our run M3 is efficient particle crossing of the $\mu=0$ barrier.
We again first examine whether this is consistent with mirror reflection.
From Figure \ref{fig:Ehst}, we see that by the time of $t=6\times10^4\Omega_c^{-1}$, the
magnetic energy density approaches saturated value ($\delta B^2/B_0^2\sim2\times10^{-3}$),
corresponding to a $\delta B/B_0\sim0.045$. This gives the mirror reflection threshold of
$\mu_{\rm mir}\sim0.032$. Within a simulation time of $\sim10^5\Omega_c^{-1}$, we need
$kI(k)\sim10^{-5}$ for effective QLD. This is achieved for $k$ up to $k_{\rm cut}\sim10k_m$,
by looking at Figure \ref{fig:spectseries_m3} at $t=10^5\Omega_c^{-1}$.
According to the same calculation as in Equation (\ref{eq:pmin}), we find mirror reflection can
be achieved for particles with $p\gtrsim p_{\rm min}=(k_m/k_{\rm cut})\mu_{\rm mir}^{-1}\sim3p_0$.
However, we see that by this time, particles with $p\gtrsim p_0$ are already fully isotropized,
and lower-energy particles have also undergone partial isotropization. Again, we conclude
mirror reflection appears insufficient to explain the detailed evolution of the  momentum distributions.

\begin{figure*}
    \centering
    \subfigure{
    \includegraphics[width=86mm]{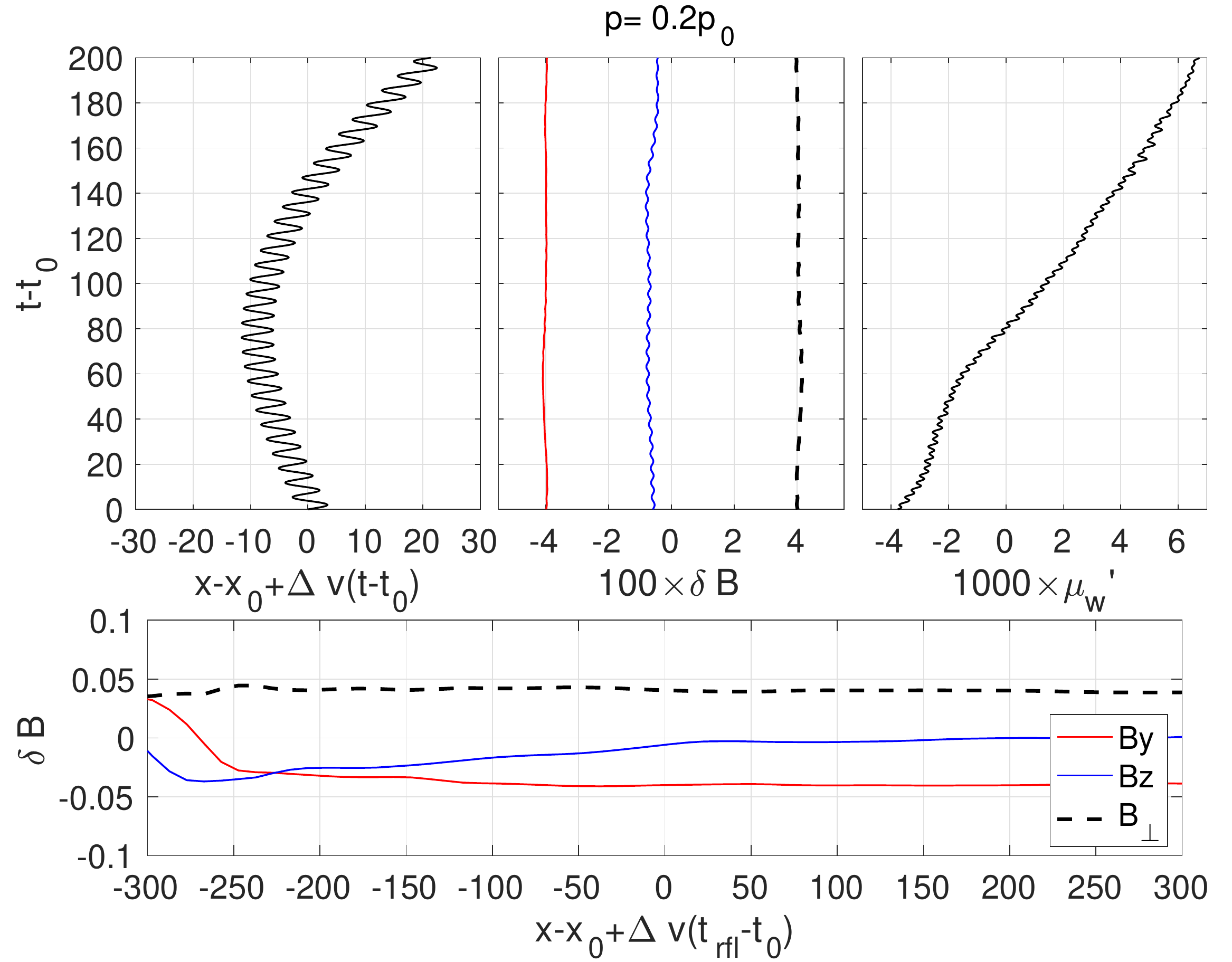}}
    \subfigure{
    \includegraphics[width=86mm]{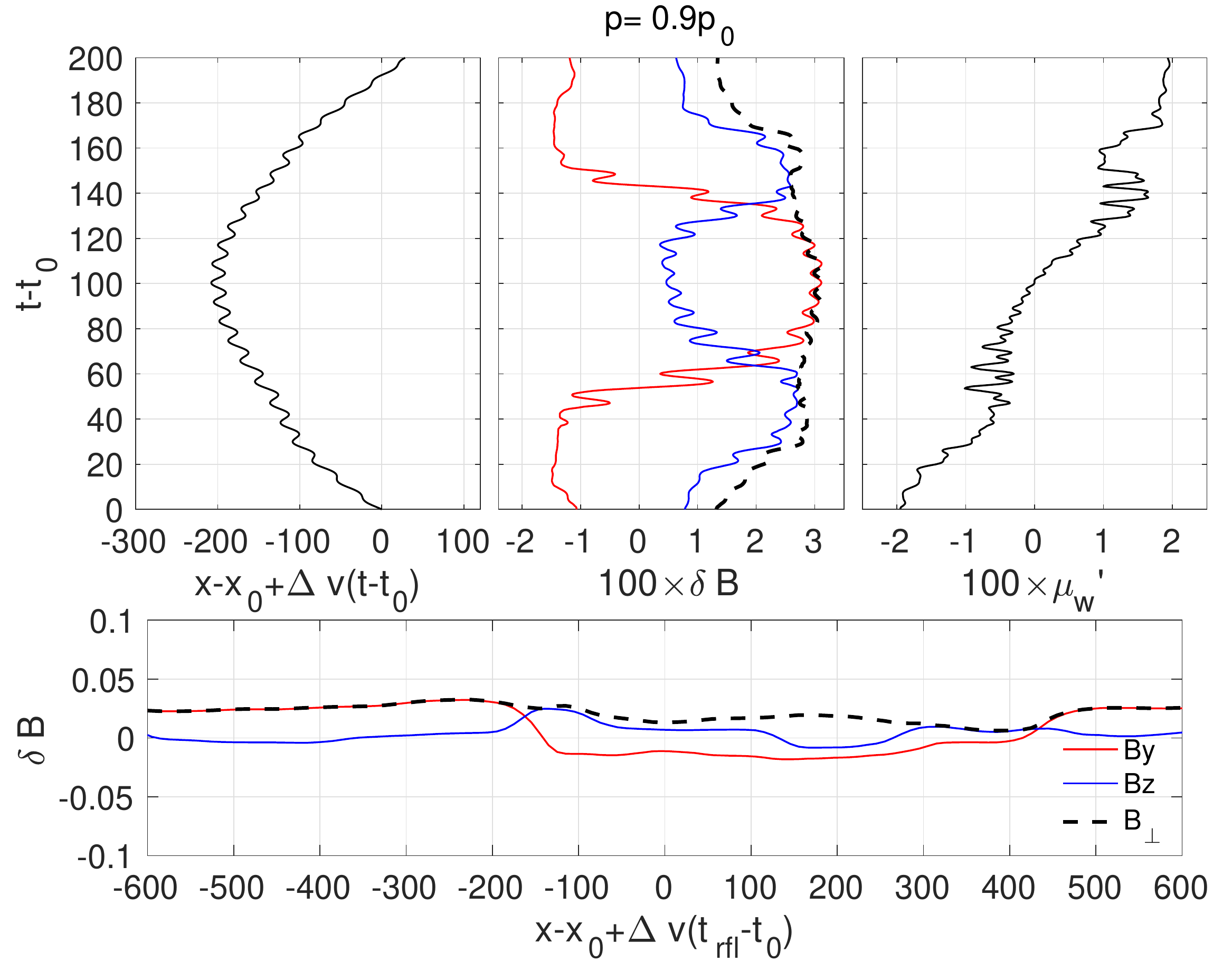}}
    \subfigure{
    \includegraphics[width=86mm]{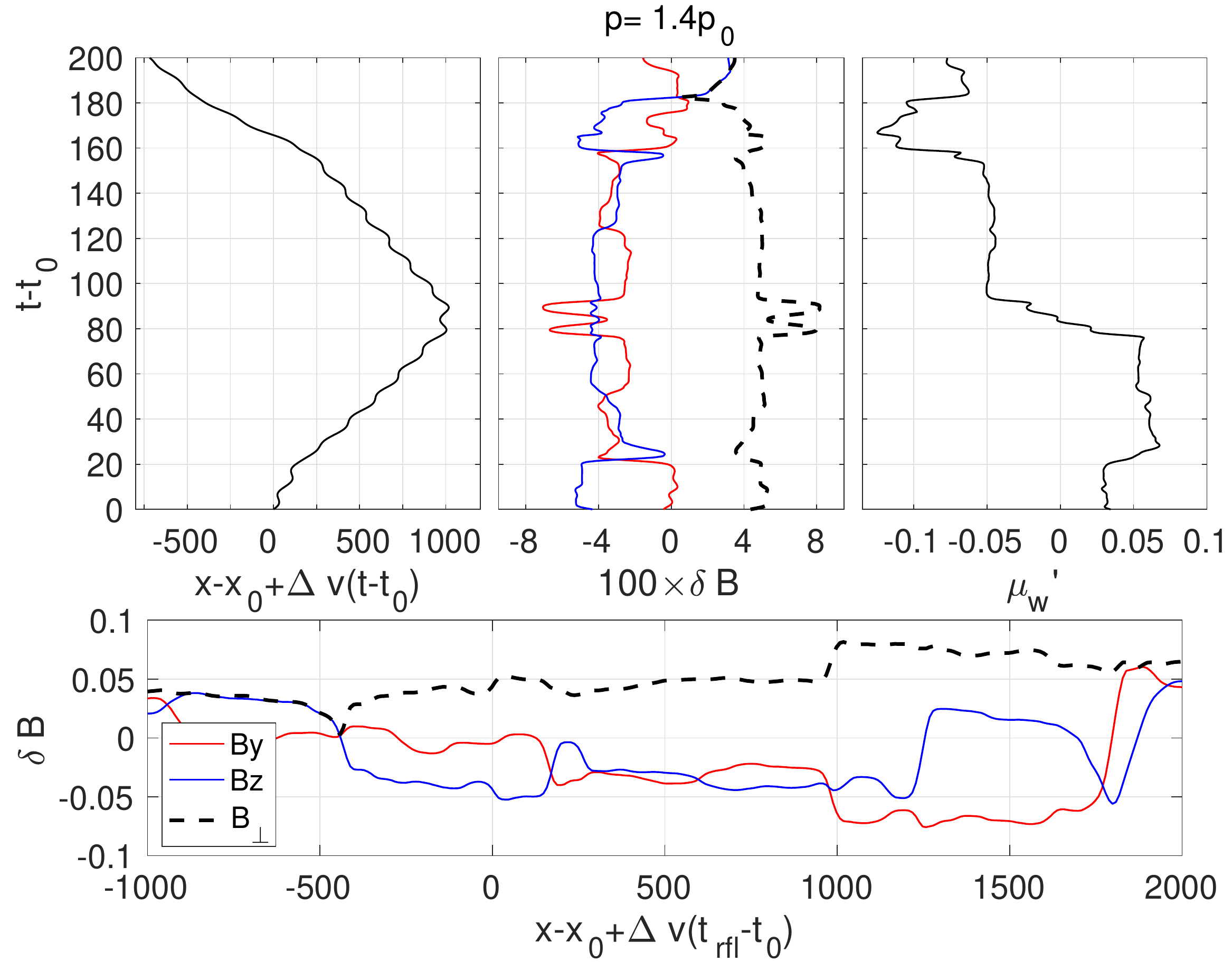}}
    \subfigure{
    \includegraphics[width=86mm]{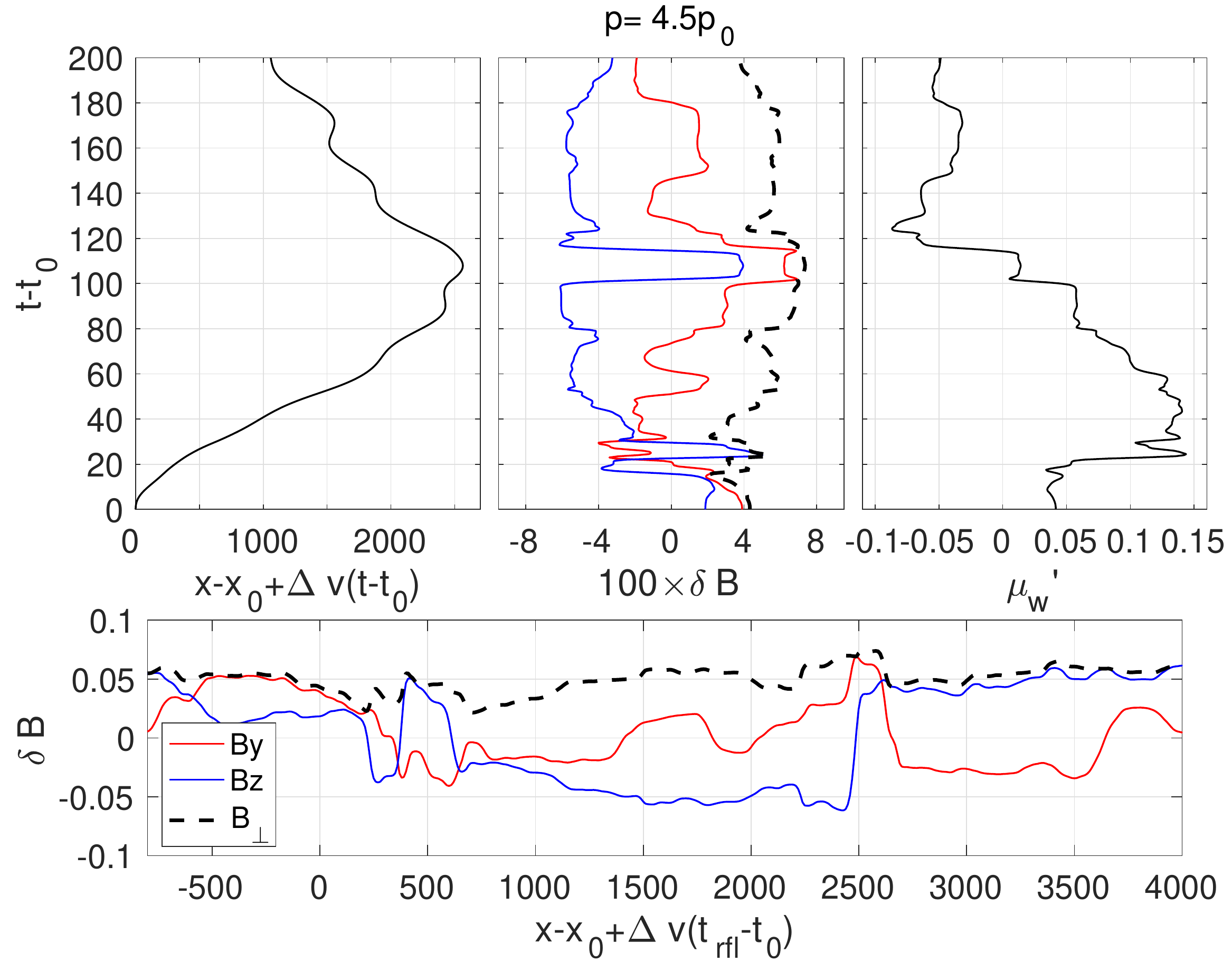}}
  \caption{Four typical reflection events, for a particle with $p=0.23p_0$ (top left),
  $p=0.92p_0$ (top right), $p=1.39p_0$ (bottom left), and $p=4.52p_0$ (bottom right).
  They are selected right after time $t_0=6\times10^4\Omega_c^{-1}$. Among each group of plots,
  the top three panels show the time evolution of particle position in the wave frame, perpendicular
  field strength (individual components in color and total in black dashed) experienced by the particle,
  and its pitch angle cosine in the wave frame {\it relative to the total field}.
  The bottom panel shows the perturbed magnetic field profiles at the time of reflection.}\label{fig:reflection}
\end{figure*}

\subsection[]{Overcoming the $90^\circ$ Barrier}\label{ssec:90deg}

To reveal the mechanism behind particle crossing the $90^\circ$ pitch angle in run M3, we have
randomly selected a subsample of particles (1280 per energy bin), and closely followed their
trajectories for a time interval of $200\Omega_c^{-1}$ starting from $t_0=6\times10^4\Omega_c^{-1}$.
We count the number of particles that have undergone reflection across $90^\circ$ over this
interval. Here, we consider particle pitch angle in the wave frame that is relative to the full
magnetic field vector (rather than background field), which is marked by $\mu'_w$.
To be counted, we require that $\mu'_w$  before and after this time interval to be beyond
$\pm v_A/\mathbb{C}\sim0.0033$ (chosen somewhat arbitrarily, but this does not qualitatively affect the
counting result). Among the 4 central momentum bins from $p=0.1p_0$ to $10p_0$, we find 9,
13, 20, 23 particles are reflected. We looked at the trajectories of all these particles, and choose to
show four particles with characteristic behaviors from each momentum bin in Figure \ref{fig:reflection}.

\subsubsection[]{Representative Cases}\label{sssec:90degcases}

For the first particle with $p\approx0.2p_0$ that undergoes a reflection, we see that surprisingly,
the field strength it experiences is almost constant. Quantitatively, the perpendicular
field strength over this time interval is $\delta B\in(0.0398, 0.0417)$. The range it spans translates to
$\mu_{\rm mir}=1.3\times10^{-3}$. While the particle does evolve quasi-adiabatically with many
gyrations (which is typical for lower energy particles given their small gyro-radii and lack of power at
small scales), its $\mu'_w$ changes by a factor of several larger than the expected $\mu_{\rm mir}$.
Checking the magnetic field profile in the bottom panel, we see that the particle is traveling against
a shallow gradient in $B_z$. The range of $B_z$ experienced by the particle $\in(-0.081, -0.041)$,
which changes the magnetic field orientation by $\sim0.04$. While this is no proof that the particle
must be reflected, it can well accommodate the change in particle $\mu'_w$, which is less than
$5\times10^{-3}$.

For the second particle with $p\approx0.9p_0$, its higher velocity allows it to traverse larger distances
over the course of reflection, and experiences more rapid magnetic fluctuations. We have again
checked that variations in total field strength are insufficient for mirror reflection. Reflection occurs
again when the particle travels against the gradient of individual field components, and in this time,
we have $B_y$ increasing rapidly and $B_z$ decreasing more gradually. These variations change the 
magnetic field orientation by $\sim0.05$ within just a few particle gyrations, which is again more than
sufficient to account for the change in particle pitch angle by $\sim0.02$.

In the remaining two examples, reflection occurs within two gyro-orbits. These cases are more common
among relatively energetic particles: because of their larger gyro-radii, they more often experience
abrupt changes in magnetic field.
These changes average out to the leading order in quasi-linear theory, except when they
lead to reflection. We see that the threshold pitch angle for reflection can be higher for these particles,
and can reach as large as $\mu\sim0.1$. This is accompanied by abrupt changes in $B_y$ and
$B_z$ at comparable levels within the distance that the particle travels in one gyration.

\subsubsection[]{Non-linear Wave-Particle Interaction}\label{sssec:nlwp}

Overall, the four examples roughly cover the range of behaviors of particles during reflection. They
can be interpreted under the framework of non-linear wave-particle interaction \citep{Volk73}.

Reflection is most effective (almost instant) for the last two cases in Figure \ref{fig:reflection},
for which abrupt changes in fields occur on a scale shorter than the distance that particle travels in one
gyro time. In this regard, we can discuss the critical pitch angle below which non-linear wave-particle
interaction takes over to reflect particles as follows. For a particle with momentum $p$ (and velocity
$v$), let $\mu_{\rm crit}$ be this critical pitch angle. It travels by
$L_{\rm crit}\sim 2\pi v\mu_{\rm crit}/\Omega$ in one gyro time, corresponding to a wave number
$k_{\rm crit}=\Omega/(v\mu_{\rm crit})=\Omega_c/(p\mu_{\rm crit})$. Reflection occurs when an
abrupt change in magnetic field orientation is comparable to $\mu_{\rm crit}$. Note
this depends on the spectrum of waves at scales below some $\zeta(k_{\rm crit})^{-1}$,
we arrive at a relation
\begin{equation}\label{eq:mu_crit}
\mu_{\rm crit}^2\sim C\int_{k_{\rm crit}/\zeta}^\infty I(k)dk\ ,
\end{equation}
where there is a square in the left hand side because $\mu_{\rm crit}\sim\delta B/B_0$, while
the right hand side corresponds to its square. We have also include dimensionless factors $\zeta$
and $C$ that encapsulate the roughness of the model.

\begin{figure*}
    \centering
    \subfigure{
    \includegraphics[width=85mm]{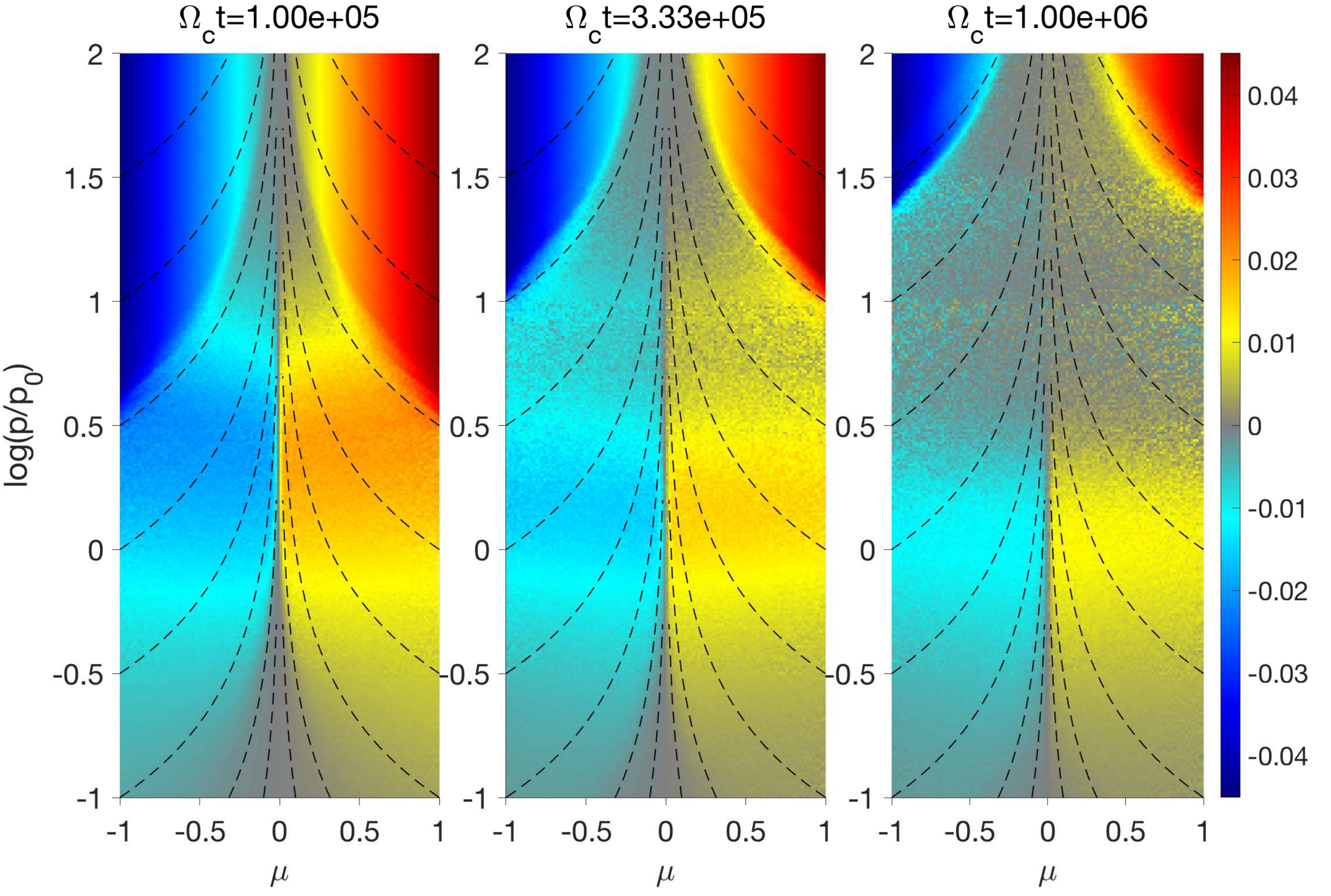}}
    \subfigure{
    \includegraphics[width=85mm]{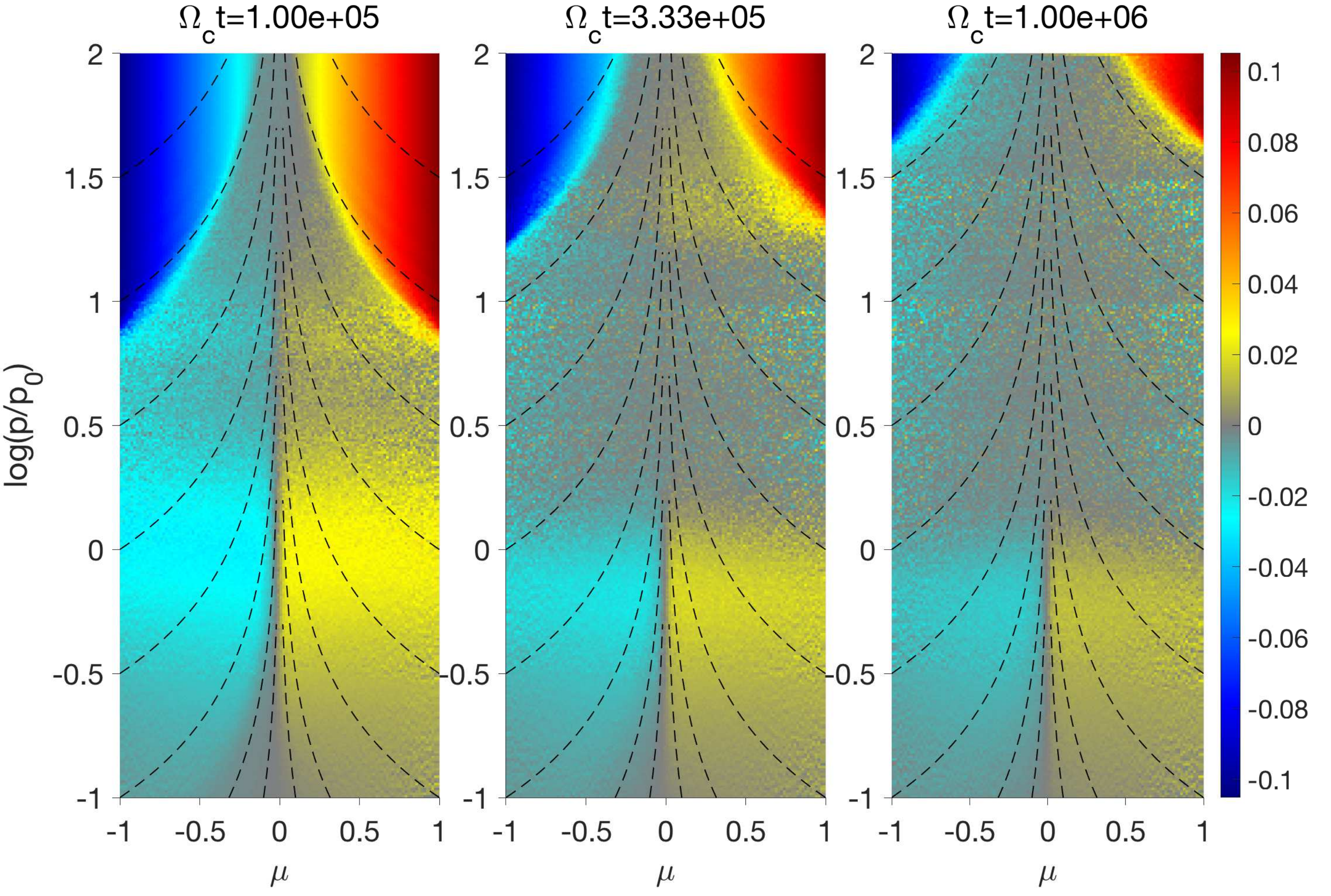}}
    \subfigure{
    \includegraphics[width=85mm]{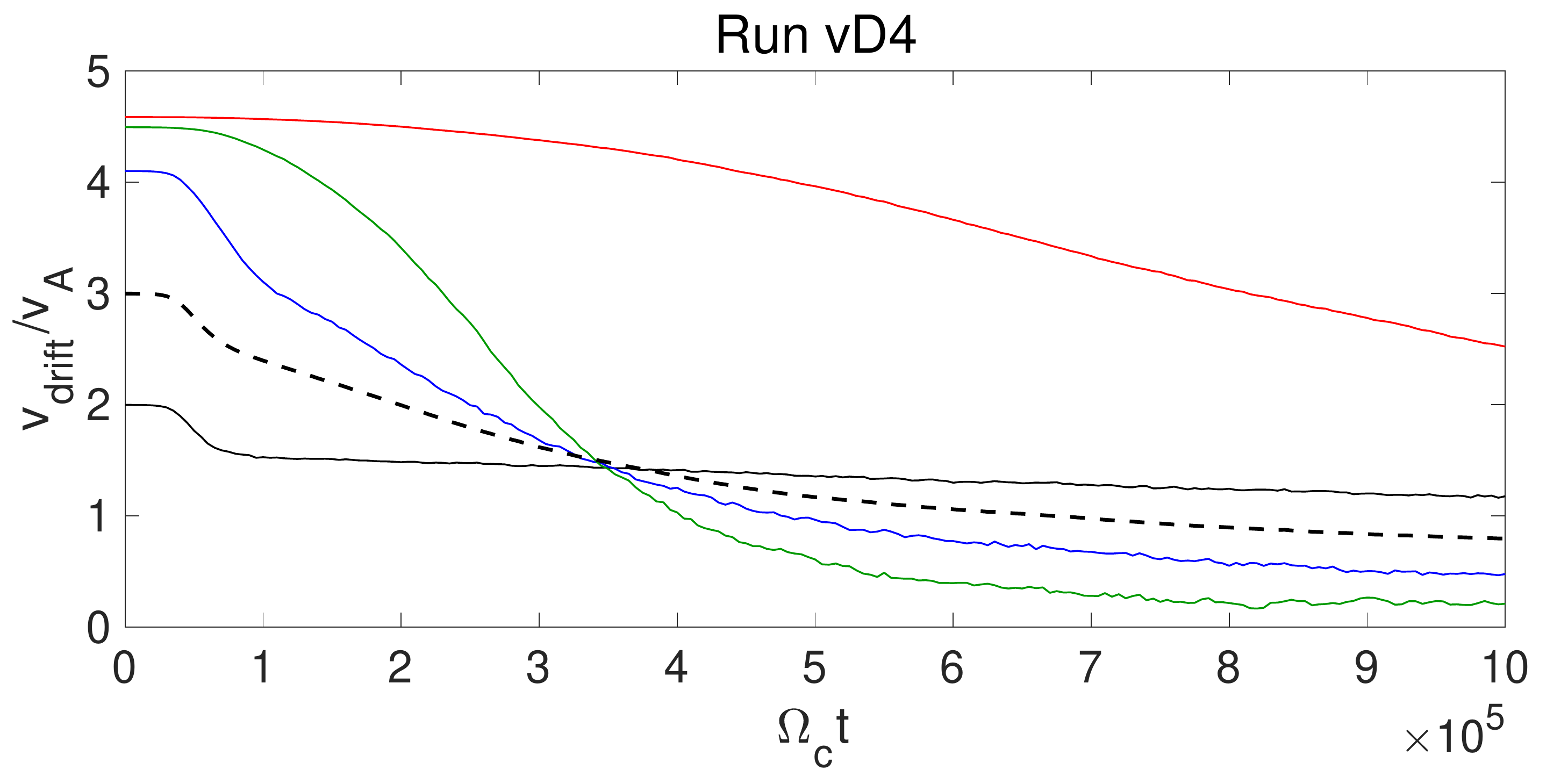}}
    \subfigure{
    \includegraphics[width=85mm]{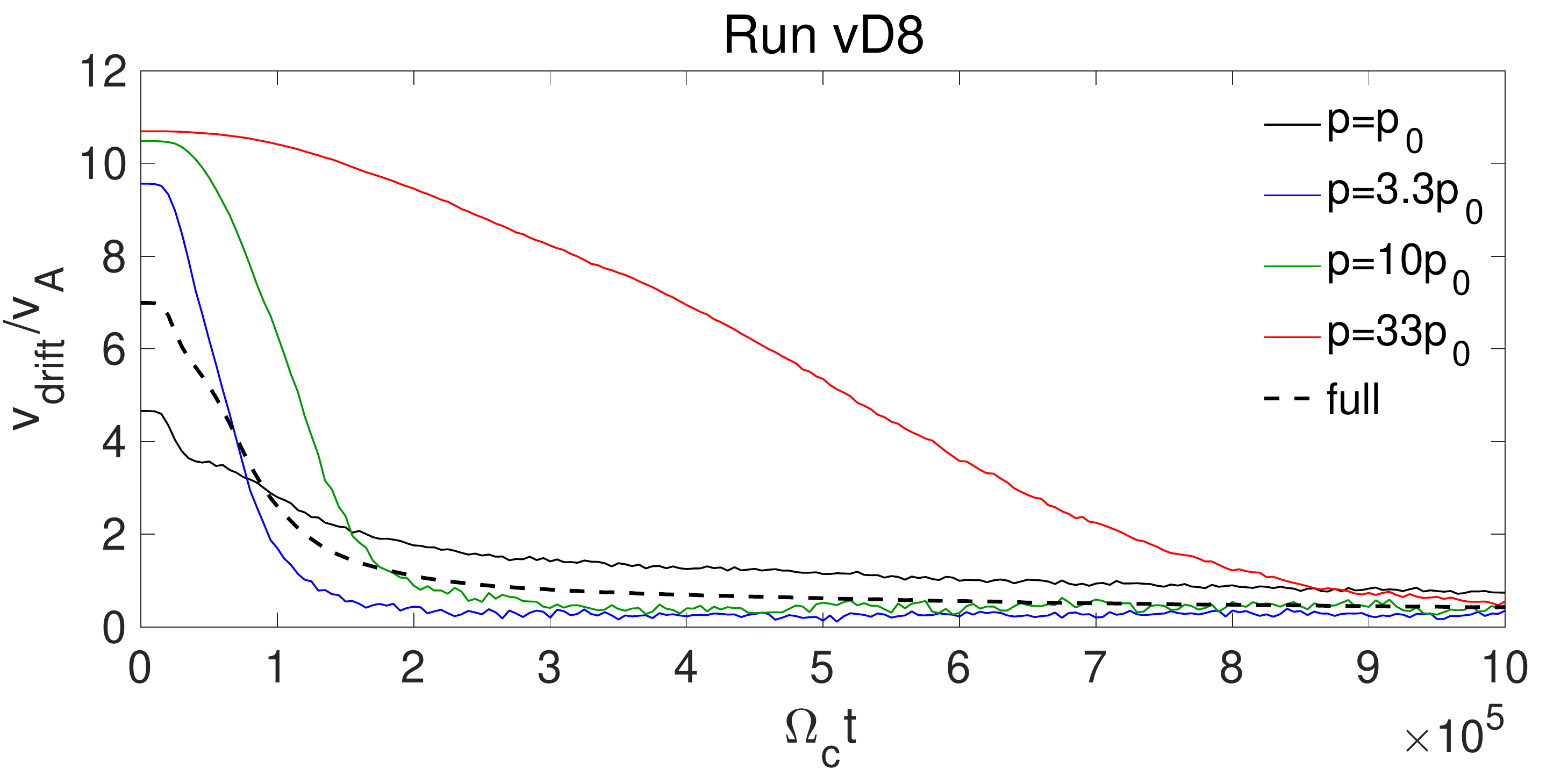}}
  \caption{Distribution function and drift velocities for runs vD4 (left) and vD8 (right).
  Top two panels: 2D distribution function $\delta f_w/f_0$ in the wave frame
  at three snapshots. Dashed lines are contours in momentum space with the same
  resonance condition.
  Bottom two panels: time evolution of particle drift velocity in the wave frame for 
  different momentum bins, as well as the full drift velocity (see legend).
  }\label{fig:dfwof_param}
\end{figure*}

The solution to this equation gives $\mu_{\rm crit}$, which exhibits the following features
\begin{itemize}
\item Higher wave intensity leads to smaller $\mu_{\rm crit}$, because more wave
power at smaller scales makes reflection easier.
\item Higher particle momentum leads to larger $\mu_{\rm crit}$, because larger gyro scale
accommodates more wave power to build up abrupt field profiles.
\item If there is a lack of power at small scales, there can be no solution, meaning that direct
reflection in a couple gyro orbits is unlikely.
\end{itemize}
Clearly, all these features qualitatively agree with our findings. They are also justified
through our discussions in Section \ref{sssec:90degcases}. 

To be more quantitative, we note that while the integral of $I(k)$ over $k$ gives the rms
fluctuations, actual fluctuations can easily exceed the rms value, and deviations by up to
$3\sigma$ are not rare. We thus take $C\sim3^2$ to account for the upper end of
fluctuations. We may further take $\zeta\sim4$ considering that waves with wavelengths
up to $4L_{\rm crit}$ can contribute significantly to the local slope within this length scale of
$L_{\rm crit}$ (i.e., sinusoidal variations have roughly constant maximum slope in just 1/4 of
its periodic domain consecutively). With these numbers, and taking $p=4.5p_0$ for run M3,
we can derive $\mu_{\rm crit}\sim0.1$ (which gives
$k_{\rm crit}/\zeta\sim1.8\times10^{-3}$, and the integral gives $0.0011)$.

Equation (\ref{eq:mu_crit}) closely resembles Equation (54) of \citet{Volk73}, with their
$v_{\rm osc}\sim 0.25-0.3(v\mu_{\rm crit})$, interpreted as the rms velocity oscillation
amplitude of a particle in ``the fluctuation field that contributes to its diffusion". The
effective diffusion coefficient at $\mu=0$ is then given by
$\sim3\Omega v_{\rm osc}^2$.

The discussions above apply to instant reflection upon encountering abrupt field gradients.
This corresponds to the $n=1$ term in Equation (48) of \citet{Volk73}.
As one lowers particle momentum, $\mu_{\rm crit}$ decreases, and eventually no
solution can be found. This situation applies to lower-energy particles in Figure
\ref{fig:reflection}. More generally, higher order effects enter (corresponding to the
$n>1$ terms), allowing particles to be reflected over multiple gyrations in a fluctuating
field. Being less effective, it is consistent with fewer crossing particles found in our
particle counting in lower-energy bins.

\section[]{Simulation Results: Parameter Study}\label{sec:param}

Our fiducial run selects a parameter range that is physically more
realistic but numerically challenging, with small $n_{\rm CR}/n_i$ and low CR
drift velocity $v_D$, leading to relatively small growth rate and relatively low
wave saturation level. In this Section, we vary parameters around our fiducial run,
and conduct a two-part parameter study.
In Section \ref{ssec:vd}, we examine
the saturation of the CRSI when $v_D$ is increased to higher values.
In Section \ref{ssec:numparams}, we discuss results of 
varying solely numerical parameters for our run Fid. 

\subsection[]{Dependence on Drift Speeds}\label{ssec:vd}

With $v_D$ increased to $4v_A$ and $8v_A$, we find that we recover the
linear growth rates very well (not shown), with left and right handed modes growing at
comparable rates (as expected).
Figure \ref{fig:Ehst} includes the time evolution of wave energy for runs vD4 and
vD8 in dotted lines. They initially follow very similar evolutionary paths as our run Fid,
except that everything (linear growth and QLD) proceeds faster.
More notably, later evolution of these two runs are more similar to run M3, where
wave energy eventually plateaus, indicating saturation is close to completion.

In Figure \ref{fig:dfwof_param}, we show results related to quasi-linear evolution of runs vD4 and vD8. The overall outcomes are in between those of run Fid and run M3. Crossing
across $90^\circ$ becomes easier as $v_D$ increases. By the end of the simulations,
particles with $4p_0\lesssim p\lesssim20p_0$ are roughly fully isotropized in run vD4.
For run vD8, the range enlarges to $p_0\lesssim p\lesssim40p_0$.
\footnote{Two numerical artifacts can be identified in Figure \ref{fig:dfwof_param}.
The banded horizontal feature in the top panels is due to the division into 8 particle
momentum bins. Within each bin, there is more noise towards higher energies because
there are fewer particles. One may also find that particle mean drift velocities at various
energies do not reach zero even they are theoretically expected to be fully isotropized
(bottom panel). This is related to coarse particle binning during outputs (mentioned
earlier in Section \ref{sssec:satM3}).}
Lower-energy particles are in the process of (but have not yet completed)
isotropization. Isotropization is limited by the rate of $90^\circ$ crossing, which is much slower for these lower-energy particles (see discussion in Section \ref{ssec:90deg}).

\begin{figure}
    \centering
    \includegraphics[width=88mm]{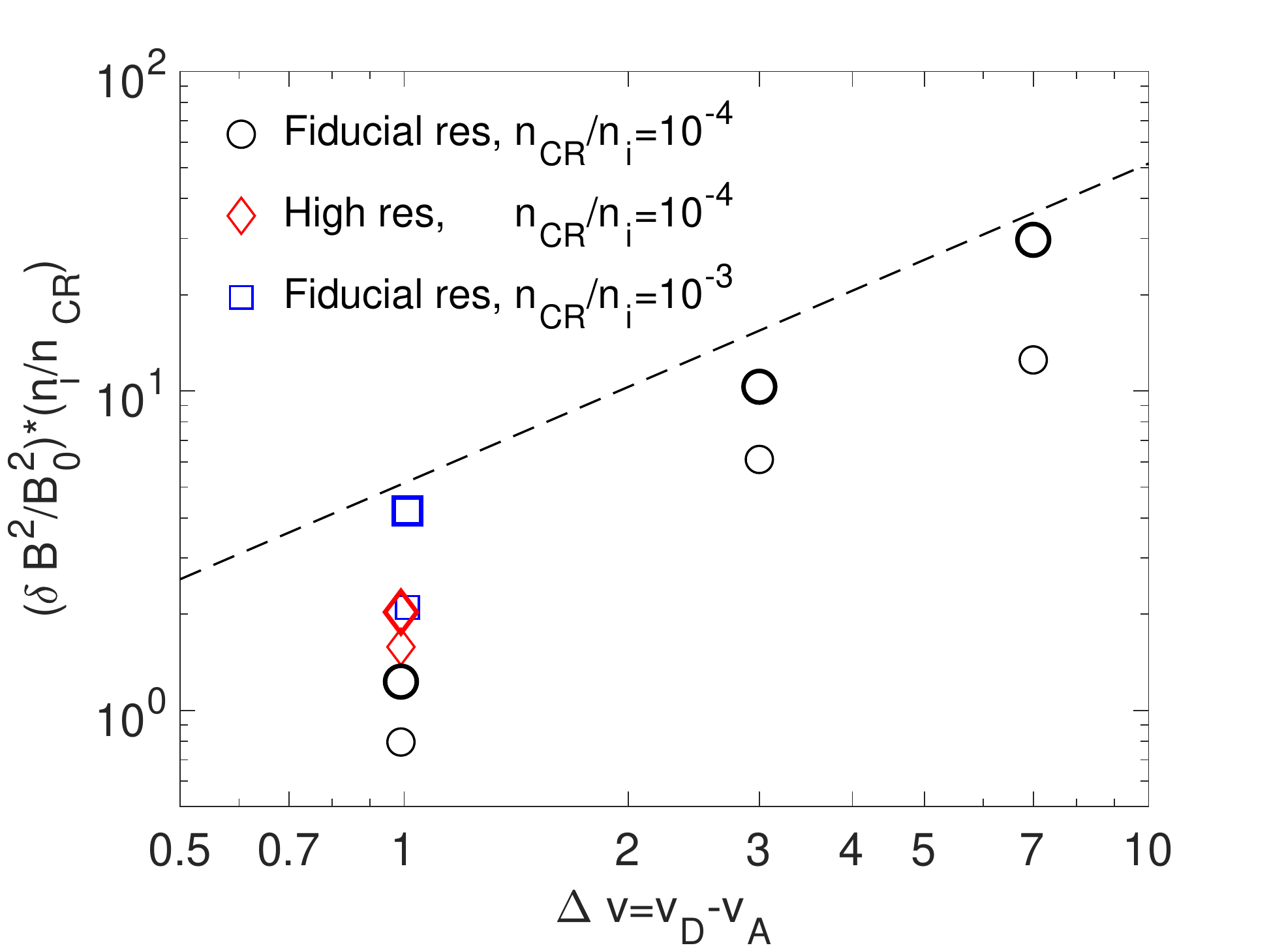}
  \caption{Saturation level measured from our simulations, in terms of wave energy density
  normalized by CR fraction. Black circles are from runs Fid, vD4 and vD8, red diamonds are
  from run Fid-hires, and blue squares  are from run M3. Small/thin symbols are measured
  based on the total wave energy density, while large/thick symbols are measured from the
  same runs, but corrected from numerical dissipation. The dashed line indicates the expected
  saturation level assuming all CRs are fully isotropized, obtained by setting Equations
  (\ref{eq:dPCR}) and (\ref{eq:Pwv}) equal.}\label{fig:saturation}
\end{figure}

\subsubsection[]{The Saturation Level}\label{sssec:satleveel}

At this point, we can gather all simulation runs discussed so far to address the saturation
level of the CRSI as a function of $v_D$. The saturation level is measured in terms of
wave energy density. With equipartition of magnetic and kinetic energies, it can simply
be represented by $\langle\delta B^2/B_0^2\rangle$ in dimensionless form.
Theoretically, the predicted wave energy saturation level is obtained by equating Equations (\ref{eq:dPCR}) and (\ref{eq:Pwv}),
and is proportional to $(n_{\rm CR}/n_i)\Delta v$. Integrating the derivative of the
distribution function numerically, the result is shown as the dashed line in Figure
\ref{fig:saturation}.

The final wave energy density obtained from our runs Fid, vD4, vD8, M3, and
Fid-hires (discussed in the next subsection) are also shown in Figure \ref{fig:saturation}.
Even though runs M3 and vD8 are close to complete saturation, their final
wave energy densities appear to be far from the expected level. This is because 
numerical dissipation, while very low in the Athena MHD code, is still non-negligible
for the extremely long simulation time necessary to reach saturation.
\footnote{ Dissipation of low-amplitude waves imposed in the initial conditions is also
evident at early times in Figure \ref{fig:Ehst}.}
With an adiabatic equation of state, we can capture the change in gas internal energy
over time (thanks to the excellent energy conservation properties in the gas component),
and find that it increases steadily as result of numerical dissipation, and reaches a level
that is comparable to the overall wave energy density by the end of the runs. When we add
this energy lost to dissipation to the direct wave energy, we find much more reasonable
saturation levels, as shown in thick/larger symbols in Figure \ref{fig:saturation}. In particular, 
run Fid remains at a level below the full saturation prediction because particles become
``stuck'' at $90^\circ$, whereas runs vD4 and vD8 are more and more close to the expected
saturation level, as more particles are fully isotropized. Run M3, which based on its distribution
function is clearly isotropized, is excellent agreement with theoretical expectations for the
wave amplitude.

\begin{figure}
    \centering
    \includegraphics[width=88mm]{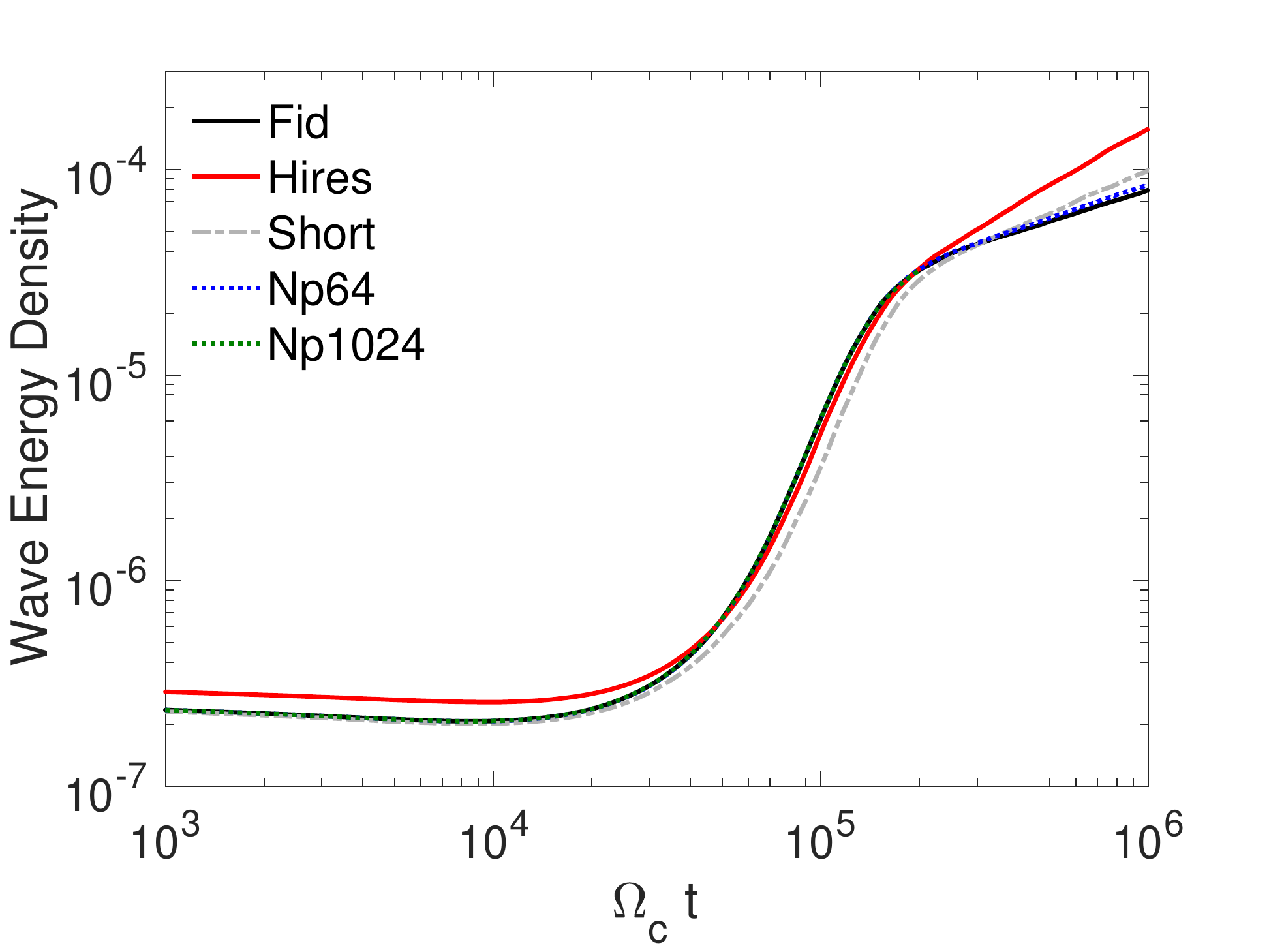}
  \caption{Time evolution of total wave intensity for runs with different numerical
  parameters.}\label{fig:Ewave_param}
\end{figure}

\begin{figure}
    \centering
    \includegraphics[width=88mm]{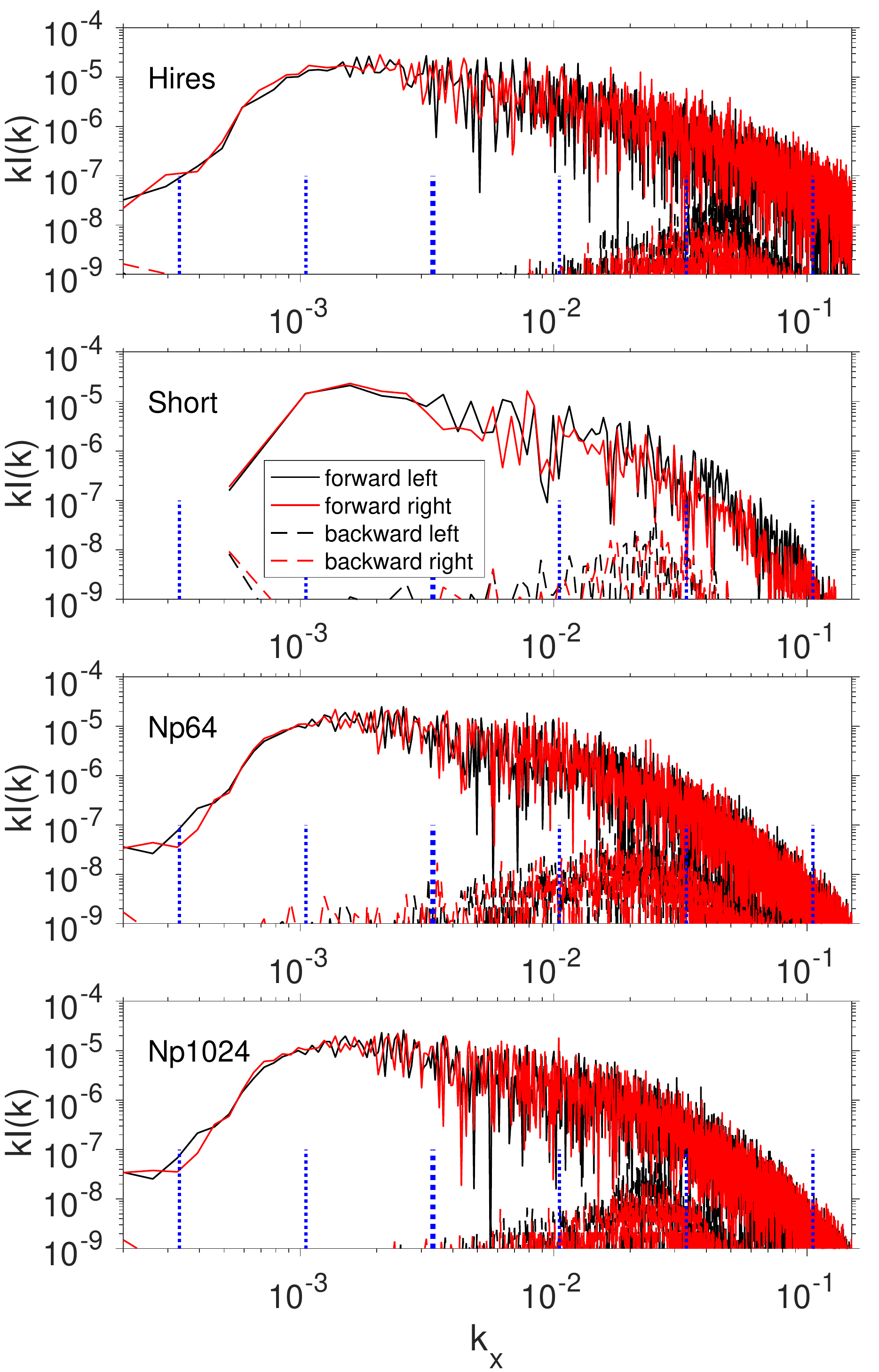}
  \caption{Wave spectrum at time $t=2\times10^5\Omega_c^{-1}$ from runs Fid-hires,
  Fid-Short, Fid-Np64 and Fid-Np1024. Vertical dotted lines are identical to those in Figure
  \ref{fig:spectseries_fid} (the location of the thick line indicates the most unstable
  wavenumber).}\label{fig:specs_param}
\end{figure}

\begin{figure*}
    \centering
    \includegraphics[width=180mm]{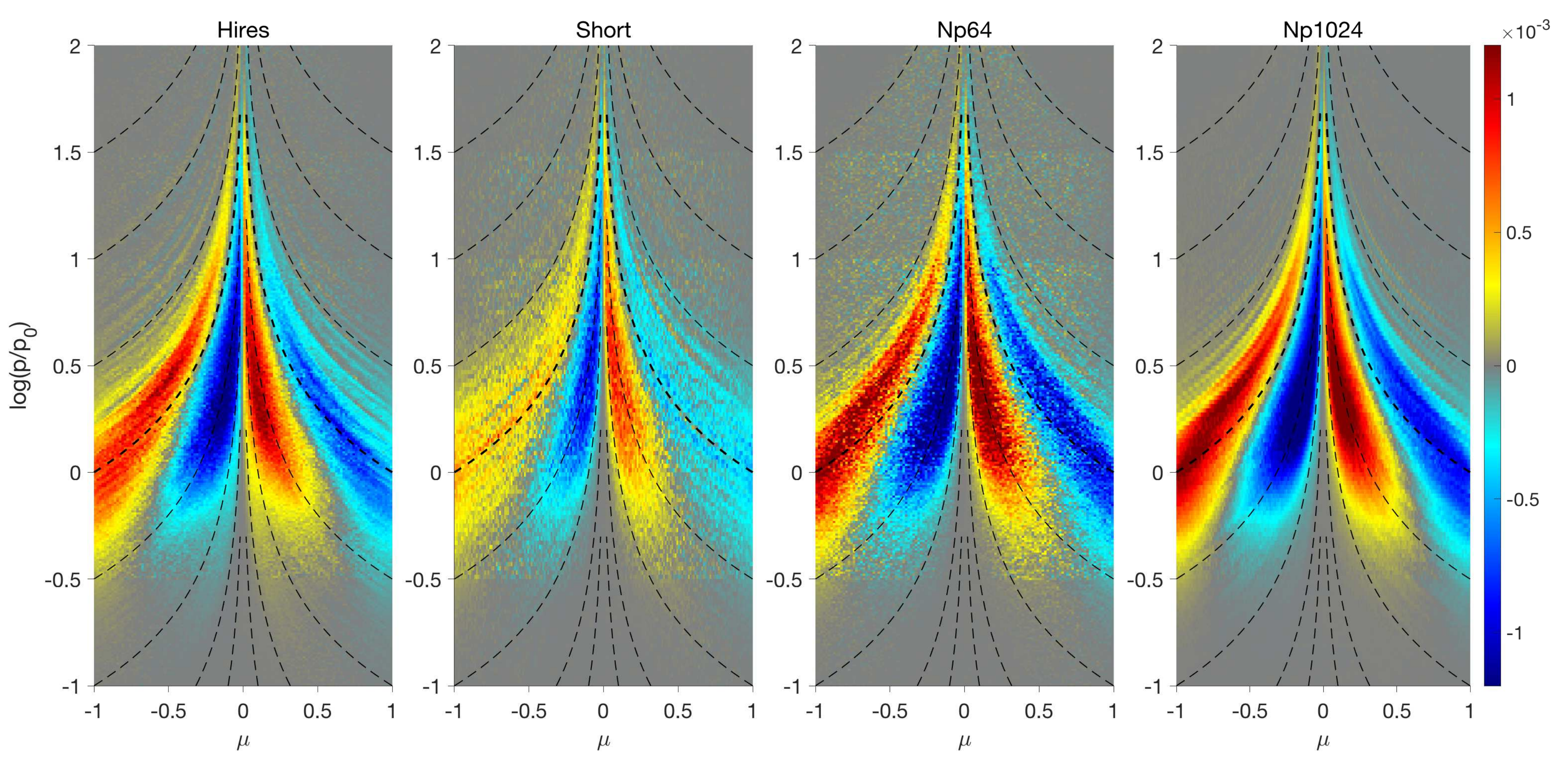}
  \caption{The 2D distribution function $\delta f/f_0$ in the lab (simulation) frame at time
  $t=10^5\Omega_c^{-1}$ for runs Fid-Hires, Fid-Short, Fid-Np64 and Fid-Np1024,
  respectively. The dashed lines are contours in momentum space that are resonant with
  the same wave ($p\mu=$const), with the thick line marking $p\mu=p_0$. The color limits
  are identical among the four panels.}\label{fig:dfof_param}
\end{figure*}

\begin{figure*}
    \centering
    \subfigure{
    \includegraphics[width=128mm]{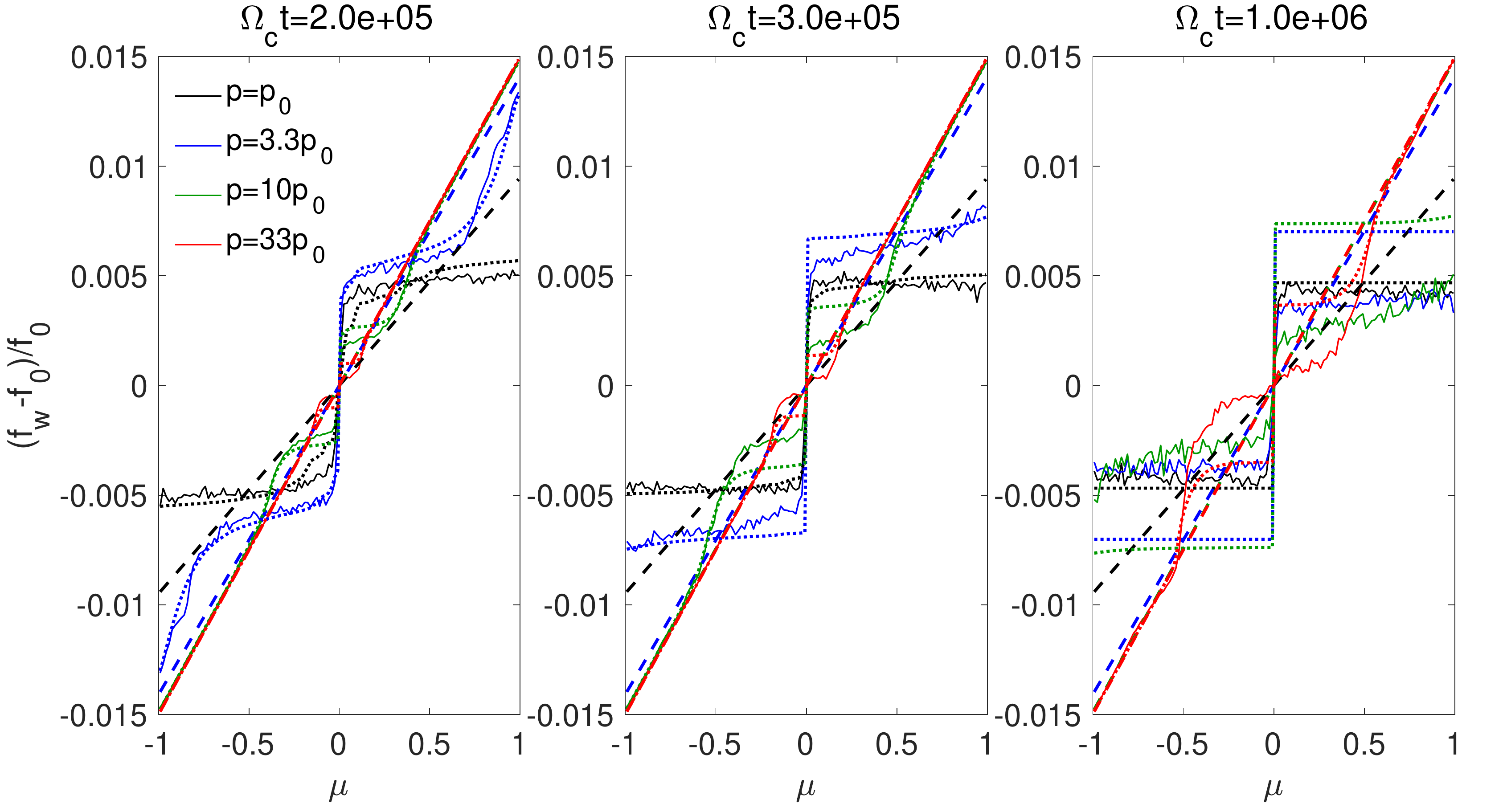}}
    \subfigure{
    \includegraphics[width=48mm]{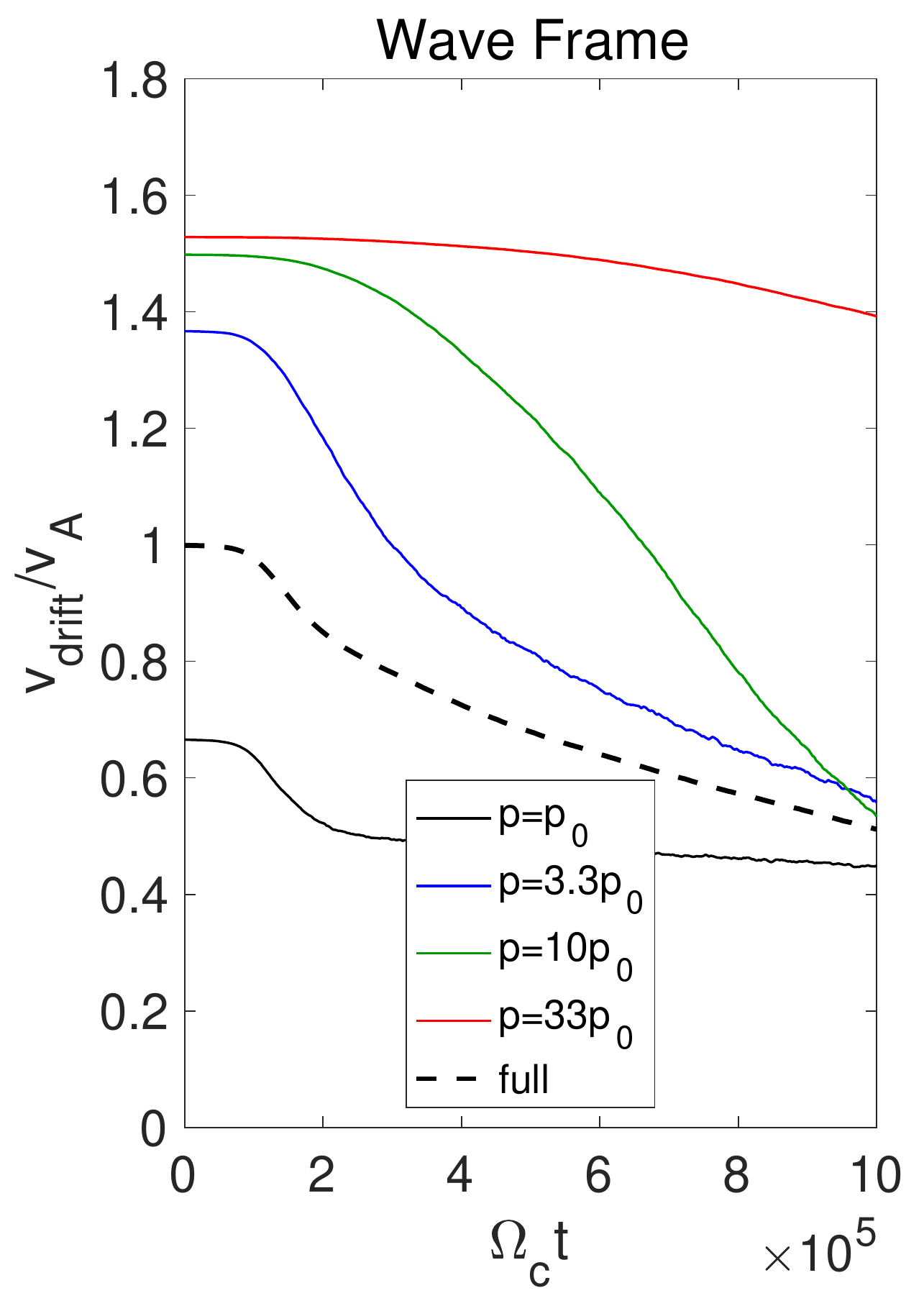}}
  \caption{Results from run Fid-Hires.
  Left three panels: snapshots of particle distribution function in the wave frame
  $(f_w-f_0)/f_0$ for four representative momenta as a function of pitch angle cosine
  $\mu_w\approx\mu$. Solid lines are measured from the simulations, thick dashed lines
  indicate the initial distribution, while the dotted lines are reconstructed distribution
  functions obtained by solving the QLD Equation (\ref{eq:qld}) in time. Right panel: time evolution
  of particle drift velocity in the same four momentum bins in the wave frame (solid lines).
  The thick dashed line represents the mean drift velocity averaged over the entire
  particle population.}\label{fig:qld_hires}
\end{figure*}

\subsection[]{Numerical Parameters}\label{ssec:numparams}

Here, we examine our runs Fid-Hires, Fid-short, Fid-Np64 and Fid-Np1024, in which we keep physical parameters the same and vary only  
numerical parameters (note that for run Fid-Hires, we have slightly reduced
the box size to save computational cost). In Figure \ref{fig:Ewave_param}, we show the time
evolution of the wave energy density of these runs. Overall, these three runs proceed very
similarly to run Fid.\footnote{All these runs have the same initial wave amplitude $A=10^{-4}$ (see Equation \ref{eq:initI}). Run Fid-Hires thus has more initial wave energy because of more
spectral range it covers.} QLD proceeds slightly slower in run Fid-Short, but it catches up to the
other runs shortly after. Run Fid-Hires eventually grows to reach larger wave amplitudes, which we
will show later is due to more efficient crossing of $90^\circ$ pitch angles.

In Figure \ref{fig:specs_param}, we show the wave spectrum from these runs at time
$t=2\times10^5\Omega_c^{-1}$. The overall shape of the spectra are comparable to each other.
The high-resolution run has less numerical dissipation, allowing the wave spectra to extend to
larger $k$. Run Fid-short has a box size that is only $1/3$ of the fiducial box (but still long
enough to fit $\sim17$ most unstable modes), and hence modes are a lot more sparsely sampled,
especially at low-$k$. This is likely related to its initially slower quasi-linear evolution. We have
also tried simulations with $1/8$ the fiducial box size (not shown); while evolution is smooth, the
lack of power at larger scales start to adversely affect the evolution of higher-energy particles.

Figure \ref{fig:specs_param} shows that there are backward-propagating modes excited in all these runs. The amplitude
of these modes reflects the level of noise. Clearly, inserting more particles helps reduce the noise.
The bottom line for properly choosing particle number is that the amplitude of such
backward-propagating modes should always be much smaller than the forward-propagating
modes driven by the CRSI. In this regard, our run Fid-Np64 is acceptable, but more caution
should be exercised if one were to use even fewer particles.

In Figure \ref{fig:dfof_param}, we show the particle distribution functions in the wave frame at
time $t=10^5$. These may be compared with Figure \ref{fig:dfof_fid}. Overall, runs
Fid-short and Fid-Np64 are more noisy, whereas run Fid-Np1024 appears exceptionally smooth.
Our run Fid lies in between. Despite these  differences in noise levels, 
 subsequent QLD does not appear to be affected.   

Of particular interest is the high-resolution run Fid-Hires, where waves grow to substantially
higher amplitudes. In Figure \ref{fig:qld_hires}, we show three snapshots of the particle
distribution function in the wave frame for four representative momentum bins. These frames are to
be contrasted with Figure \ref{fig:qld_fid} for the same snapshots. Clearly, in this run, a
substantial fraction of particles with $p\gtrsim3p_0$ have managed to cross the $90^\circ$
barrier. This fraction steadily increases with time and, were we to run the simulation longer
(e.g., to $t=2\times10^6$), particles in these momentum bins would essentially be fully
isotropized. This trend can also be seen from the last panel, which shows the time evolution
of the mean CR drift velocity: $v_d$ for particles with $p\gtrsim3p_0$ are steadily decreasing
towards zero. Clearly, crossing the $90^\circ$ barrier is easier compared to run Fid. This is
natural because higher resolution allows more small-scale modes to be better preserved,
thus enhancing the chance of non-linear wave-particle interaction to reflect the particles
(see Section \ref{sssec:nlwp}). In Figure \ref{fig:saturation}, we see that the wave energy
density in our run Fid-Hires is also a factor of several closer to the expected saturation
level.  We expect that increasing the resolution further (which would be much more computationally
costly) would eventually lead to full isotropization of the bulk CR populations.

In sum, our exploration of numerical parameters suggests that full saturation of the CRSI
(ideal case, no damping) can be achieved as long as there is sufficient numerical resolution.
The required resolution is mainly limited by the demands imposed by crossing the $90^\circ$ pitch angle,
which is sensitive to wave amplitudes at small scales. Similar results to our fiducial model can be achieved
when one reduces the number of particles and/or the box size at some modest level ($\sim$ a
factor of $\sim3$).

\begin{figure}
    \centering
    \includegraphics[width=88mm]{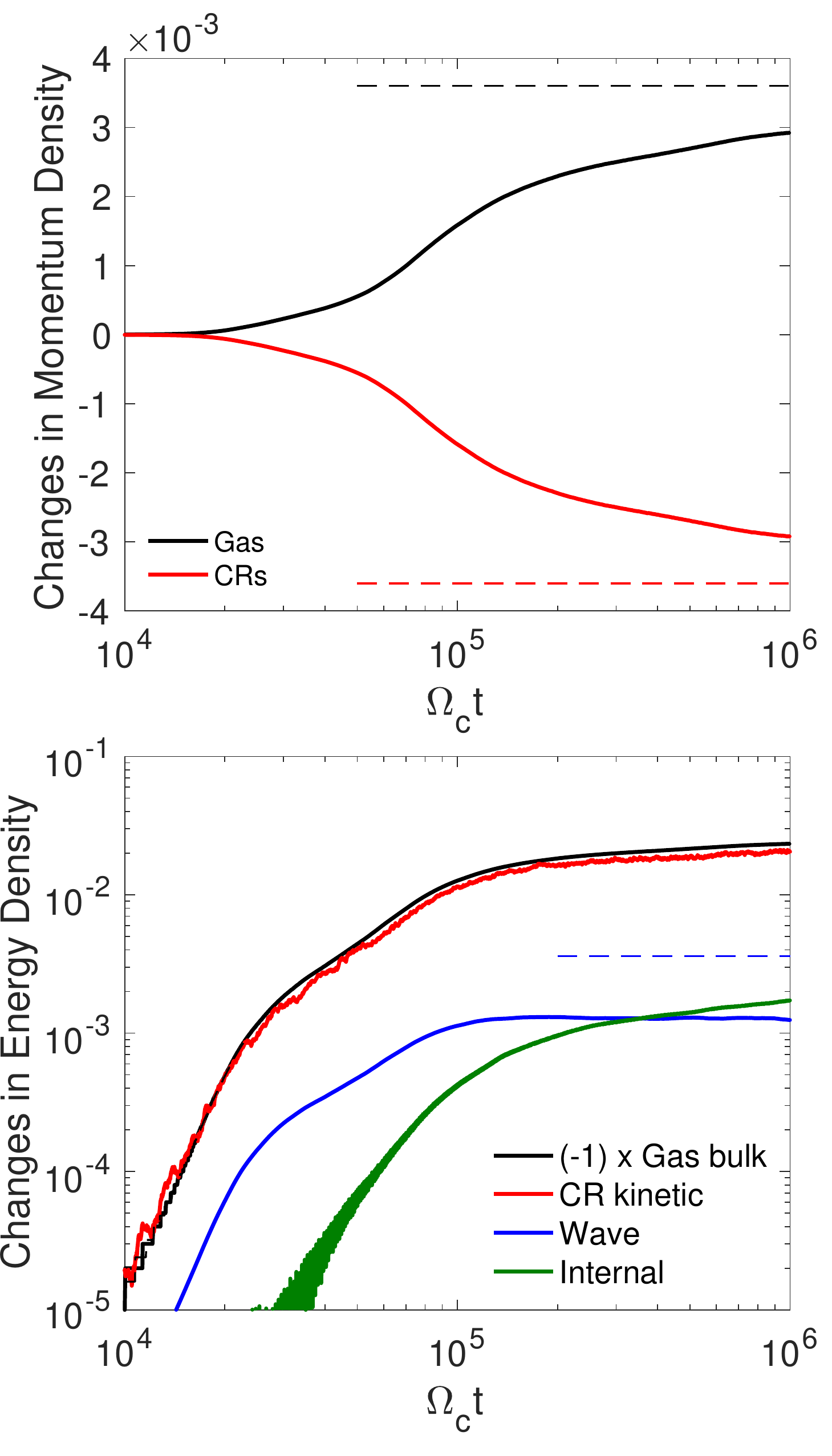}
  \caption{Time evolution of changes in momentum density (top) and energy density (bottom)
  in various components (labeled in the legends) in run vD8. The horizontal dashed lines mark
  the expected saturation levels (top: momenta; bottom: wave energy). See Section
  \ref{ssec:feedback} for more details.}\label{fig:cons_vD8}
\end{figure}

\section[]{Discussion}\label{sec:discussion}

\subsection[]{CR Feedback to Background Gas}\label{ssec:feedback}

Through the CRSI and QLD, CRs transfer momentum to background gas, which serves as
an important dynamical feedback mechanism in broad range of astrophysical 
environments.
In the following, we take run vD8 as an example and discuss CR feedback in terms
of momentum and energy.

Momentum feedback is manifested in the top panel of Figure \ref{fig:cons_vD8}, which
shows changes in the mean gas and CR momentum density 
($\Delta {\cal P}_g$ and $\Delta {\cal P}_{\rm CR}$).
In our simulation frame, the initial CR momentum is zero, and initial
gas momentum is negative. Over time, the CR momentum density decreases, accompanied by an
increase in gas momentum density of exactly the same amount.
Physically, the backreaction from the CR isotropization process effectively gives rise to a ``parallel force"
$F_{\parallel}$. 
This parallel force is mediated by forward-propagating Alfv\'en waves,
and transfers CR momentum to the gas (e.g., \citealp{ThomasPfrommer19}).
The amount of momentum transferred to the gas is in fact the same as the ``effective'' wave
momentum defined in Equation (\ref{eq:Pwv}), which is the momentum needed to excite such
forward-propagating waves \citep{Kulsrud05}.
By the end of our run vD8, although the evolution of gas and CR momenta are not yet
fully saturated, both are approaching the expected saturation level indicated by the
dashed lines, obtained from integrating Equation (\ref{eq:dPCR}). In the case of
run M3 (not shown here), we do observe that saturation of gas and CR momenta is
achieved.  
That is, the mean CR streaming velocity relative to the gas is reduced from $2 v_A$ to $v_A$ (Figure \ref{fig:vd_m3} shows that the final mean CR velocity is zero in the wave frame), and the gas momentum density is increased to balance the loss from CRs.

Energy feedback can be discussed by decomposing the energy budget into various
components. These include gas bulk kinetic energy
$E_{k,{\rm bulk}}=\langle\rho v_x^2/2\rangle$, representing gas motion parallel to
background field, wave energy $E_{\rm wave}$, CR kinetic energy $E_{\rm CR}$,
and internal energy of the gas $E_{\rm internal}=P_g/(\Gamma-1)$. 
The time evolution of these quantities is shown in the bottom panel of Figure \ref{fig:cons_vD8}, where
we only plot their changes relative to the initial condition.
The overall energy conservation is expressed as
\begin{equation}
E_{k,{\rm bulk}}+E_{\rm CR}+E_{\rm wave}+E_{\rm internal}={\rm const}\ .
\end{equation}
Note that $E_{k,{\rm bulk}}$ and $E_{\rm CR}$ are frame-dependent.
Here it suffices to report results in our simulation frame.

In the simulation frame, the gas bulk energy density decreases over time (as the bulk gas velocity becomes less negative). This is associated by the work done by
the same parallel force $F_{\parallel}$ that changes the gas momentum. In fact, we can
exactly reproduce the black solid line in the bottom panel of Figure \ref{fig:cons_vD8}
using $F_{\parallel}$ inferred from acceleration in the gas (from the top panel), with
$\Delta E_{k,{\rm bulk}}\approx v_g\Delta {\cal P}_g$.
(Since $v_g$ only changes by a tiny fraction over the process, it is approximately taken to be constant here.)
This same force also does opposite work on the CRs, increasing their total kinetic energy.
Some of this energy is used to excite and amplify the waves, leading to their continued
growth. In the mean time, the waves are slowly dissipated numerically, increasing the
internal energy of the gas.\footnote{While damping in our simulations is purely numerical,
we can see that it is negligible in early phases of linear and quasi-linear evolution. Only
towards later stages where saturation is largely achieved (and hence there is little wave growth)
does numerical dissipation become more noticible. Thus, our results are largely unaffected
by numerical dissipation.}

We emphasize that the CRSI itself does not lead to heating of background gas. Heating
only occurs through internal dissipation in the gas, which is generally accompanied by
wave damping. This fact is reflected in the more recent formulation of
CR-(magneto)hydrodynamics of \citet{ThomasPfrommer19}, though not necessarily
correctly modeled in some earlier studies. 
If wave growth is balanced by wave damping in a quasi-steady state (not achieved in
our present simulations), then the heating rate is simply determined by the rate of wave
growth/damping. 

\subsection[]{Towards More Realistic Simulations}\label{ssec:realistic}

There are several directions to further improve the realism of our simulations.
In particular, there are several wave damping mechanisms to be incorporated,
including ion-neutral damping, non-linear Landau damping, and turbulent damping of waves. 

Ion-neutral damping, or ampibolar damping, is well known (e.g.,
\citealp{KulsrudPearce69,Soler_etal16}). It is present when background gas is partially
ionized. The ions are directly coupled and respond to magnetic fields, but the
neutrals hardly feel the magnetic fields and can possess different bulk velocity from the
ions. Ion-neutral collisions give friction, and lead to damping of Alfv\'en waves
in the ion fluid, especially at small scales. Given that crossing the $\mu=0$ barrier
is sensitive to the spectrum of waves at small scales, ambipolar damping may cause
severe challenge to the isotropization of the CRs.
This effect will be incorporated and studied in greater detail in a follow-up paper.

Non-linear Landau damping is a result of wave-particle interaction at Landau resonance
between background thermal particles and a beat wave formed by two propagating
waves \citep{LeeVolk73}, and is more effective in high-$\beta$ plasmas. Our MHD-PIC
formulation do not capture the kinetic physics of the background plasma, and hence can
not capture non-linear Landau damping. However, this is not necessarily a limitation: we
would lose the advantage of MHD-PIC by resolving the kinetic physics, and in principle,
this damping mechanism can be incorporated in the form of sub-grid physics.

Turbulent damping is due to the distortion of CR-generated Alfv\'en waves when they
collide with oppositely directed wave packets originating from background turbulence.
As a result, the wave energy cascades into smaller scales and is ultimately dissipated
\citep{FarmerGoldreich04}. This damping rate has only been calculated approximately
(e.g., see more recent work of \citealp{Lazarian16}). 
Incorporating this effect is fully compatible with our MHD-PIC
framework, and can be studied in future multi-dimensional simulations.

For this first study, we focused on the classical theory with 1D simulations with uniform
background density and magnetic field, which enable us to confirm resolution and box
size requirements. Multi-dimensional extensions are straightforward and are underway,
although are much more computationally expensive. Such extensions would allow modes
propagating obliquely with respect to background field to be investigated, and open up
more parameter space, as the relative importance of thermal, magnetic and CR pressures,
and background turbulence can play certain roles, 

Closely related to the CRSI is another type of gyroresonance instability, which is
driven by CR pressure anisotropy (e.g., \citealp{LazarianBeresnyak06}),
and can be easily triggered in compressible MHD flows (when CR pressure anisotropy 
exceeds $\sim v_A/c$). It excites circularly polarized Alfv\'en waves propagating to
opposite directions, which then isotropize the CRs via QLD. Recently, \citet{Lebiga18}
studied this instability numerically based on the MHD-PIC framework in full $f$
approach (with relatively strong anisotropy), and found good agreement with quasi-linear
theory. In the absence of CR streaming, the CR distribution function is always symmetric
about $\mu=0$, and hence crossing $90^\circ$ pitch angle is not required for CR
isotropization. In reality, CR streaming and pressure anisotropy likely co-exist. This
would excite a wide ranges of waves propagating in both directions, calling for further
investigations.

Finally, in any simulation with periodic boundary conditions (even with arbitrarily high resolution such that crossing $90^\circ$ pitch angle is not a limitation), there is a finite free energy of the system based on the initial energy and momentum densities of gas and CRs.  In the absence of damping (numerical or physical), growth of Alfv\'en waves would halt when QLD has reduced the mean velocity difference between the gas and the CRs to $v_A$.  
 In reality, at a given location the CRSI is typically driven by a large-scale CR pressure gradient. For instance, CRs are continuously generated from the galactic midplane region, and diffusively propagate downhill through the gradient to escape from the galaxy.  Energy and momentum are continually added to the system.  
This configuration likely yields continued wave growth, balanced by damping to
reach a steady state, a situation that has yet to be addressed with more realistic
boundary conditions.

\subsection[]{Comparison with conventional PIC simulations}\label{ssec:compare}

Our MHD-PIC approach to study the CRSI strongly contrasts with conventional full-PIC or hybrid-PIC approaches that are widely used in modeling highly non-linear plasma problems.
The full-PIC approach treats charged particles of all kinds as particles. It resolves all
scales down to the electron skin depth (eventually the Debye length), and hence all plasma
wavemodes are self-consistently captured along with relevant wave-particle interactions.
The hybrid-PIC approach
treats electrons as a massless conducting fluid. By compromising the electron-scale physics, it alleviates the stringent resolution requirement of the full-PIC method, while
it can still capture the ion-scale physics well.
For the CRSI problem, employing full-PIC or hybrid-PIC methods means that background ISM
fluid must be represented by particles, hence resolving the corresponding microscopic
scales. This causes two major drawbacks: substantially enhanced numerical cost, and PIC
noise due to the finite number of particles representing the background plasma. 

As already discussed in Section \ref{sec:method}, there is substantial scale separation in
the CRSI problem between the CR gyro-radii and the ion skin depth (and below). Conducting
a hybrid-PIC simulation using the same parameter settings as our fiducial simulations
would require more than an order of magnitude more cells, and another order of magnitude
more cells would be needed in the full-PIC case.
The numerical cost in the full-PIC case is further exaggerated by the ion-to-electron mass
ratio, requiring a factor of $m_i/m_e$ more timesteps per CR gyro period.
Moreover, the number of background particles per cell needs to be sufficiently high in
order to reduce the noise floor as compared to the signal level of growing waves and
avoid artificial QLD of the CRs by such noise.

Despite these difficulties, \citet{HolcombSpitkovsky19} have recently presented a 1D study
of the CRSI using full-PIC simulations. Typical parameters of their simulations are: $m_i/m_e=100$, $\upsilon_A/c=0.1$ (note in full-PIC, $c={\mathbb C}$ in our notation),
10/100 cells per electron/ion skin depth, $N_{\rm ppc}=50-1000$ (all plasma particles
equally divided among background and CR species),
$n_{\rm CR}/n_i \in [2\times 10^{-4}, 2 \times 10^{-2}]$ and  $\upsilon_D/\upsilon_A$ from
1.4 (low-anisotropy) to 7.9 (high-anisotropy) \footnote{Let us recall that in present study
({\it Fid} run) the parameters are:
$m_i/m_e \to \infty$, $\upsilon_A/\mathbb C=3.3 \times 10^{-3}$,
10 ion skin depths per cell, $N_{\rm ppc}=2048$ (CRs only),
$n_{\rm CR}/n_i=10^{-4}$ and  $\upsilon_D/\upsilon_A=2$.}.
They considered two different CR distribution functions: ring and power-law, with the former
representing a mono-energetic CR distribution in the drift frame.
They have successfully reproduced the peak growth rate in runs with the ring distribution,
and similar to our results, achieved full isotropization in runs with relatively large
$n_{\rm CR}/n_i$ and/or CR drift velocity.

We note that the large ratio of $v_A/c$ adopted in \citet{HolcombSpitkovsky19} makes the
resonant condition (\ref{eq:res}) asymmetric with respect to forward and backward traveling
particles, thus modifying the standard CRSI dispersion relation for left/right handed modes
in different ways. Together with the choices of large $n_{\rm CR}/n_i$ ratio, drift
velocity being mildly relativistic, and the use of a ring distribution, these parameters
represent substantial compromises to alleviate the issue of scale separation and particle
noise. As a result, reasonable agreement with linear theory was found mainly in the special
case of the ring distribution, and power-law distribution with exaggerated high anisotropy
and high $n_{\rm CR}/n_i$ values. These more extreme run parameters (approaching the Bell
regime), together with the asymmetry in left/right handed modes caused by large $v_A/c$, make
the left handed modes in these simulations grow at much reduced rates compared with right
handed modes, leading to asymmetric scattering for forward/backward traveling CRs. 
While this situation might apply near strong shocks where the Bell instability operates, the
parameter regime is not representative of typical conditions in the ISM.
Additionally, they have suggested mirror reflection as being responsible for crossing
$90^\circ$ pitch angle. However, the reflection event shown in their Figure 10 in
fact appears more consistent with non-linear wave-particle interaction, as we have
demonstrated in this work.
 
Perhaps the chief advantage of full-PIC and hybrid-PIC methods over the MHD-PIC approach
is the ability to capture non-linear Landau damping, as the background plasma is treated
kinetically. However, the physics of non-linear Landau damping is highly subtle, and
damping relies on second-order longitudinal electric fields resulting from a beat wave
formed from two circularly polarized Alfv\'en waves \citep{LeeVolk73}.
It is conceivable that capturing such effect must require extremely low level of
background noise, and hence exceedingly large number of background plasma particles:
another major challenge yet to be resolved.

Finally, we comment that several of the difficulties for simulating the CRSI outlined in
Section \ref{sec:method} are common for all variations of PIC methods. The $\delta f$
approach coupled with a $\kappa$ distribution that we have adopted could also be
incorporated in full-PIC and hybrid-PIC simulations, likely leading to substantial
improvements in future studies of the CRSI.

\section[]{Summary and Conclusions}\label{sec:conclude}

In this paper, we have presented the first numerical studies of the CRSI using the MHD-PIC
method. We use 1D simulations in a periodic box and an idealized setup (no damping, 
a reduced speed of light $\mathbb C$), for a range of model and numerical parameters.  We have
been able to accurately reproduce the theoretically predicted linear growth rate, follow quasi-linear evolution as CR particles interact with CSRI-driven Alfv\'en waves, and
identify the mechanism responsible for CR particles to cross the $90^\circ$ pitch angle. In the course of this study, we have developed techniques to investigate CRSI and QLD with substantial precision,
paving way for future studies with more realistic setups and incorporating more physics.

Unlike conventional PIC methods, the MHD-PIC method has unique advantages in
bypassing the microscopic scales of the background thermal plasma, allowing one to focus on
the interaction between the CRs and background gas around the  gyro-resonant scale. However,
the CRSI still involves substantial scale separation, ranging from scales much smaller than,
to scales much larger than, the CR gyro-resonant scale. Covering this large range of scales is necessary
to properly capture the crossing of the $90^\circ$ pitch angle and QLD.
Moreover, the extremely weak level of anisotropy ($\sim v_A/c$) in the CR distribution function
over a wide energy range makes it challenging to properly represent with a sufficiently large number of particles. 

To address these challenges, we have extended the MHD-PIC formulation to use a $\delta f$
(instead of full $f$) representation of the CR particles on top of a known initial equilibrium
distribution function $f_0$, which dramatically reduces the Poisson noise. While our fiducial
simulations employ as many as 2048 particles per cell to achieve
better precision, similar results can be achieved with fewer particles, and we recommend
testing the level of Poisson noise in a non-CR-streaming simulation to determine the optimal
number of particles. 
Using the conventional full $f$ representation, while
the signature of the CRSI can be identified, the high noise level precludes any quantitative
measurement of its overall properties.

The $\delta f$ method requires the background distribution function $f_0$ to be continuous,
which is incompatible with $f_0$ being a truncated power-law. We have therefore chosen $f_0$ to be a
$\kappa$ distribution, and the corresponding dispersion relation of the CRSI is analytically
derived. We note that under realistic large-scale ISM conditions ($n_{\rm CR}/n_i$ extremely small and moderate $v_D/v_A$), the left and right handed Alfv\'en waves
grow at equal rates. The Bell-type modes, in which the right-handed modes grow
faster, emerge when $n_{\rm CR}/n_i$ and/or $v_D/v_A$ become large. We have carefully chosen
parameters to ensure sufficient scale separation $v_A/{\mathbb C}=1/300$, and that the left
and right handed modes growth at about the same rates (except for run M3 where there is
a small difference), so that the conditions are as close to ISM as possible. Our particle
population also covers a wide range of CR momenta from $0.01p_0$ to $100p_0$, with
$p_0={\mathbb C}$ being the characteristic momentum of the $\kappa$ distribution.

We have found that when employing a standard periodic boundary condition,
properly capturing the linear growth rate requires an excessively long simulation box, corresponding to the
box-crossing time of a typical particle approaching the wave growth time. 
We attribute this
to the fact that the growth of the CRSI is driven by QLD, which requires particles to
encounter sufficient number of independent wave packets, and is related to the
random-phase approximation commonly employed in describing QLD. 
To address this issue, we implemented a phase-randomization technique for particles that cross the periodic
boundaries, which allows us to achieve even better results using much smaller
simulation boxes (size $\lesssim50$ instead of $\gtrsim1000$ most unstable wavelength).

Equipped with these techniques, we are able  not only to accurately reproduce predicted CRSI linear
growth rates over a broad range of wave spectrum, but also to accurately follow the
quasi-linear evolution of the system. We observe that QLD proceeds to isotropize the
CR distribution function $f_w$ in the frame of forward-propagating Alfv\'en waves, so
that $\pa f_w/\pa\mu_w$ approaches 0 on either side of the pitch angle cosine
$\mu_w\approx\mu=0$. With an increasingly flat distribution of particle pitch angles, wave growth slows down and eventually stops.
As longer-wavelength modes that resonate with higher-energy CRs grow more slowly,
wave growth proceeds from the most unstable to longer wavelengths (and in principle also
to shorter wavelengths, though we lack resolution for these).  In parallel,  quasi-linear evolution
proceeds from low-energy to higher-energy CRs as waves become available.

Whether full isotropization of CRs in the wave frame is achieved depends on whether
particles manage to overcome the $\mu=0$ barrier (crossing the $90^\circ$ pitch angle).  The mechanisms involved go beyond quasi-linear theory. While mirror reflection is commonly invoked
as the main mechanism enabling crossing $\mu=0$, we have found that reflection generally occurs
via non-linear wave-particle interaction, with particles encountering abrupt changes in
field orientation. Reflection is more easily achieved for higher-energy particles, and is
facilitated by the presence of waves at wavelength much smaller than the
gyro-resonance scale (see Section \ref{ssec:90deg} for details). 

In our study, efficient crossing of the $\mu=0$ barrier is achieved in simulation runs
with strong wave growth (runs M3 and vD8). For these runs, full saturation is reached, with the mean CR drift velocity reduced to Alfv\'en speed.  The final particle distribution function,
as well as saturation amplitudes of the waves, are in excellent agreement with
theoretical expectations.
Our fiducial run, with a relatively low $n_{\rm CR}/n_i=10^{-4}$ and small initial drift
speed $v_D=2v_A$, becomes  stuck in a state of pre-mature saturation: while the distribution is flat on each side of $\mu=0$ within most momentum bins, the bulk of the CR
population fails to overcome the $\mu=0$ barrier. The inability of particles to breach the $\mu=0$ barrier is primarily due 
to numerical dissipation of low-amplitude waves at small scales, together with absence of even shorter-wavelength waves due to limited resolution.   We have further found that enhancing
numerical resolution can substantially alleviate the situation. Thus, proper choice of
numerical resolution for given physical parameters is also crucial for saturation.

The CRs provide direct momentum feedback to background gas, which accelerates
in accordance with momentum conservation of the gas-CR system. Wave power is directly 
proportional to the momentum transferred from the CRs to the gas.
In terms of energy feedback, the work
done by the CRs leads to gas acceleration and wave growth, but no direct
heating. Gas heating is only achieved through wave damping, which is
purely numerical in our simulations.

As a first study, we have restricted ourselves to 1D simulations in a periodic box in
the ideal MHD limit. Our explorations 
have identified and clarified the
numerical methods necessary to properly capture the essential 
mechanisms of the CRSI, and have
paved the way for future investigations to incorporate more realistic physics. These
may include implementing various damping mechanisms, extensions to
multi-dimensions, and employing more realistic boundary conditions.

\acknowledgments

We thank Cole Holcomb, Matt Kunz, Peng Oh, and Anatoly Spitkovsky for useful discussions.
XNB acknowledges support by the Youth Thousand Talent program.  The work of ECO  was supported by grant 510940 from the Simons Foundation. 
The work of IP was supported by NSF grants PHY-1804048 and PHY-1523261 and facilitated by the Max-Planck/Princeton Center for Plasma Physics.
Computation of this work has been conducted at computational facilities at Princeton
Institute for Computational Science and Engineering (PICSciE), and at the YZ cluster
from GyroTech Co. Ltd.

\appendix 

\section[]{A. Decomposition of Alfv\'en Waves}\label{app:wavedecomp}

Given that modifications from the CRs to the Alfv\'en waves are tiny in this context, it
suffices to consider pure linear Alfv\'en waves. At fixed wavelength, there are four independent
Alfv\'en modes, depending on left/right-handed polarization, and direction of propagation relative
to fluid rest frame.
We here take the perspective that the left/right handedness is defined from the receiver's
point of view. That is, for a wave propagating toward an observer (relative to the fluid
rest frame), the perturbed electric (and magnetic) field vector rotates in a anti-clockwise (clockwise)
sense for a left (right) handed wave.

Let $\delta{\mb v}$ and $\delta{\mb B}$ be perturbed transverse velocities and magnetic
fields, and take the perturbations in the form of $\exp{[{\rm i}(kx-\omega t)]}$. Note that $k$ can
have both signs indicating direction of propagation, whereas $\omega$ always has positive real part.
The eigenvectors for Alfv\'en waves satisfy
\begin{equation}
\frac{\delta{\mb v}}{v_A}=-{\rm sgn}(k)\frac{\delta{\mb B}}{B_0}\ .
\end{equation}
The sign of $B_0$ is implicitly contained in the above, whereas $v_A$ is considered to be
positive. In the following, we always take $B_0>0$ without loss of generality (for $B_0<0$, sign chages
must be employed in many of the subsequent descriptions), and consider forward propagation to
be along the direction of ${\mb B_0}$ (positive $\hat x$). Under this convention, 
the $+/-$ sign of $k$ corresponds to  forward/backward propagation. The left/right-handed modes have
$\delta B_y=\mp{\rm i}\delta B_z$ and $\delta v_y=\mp{\rm i}\delta v_z$ for forward-propagating waves.
The opposite sign should be adopted in the above for backward-propagating waves.

Based on the above description, the natural way to decompose circularly polarized
Alfv\'en modes is to consider the following combination:
\begin{equation}
C^\pm(x)=\frac{1}{2}\bigg[\frac{\delta v_y\pm{\rm i}\delta v_z}{v_A}
-{\rm sgn}(k)\frac{\delta B_y\pm{\rm i}\delta B_z}{B_0}\bigg]\ .
\end{equation}
For instance, for forward propagating ($k>0$), right-handed ($\delta B_y={\rm i}\delta B_z$)
waves, we choose the plus sign. A discrete Fourier transform of $C^\pm(x)$,
\begin{equation}
W^\pm(k_i)=\frac{1}{N}\sum_{n=0}^{N-1}C^\pm(x_n)e^{-{\rm i}k_ix_n}\ ,
\qquad{\rm where}\ k_i=\frac{2\pi}{L_x}i\ ,
\end{equation}
gives the wave amplitude and phase (which is what is being plotted in the Figures).
The wave intensity is then given by
\begin{equation}
I^\pm(k)=|W^\pm(k)|^2\frac{L_x}{2\pi}\ .
\end{equation}
To initialize a wave spectrum with $I^\pm(k)=A^2/k$, we can simply choose
$|W^\pm(k)|=A\sqrt{\Delta k/|k|}$, where $\Delta k\equiv 2\pi/L_x$.

\begin{figure*}
    \centering
    \subfigure{
    \includegraphics[width=87mm]{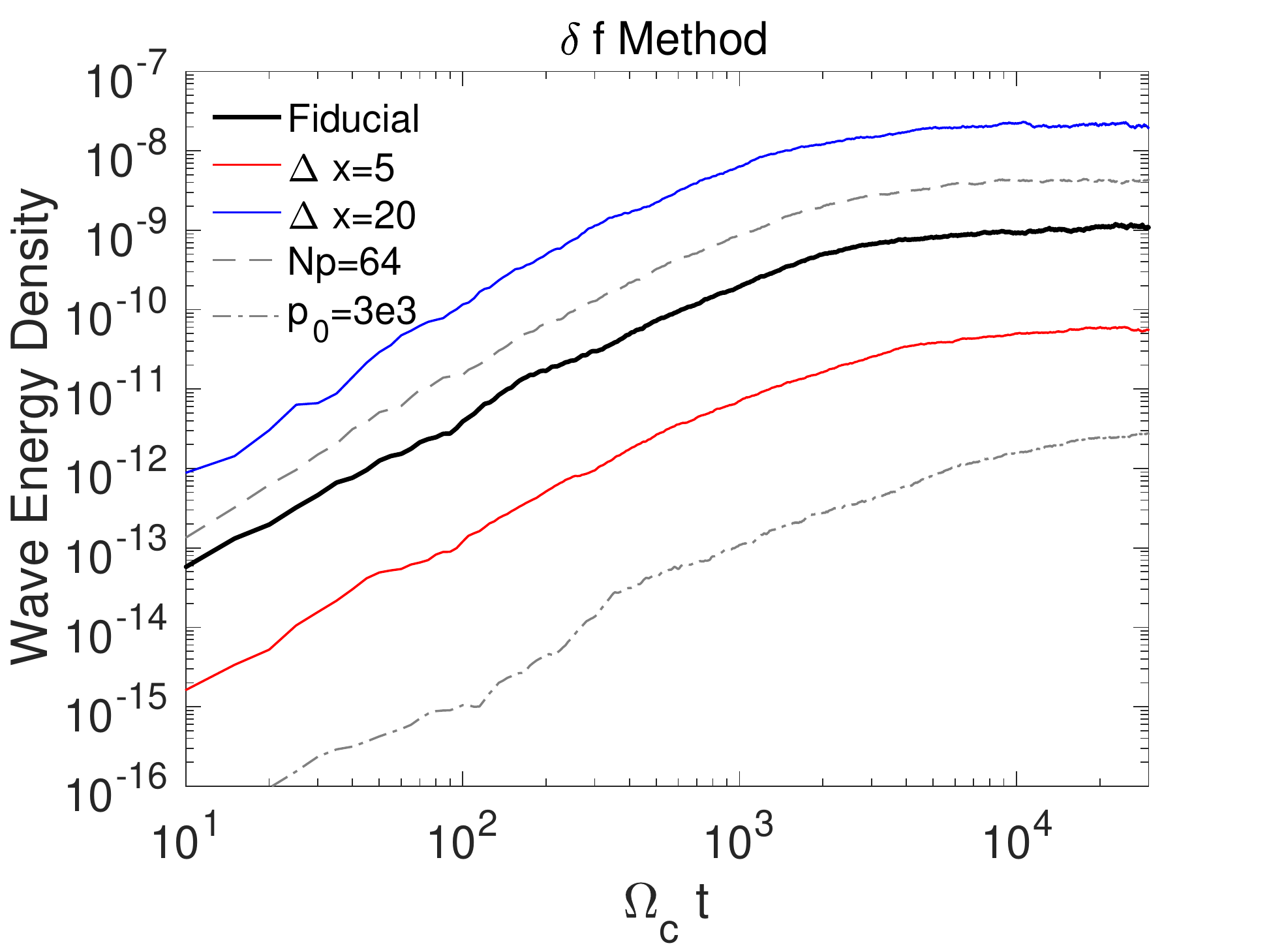}}
    \subfigure{
    \includegraphics[width=87mm]{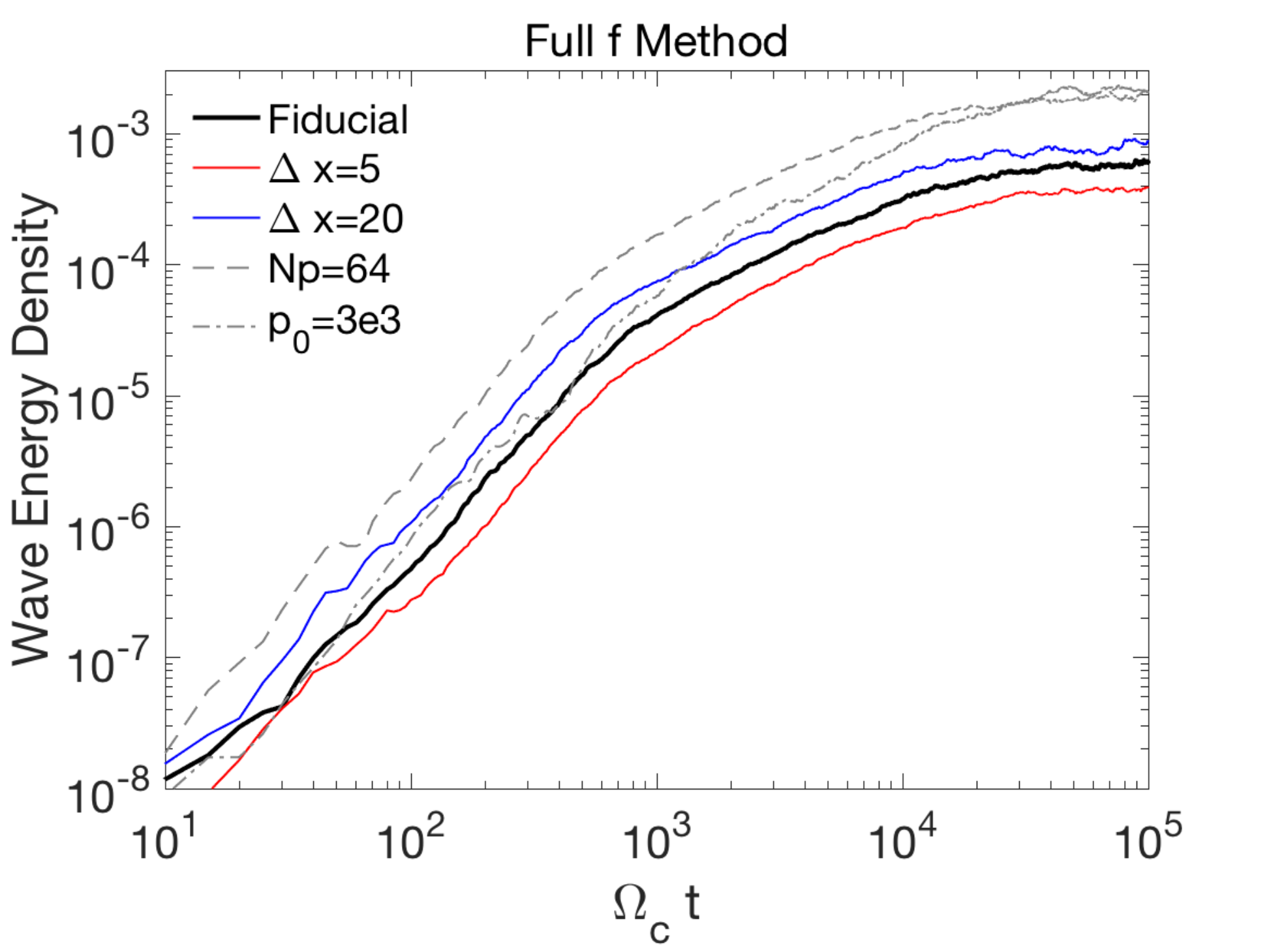}}
  \caption{Time evolution of system noise (in terms of total wave energy density)
  in zero-drift simulations described in Appendix \ref{app:noise}. Left and right
  panels show the results from the $\delta f$ method and the full $f$ method, respectively.
  Bold black lines in both plots correspond to simulations with identical setup to the
  fiducial simulation except that $v_D=0$, and $n_{\rm CR}/n_i=10^{-3}$ for
  the $\delta f$ case while $n_{\rm CR}/n_i=10^{-4}$ in the full $f$ case. Other lines marked in the
  legend indicate runs where one of the parameters is varied while all others are unchanged.}\label{fig:noise}
\end{figure*}

\section[]{B. Level of Noise in the $\delta f$ and full $f$ Methods}\label{app:noise}

In this Appendix, we study the noise properties in our simulations. We do so by conducting
simulations with almost identical setup to our fiducial runs, except that we change the CR
drift velocity to $v_D=0$, and that we start from an initial amplitude that is much smaller
with $A=10^{-12}$. We also use a shorter simulation box $L=24000$ (which we verify has no
influence on the result).
In this case, we only expect wave damping to occur, but Poisson noise from particle sampling
will stand out and causes fluctuations that overwhelm the initially low-amplitude waves.

In Figure \ref{fig:noise}, we show the time evolution of mean fluctuation energy density in the
system, which is given by $(\rho v_\perp^2+B_\perp^2)/2$. This is done for both the $\delta f$
method (left), and the full $f$ method (right). Note that we have chosen $n_{\rm CR}/n_i=10^{-3}$
in tests using the $\delta f$ method (a smaller value would take longer time to reach steady state),
whereas we use $n_{\rm CR}/n_i=10^{-4}$ in the full $f$ case (larger value would make the noise
somewhat too large). Clearly, we see that even with a higher $n_{\rm CR}/n_i$, the noise with
the $\delta f$ method is dramatically smaller than that in the full $f$ method by more than five
orders of magnitude (this would be more than six orders of magnitude if they had the same
$n_{\rm CR}/n_i$).

The energy density of the waves in this test results from the Poisson noise in the particles. We
can see that in both cases, reducing the number of particles by a factor of $4$ leads to a factor
of $\sim4$ increase in wave energy density (or a factor $\sqrt{4}$ in wave amplitude). The wave
spectrum (not shown) is consistent with white noise (with $I(k)$ being flat) for all modes with 
cutoff at high $k$ due to numerical damping. The spectrum at intermediate $k$ grows and
saturates faster, whereas modes at the low-$k$ end grow slower, explaining the secular trend
at late times.

Interestingly, the noise level also depends on resolution, but the dependence is different for the
full $f$ and $\delta f$ methods. For the $\delta f$ method, we find a sensitive dependence, which
scales roughly as $\Delta x^4$, and higher spatial resolution can greatly reduce Poisson noise.
On the other hand, for the full $f$ method, the dependence is much weaker and is sub-linear.
In addition, we find that using a larger $p_0$ while keeping other parameters fixed also gives
very different noise levels. Enlarging $p_0$ by a factor of 10, noise is dramatically reduced by
a factor of $\sim300$ when using the $\delta f$ method. 
Noise is enhanced
by a factor of $\sim3-4$ when using the full-$f$ method. We do not aim at further studying and
explaining these trends, but simply report them here for reference.

\begin{figure*}
    \centering
    \subfigure{
    \includegraphics[width=87mm]{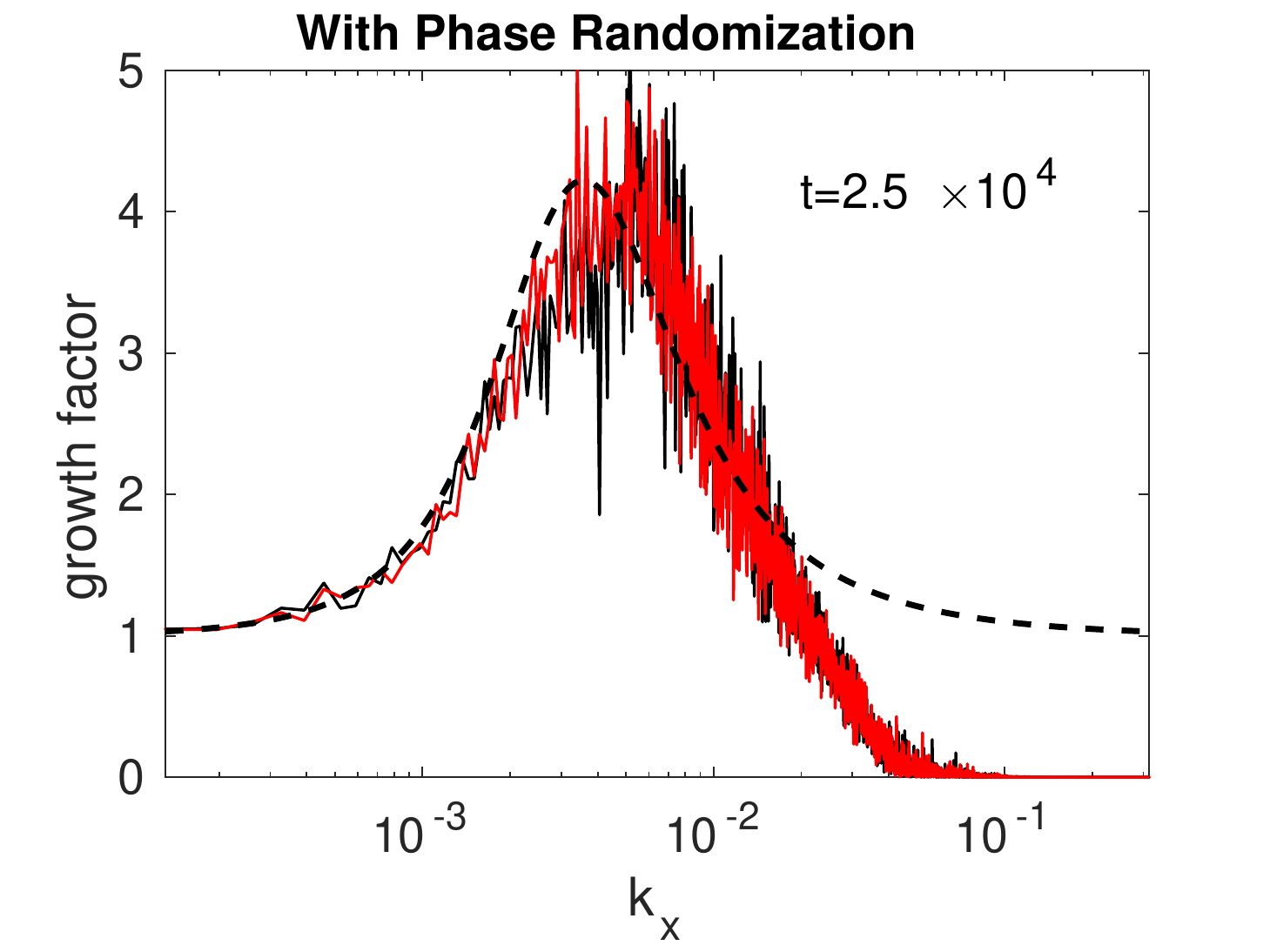}}
    \subfigure{
    \includegraphics[width=87mm]{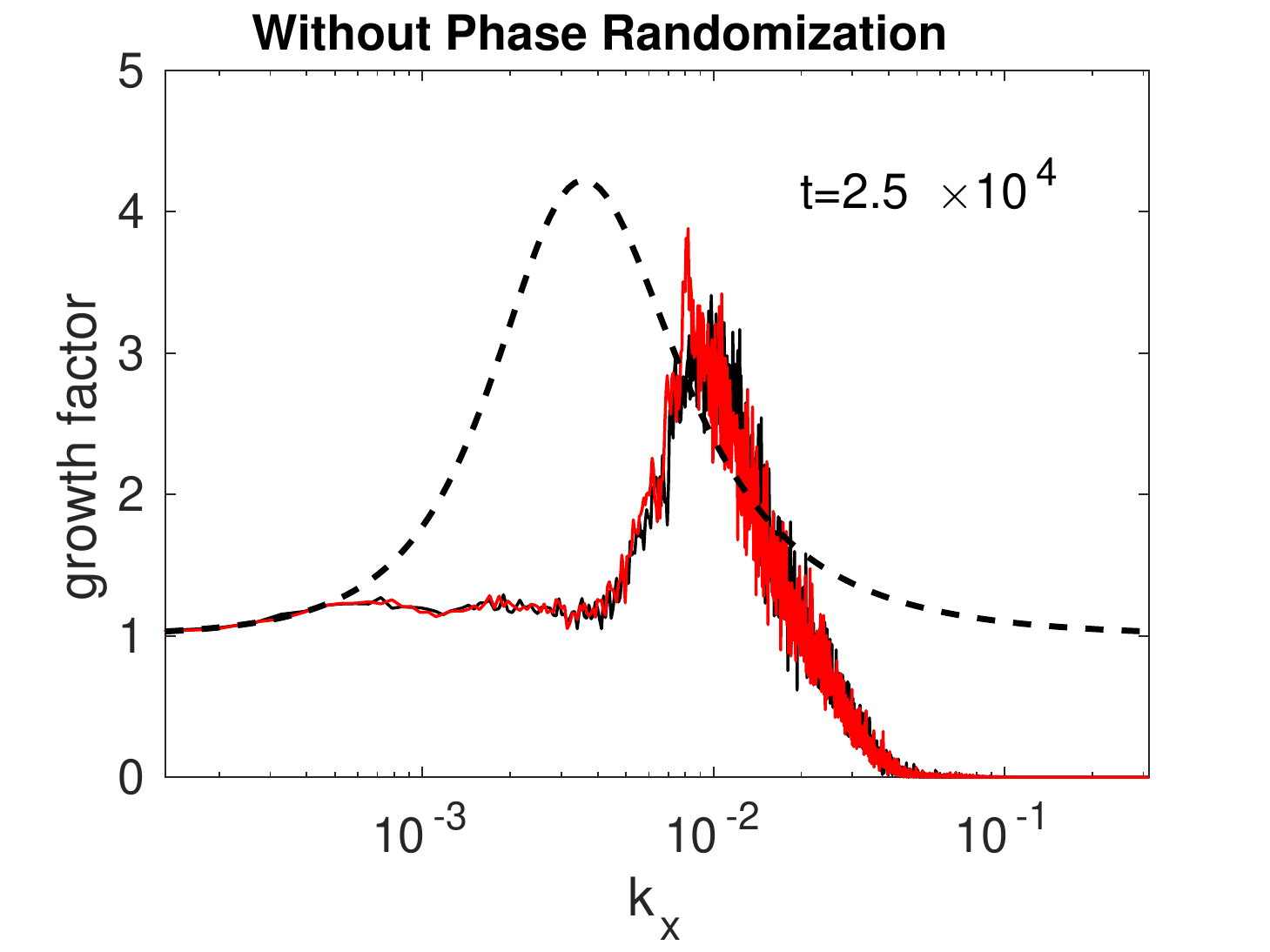}}
    \subfigure{
    \includegraphics[width=87mm]{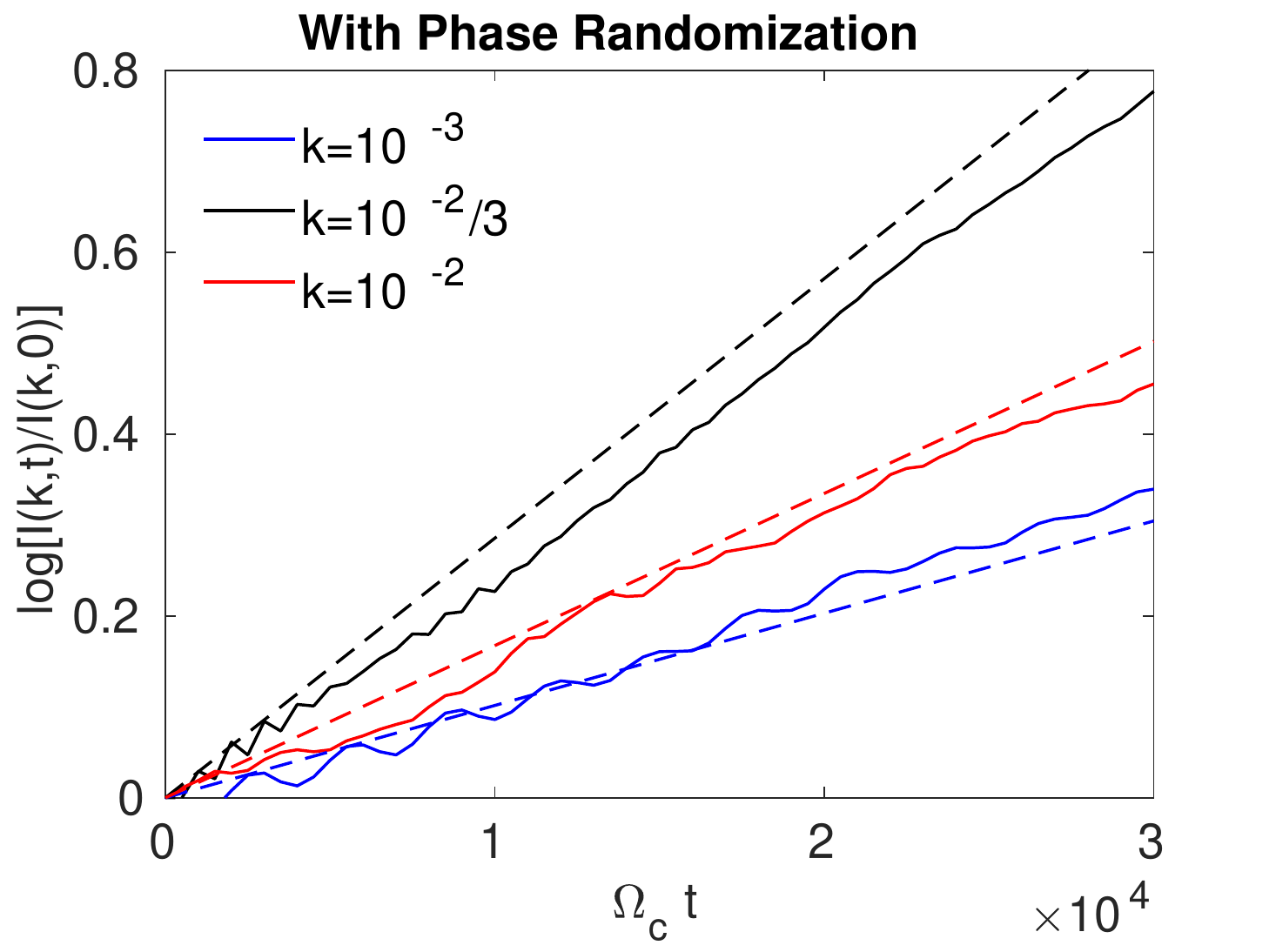}}
    \subfigure{
    \includegraphics[width=87mm]{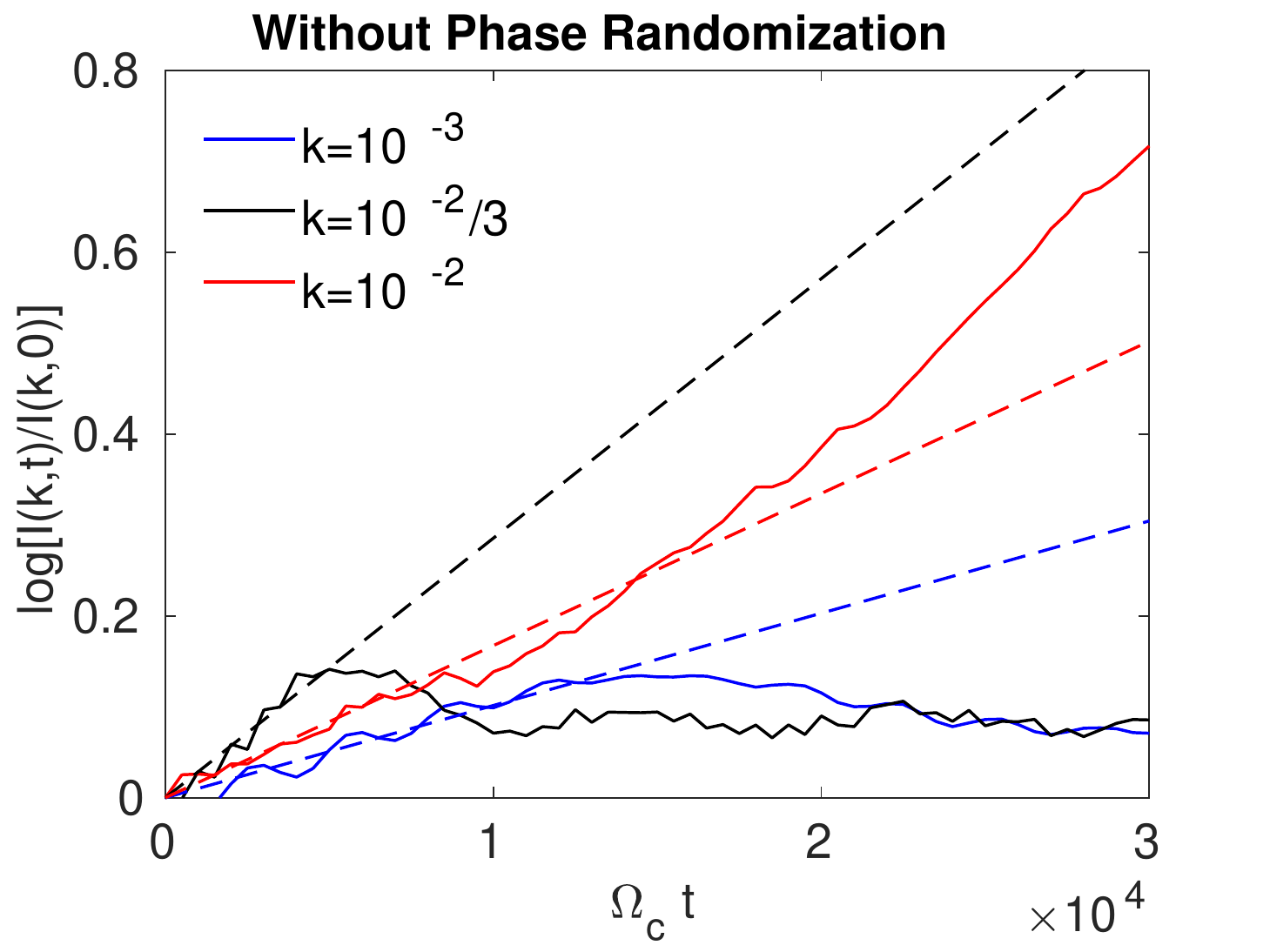}}
  \caption{Comparison between fiducial simulations with (left) and without (right) implementing
  phase randomization across periodic boundaries. The top two plots show the growth factor
  of wave intensity $I(k)$ between $t=0$ and $t=2.5\times10^4\Omega_c^{-1}$, with red/black
  lines marking left/right handed modes. The dashed line show the growth factor expected from
  linear theory. The bottom two plots show the time history of growth factor for individual modes
  with different wave numbers $k_x$ (indicated in the legend). The dashed lines of the same
  color indicate the expected growth factor evolution.}\label{fig:phrand_cmp}
\end{figure*}

\begin{figure*}
    \centering
    \subfigure{
    \includegraphics[width=58mm]{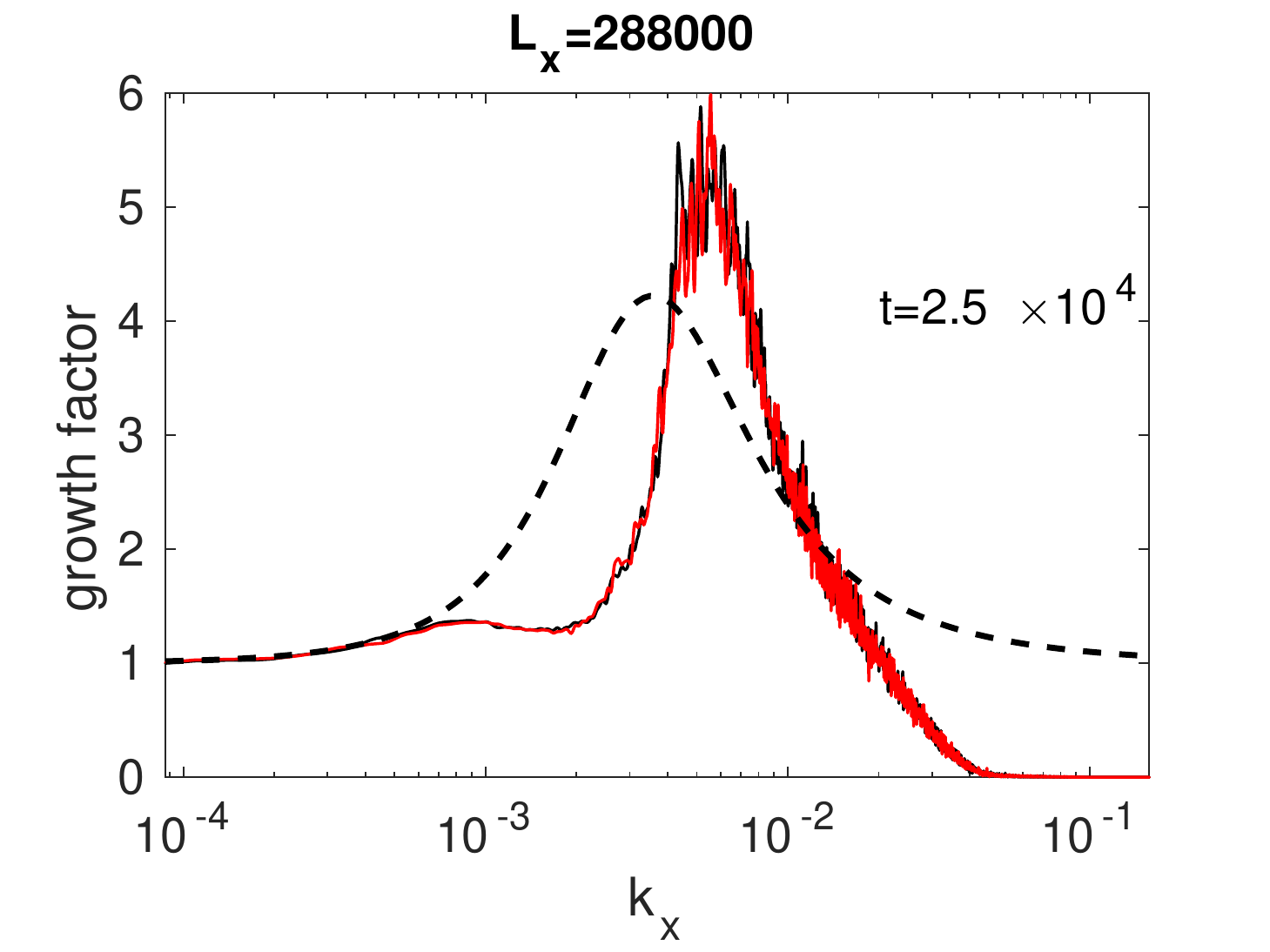}}
    \subfigure{
    \includegraphics[width=58mm]{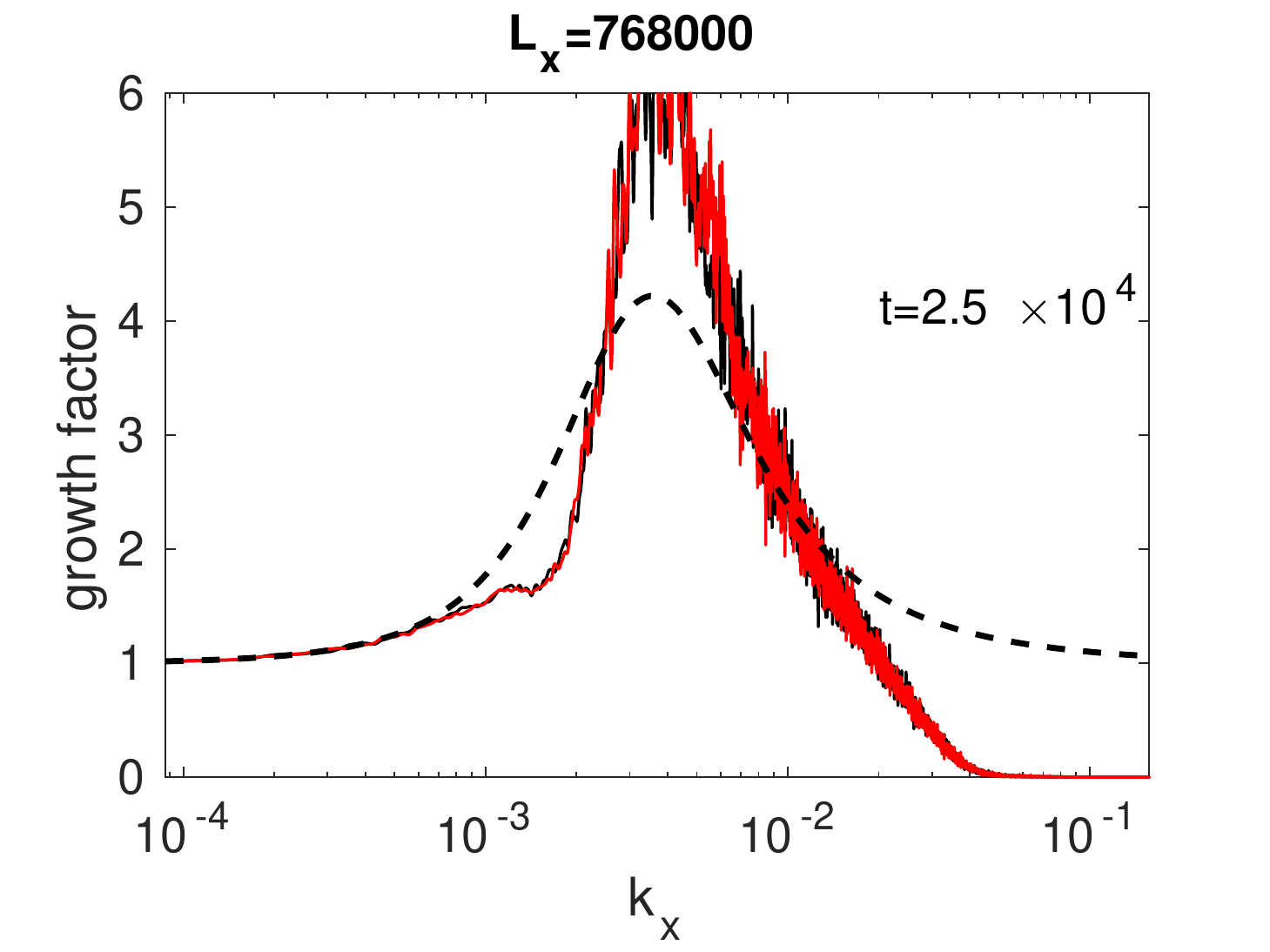}}
    \subfigure{
    \includegraphics[width=58mm]{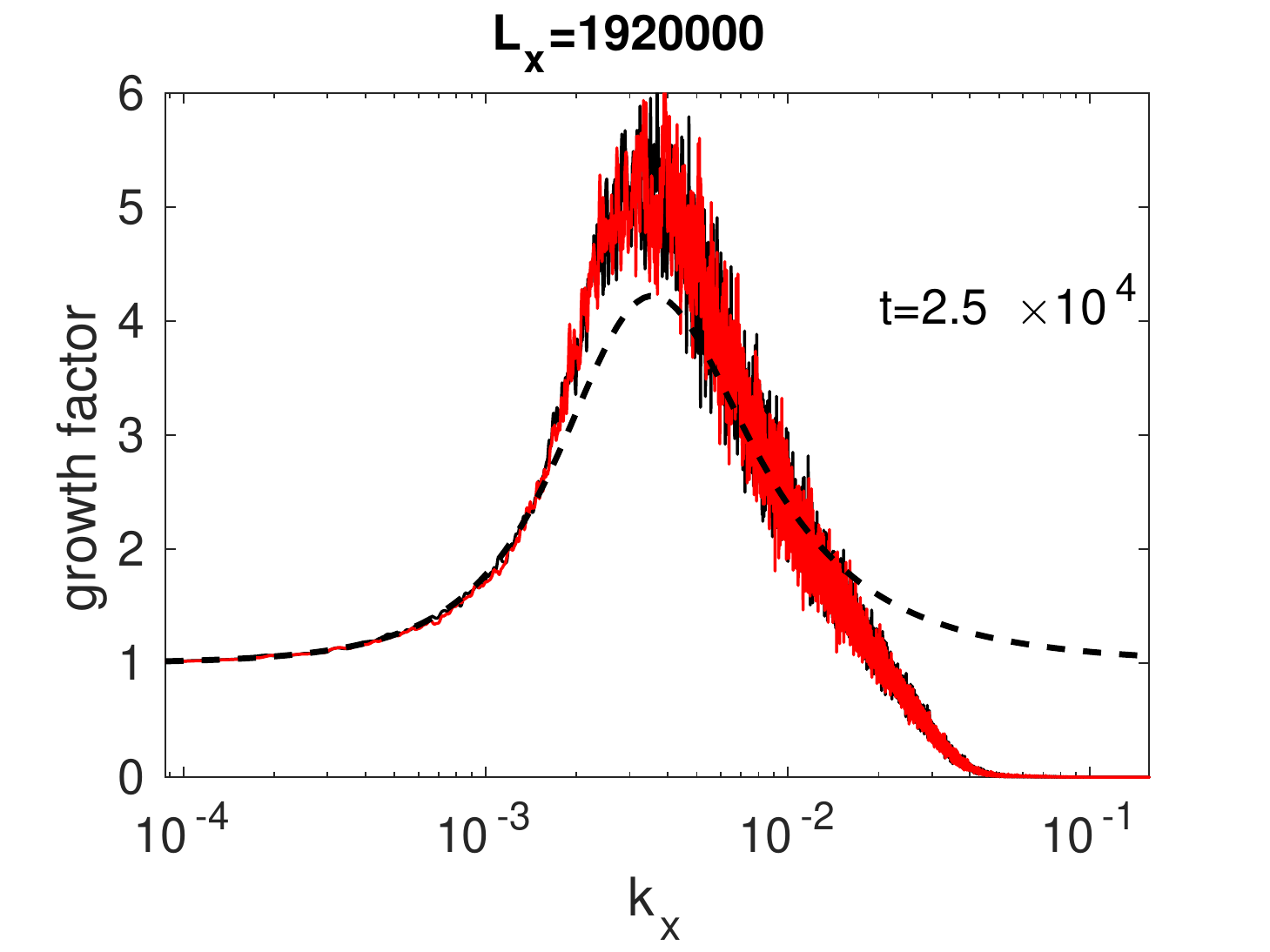}}
  \caption{Growth factor of wave intensity $I(k)$ at time $t=2.5\times10^{-4}$ for three runs
  without phase randomization. Red/black color mark left/right handed modes.
  }\label{fig:boxsize_cmp}
\end{figure*}

\section[]{C. Random Phase Approximation and Simulation Box Size}\label{app:boxsize}

In this appendix, we show that simulations without implementing phase randomization
across periodic boundaries fail to properly capture the linear grow rate of the CRSI unless
the simulation box size is extremely long. 
This failure reflects the fact that the random phase
approximation must be satisfied for particles to properly follow quasi-linear diffusion,
which drives the CRSI.

We first show in Figure \ref{fig:phrand_cmp} a comparison between our fiducial run with
and without phase randomization. At some early time (less than one e-folding time), we
plot the growth factors across the wave spectrum. Despite some slight deviation, our fiducial run
with phase randomization nicely matches theoretical expectation (except for numerical damping
at large $k$). Without phase randomization, we see that while the fastest growth rate is not
different from theoretical expectation by much, the wave spectrum has significant offset, with
most unstable wavelength being about a factor of three shorter.

Further looking at the time history of the wave intensity of three representative modes
(with the fastest growing mode $k=(1/3)\times10^{-2}$ shown in black) in the bottom two
panels, we see that with phase randomization, wave growth is clearly exponential. Without
randomizing the phases, the exponential growth is only maintained for up to
$t\sim5000\Omega_c^{-1}$, after which the wave amplitudes wildly vary.

We then conducted a serious of simulations using fiducial parameters but in different box
sizes and without phase randomization. The box sizes are $L_x=2.88\times10^5$,
$7.68\times10^5$ and $1.92\times10^6$, respectively. They are $3$, $8$ and $20$ times
the length of the fiducial box that we use ($0.96\times10^5$), while they are run for shorter
amount of time (about $2.5-4\times10^4\Omega_c^{-1}$) just within the expected linear
growth phase. Note that even for the longest box that we have attempted, the box crossing
time for a relativistic particle is $1.92\times10^6/300=6.4\times10^3\Omega_c^{-1}$, which
is about a factor $\sim4-5$ smaller than the growth time of the CRSI.

In Figure \ref{fig:boxsize_cmp}, we show again the spectrum of growth factors for these runs.
Clearly, the offset in the fastest growing wavelength from theoretical expectation gets smaller
with increasing box size, and eventually vanishes in the run with longest box size. There is
still some deviation in the absolute growth rate, though only by a small factor (the numerical
growth rate is accurate within about $\sim10\%$). We thus consider this longest box size run
to be marginally appropriate to study the CRSI.

\begin{figure*}
    \centering
    \subfigure{
    \includegraphics[width=88mm]{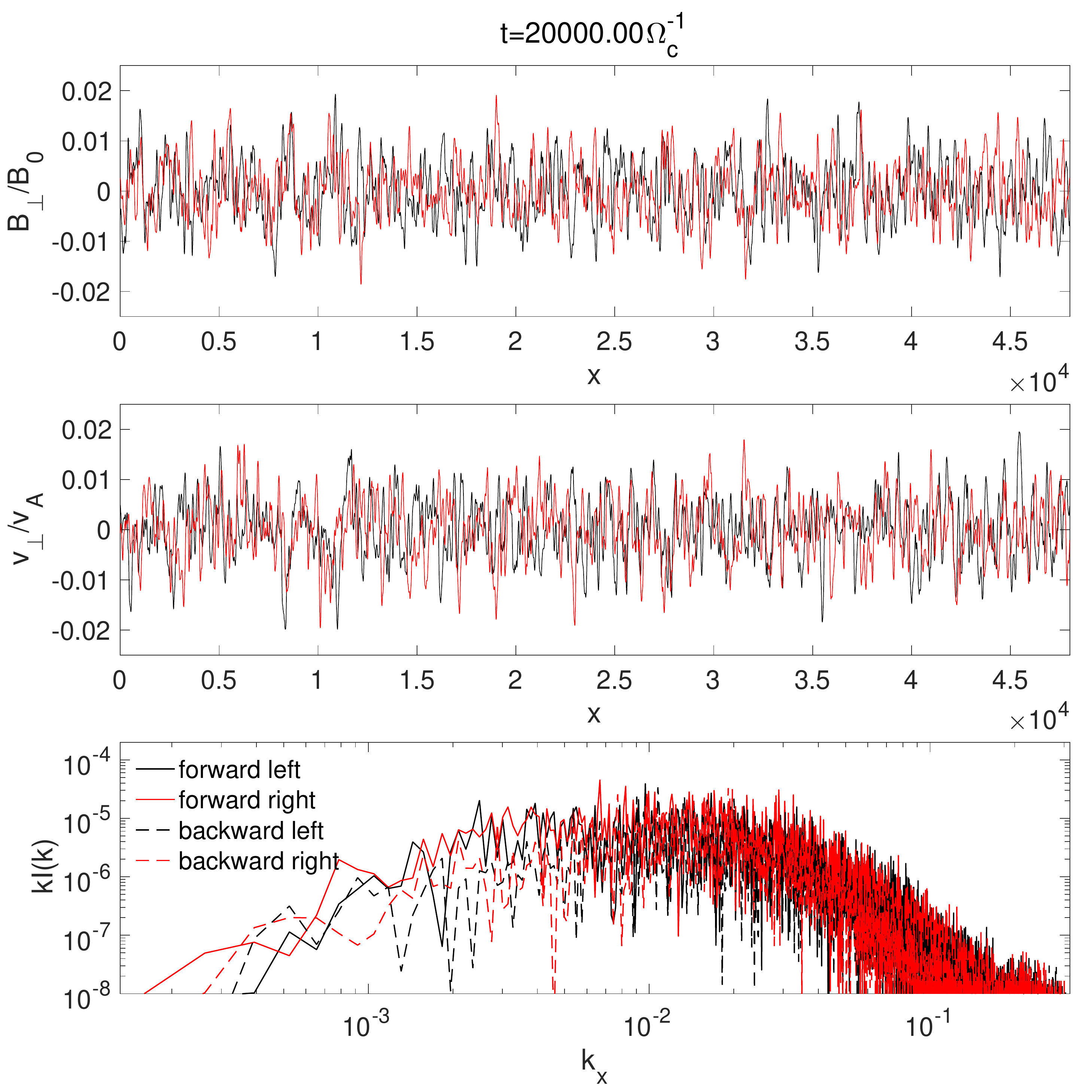}}
    \subfigure{
    \includegraphics[width=88mm]{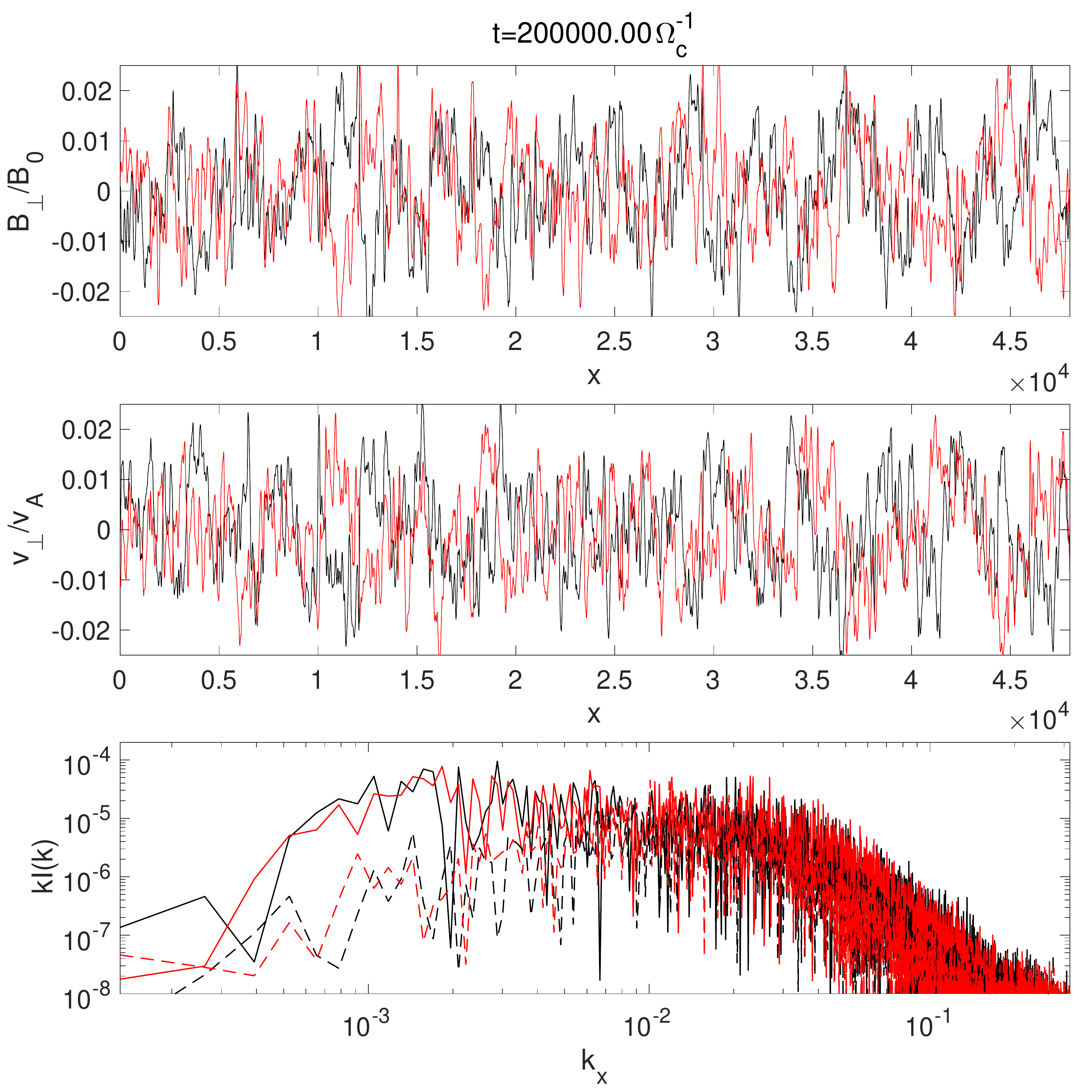}}
  \caption{Wave pattern and their Fourier power spectrum [$kI(k)$] at two snapshots of the
  fiducial simulation using the full-$f$ method: left for $t=2\times10^4\Omega_c^{-1}$ and right
  for $t=2\times10^5\Omega_c^{-1}$.}\label{fig:fullf}
\end{figure*}

\section[]{D. Comparison with the Full-$f$ method}\label{app:fullf}

To demonstrate the advantage of the $\delta f$ method, we show in this Appendix
results obtained from the full $f$ approach. The simulation setup is identical to our
run Fid except that the $\delta f$ weighting scheme is turned off and that the box size is halved. Phase randomization
is also included. To further reduce
Poisson noise, we use $N_p=2048$ particles per bin per cell (in total $16384$ per cell).
According to the study in Appendix \ref{app:noise}, the level of noise is reduced by
a factor of $8$ in wave intensity, with wave energy density $\sim10^{-4}$, or
$\delta B^2/B_0^2\lesssim5\times10^{-5}$, in the form of Alfv\'en waves propagating
in both directions. This level of noise would yield a rate of QLD, according to
(\ref{eq:nu_qld}), that is comparable to the maximum wave growth rate of
$\sim3\times10^{-5}\Omega_c$. The simulation is run to time $t=2.25\times10^5\Omega_c^{-1}$.

In Figure \ref{fig:fullf}, we show the two snapshots the wave pattern and spectrum.
We see that noise level indeed quickly reaches the level of $\delta B/B\lesssim10^{-2}$,
causing appreciable QLD of particles. 
Note that since Alfv\'en waves in both directions
are present due to noise, CR energy would no longer be conserved in the (forward-propagating)
wave frame, compromising the physics of the CRSI. As a result, we can only barely identify
the signature of the CRSI in the wave spectrum at early time of $t=2\times10^4$. At
later time, forward-propagating waves are more easily seen to be amplified. However, these
waves are at longer wavelengths (with $k\sim10^{-3}$ instead of
$\Omega_c/p_0=3.3\times10^{-3}$) where intrinsic numerical noise is smaller, allowing
the CRSI to better stand out. The growth rates at these longer wavelength are somewhat
too noisy to measure with sufficient precision to compare with theoretical values, and
through our analysis, particles responsible for wave growth must have already undergone 
some QLD from pure noise.

\begin{figure}
    \centering
    \includegraphics[width=88mm]{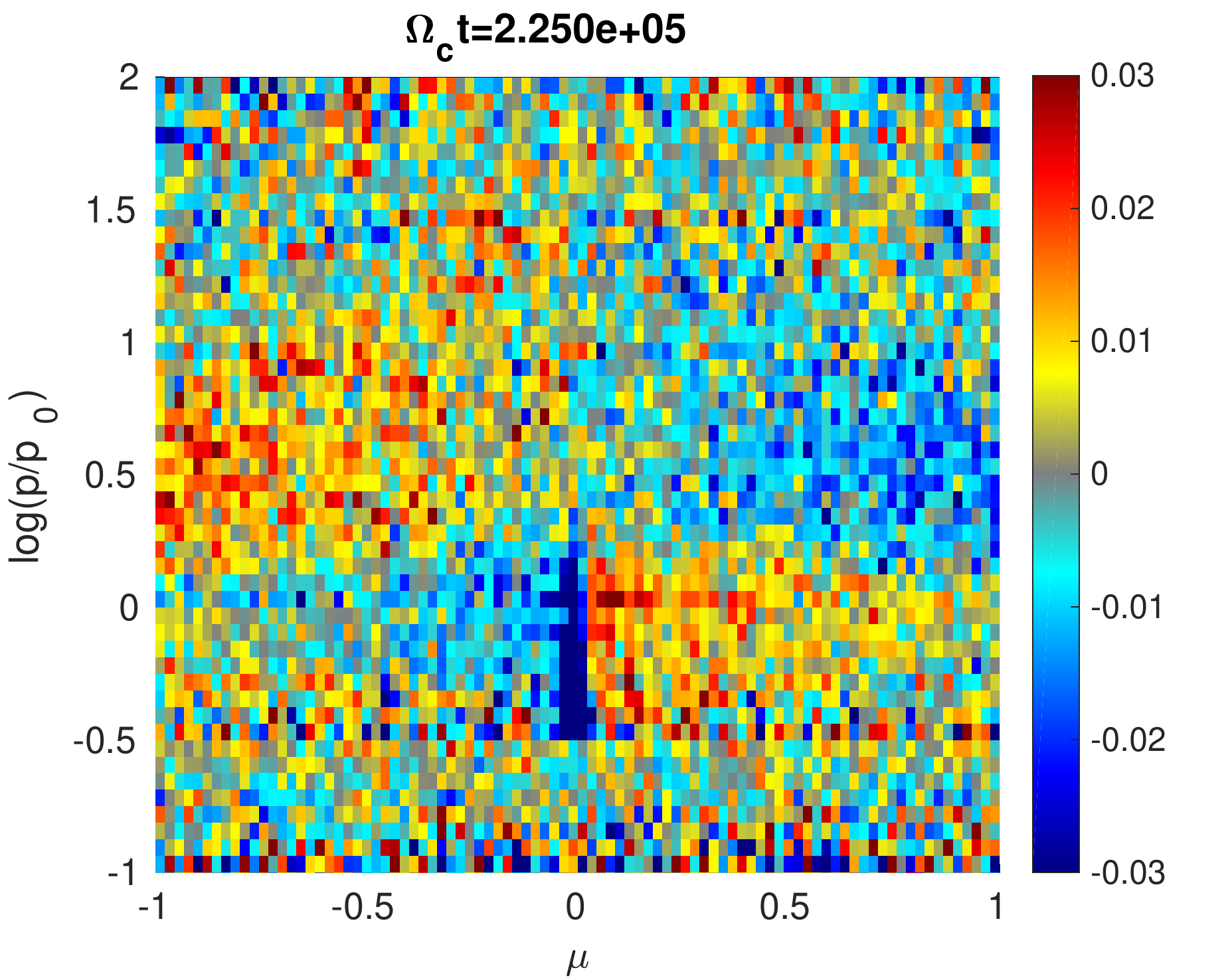}
  \caption{The distribution function $(f-f_0)/f_0$ at the last snapshot $t=2.25\times10^5\Omega_c^{-1}$
  from the fiducial simulation using the full-$f$ method. }\label{fig:fullf_f}
\end{figure}

We have further inspected the evolution of the distribution function, and show in Figure
\ref{fig:fullf_f} this distribution function for the last snapshot of the run. Note that because
the binning is based on the full-$f$ prescription (i.e., direct particle counting in each bin),
it is much noisier than we had before under the $\delta f$ prescription. On the other hand,
we can find signatures of isotropization. In fact, particles at $p\sim3p_0$ are already nearly
isotropized, and particles with $p\gtrsim10p_0$ are still in early stages of isotropization.
For particles with $p\lesssim p_0$, accumulation near $\mu\sim0$ is also clearly
visible, as they are stuck and can not cross the $90^\circ$ pitch angle. While these features
are qualitatively consistent with expectations, QLD proceeds much faster than that obtained
using the $\delta f$ method owing to substantial enhancement of scattering through noise.
In particular, recall from discussions in Section \ref{ssec:qlinFid}, isotropization is never achieved
when using the $\delta f$ method. 

Overall, we conclude that while the full-$f$ method can capture some basic properties of
the CRSI, numerical noise adversely affects its evolution at all stages, leading to severe
discrepancies at a quantitative (and sometimes even qualitative) level, which precludes making any
accurate/reliable measurement.
The situation can only get worse under more realistic conditions (low $n_{\rm CR}/n_i$
and small $v_D$).

\bibliographystyle{apj}

\begin{thebibliography}{59}
\expandafter\ifx\csname natexlab\endcsname\relax\def\natexlab#1{#1}\fi

\bibitem[{{Achterberg}(1981)}]{Achterberg81}
{Achterberg}, A. 1981, \aap, 98, 161

\bibitem[Amato \& Blasi(2009)]{Amato09} Amato, E., \& Blasi, P.\ 2009, \mnras, 392, 1591

\bibitem[Amato \& Blasi(2018)]{Amato18} Amato, E., \& Blasi, P.\ 2018, Advances in Space Research, 62, 2731 

\bibitem[{{Baade} \& {Zwicky}(1934)}]{BaadeZwicky34}
{Baade}, W. \& {Zwicky}, F. 1934, Proceedings of the National Academy of
  Science, 20, 259

\bibitem[{{Bai} {et~al.}(2015){Bai}, {Caprioli}, {Sironi}, \&
  {Spitkovsky}}]{Bai_etal15}
{Bai}, X.-N., {Caprioli}, D., {Sironi}, L., \& {Spitkovsky}, A. 2015, \apj,
  809, 55

\bibitem[{{Bell}(1978)}]{Bell78}
{Bell}, A.~R. 1978, \mnras, 182, 147

\bibitem[{{Bell}(2004)}]{Bell04}
---. 2004, \mnras, 353, 550

\bibitem[{{Birdsall} \& {Langdon}(2005)}]{BirdsallLangdon05}
{Birdsall}, C.~K. \& {Langdon}, A.~B. 2005, {Plasma Physics Via Computer
  Simulation} (Taylor \& Francis Group, 2005)

\bibitem[{{Blandford} \& {Ostriker}(1978)}]{BlandfordOstriker78}
{Blandford}, R.~D. \& {Ostriker}, J.~P. 1978, \apjl, 221, L29

\bibitem[Blasi et al.(2012)]{Blasi12} Blasi, P., Amato, E., \& Serpico, P.~D.\ 2012, Physical Review Letters, 109, 061101 


\bibitem[{{Booth} {et~al.}(2013){Booth}, {Agertz}, {Kravtsov}, \&
  {Gnedin}}]{Booth_etal13}
{Booth}, C.~M., {Agertz}, O., {Kravtsov}, A.~V., \& {Gnedin}, N.~Y. 2013,
  \apjl, 777, L16

\bibitem[{Boris(1970)}]{Boris70}
Boris, J.~P. 1970, in Proceedings of the Fourth Conference on Numerical
  Simulation Plasmas (Navel Research Laboratory, Washington, D.C.), pp. 3--67

\bibitem[{{Breitschwerdt} {et~al.}(1991){Breitschwerdt}, {McKenzie}, \&
  {Voelk}}]{Breitschwerdt_etal91}
{Breitschwerdt}, D., {McKenzie}, J.~F., \& {Voelk}, H.~J. 1991, \aap, 245, 79

\bibitem[Breitschwerdt et al.(1993)]{Breitschwerdt93} Breitschwerdt, D., McKenzie, J.~F., \& Voelk, H.~J.\ 1993, \aap, 269, 54 


\bibitem[{{Denton} \& {Kotschenreuther}(1995)}]{DentonKotschenreuther95}
{Denton}, R.~E. \& {Kotschenreuther}, M. 1995, Journal of Computational
  Physics, 119, 283

\bibitem[{{Dimits} \& {Lee}(1993)}]{DimitsLee93}
{Dimits}, A.~M. \& {Lee}, W.~W. 1993, Journal of Computational Physics, 107,
  309
  
\bibitem[Draine(2011)]{Draine11} Draine, B.~T.\ 2011, Physics of the Interstellar and Intergalactic Medium by Bruce T.~Draine.~Princeton University Press, 2011.~ISBN: 978-0-691-12214-4,  



\bibitem[{{Drury}(1983)}]{Drury83}
{Drury}, L.~O. 1983, Reports on Progress in Physics, 46, 973

\bibitem[{{Dupree}(1966)}]{Dupree66}
{Dupree}, T.~H. 1966, Physics of Fluids, 9, 1773

\bibitem[{{En{\ss}lin} {et~al.}(2007){En{\ss}lin}, {Pfrommer}, {Springel}, \&
  {Jubelgas}}]{Enblin_etal07}
{En{\ss}lin}, T.~A., {Pfrommer}, C., {Springel}, V., \& {Jubelgas}, M. 2007,
  \aap, 473, 41
  
  
\bibitem[En{\ss}lin et al.(2011)]{Ensslin11} En{\ss}lin, T., Pfrommer, C., Miniati, F., \& Subramanian, K.\ 2011, \aap, 527, A99 
  

\bibitem[{{Everett} {et~al.}(2008){Everett}, {Zweibel}, {Benjamin}, {McCammon},
  {Rocks}, \& {Gallagher}}]{Everett_etal08}
{Everett}, J.~E., {Zweibel}, E.~G., {Benjamin}, R.~A., {McCammon}, D., {Rocks},
  L., \& {Gallagher}, III, J.~S. 2008, \apj, 674, 258

\bibitem[Farber et al.(2018)]{Farber18} Farber, R., Ruszkowski, M., Yang, H.-Y.~K., \& Zweibel, E.~G.\ 2018, \apj, 856, 112 

\bibitem[{{Farmer} \& {Goldreich}(2004)}]{FarmerGoldreich04}
{Farmer}, A.~J. \& {Goldreich}, P. 2004, \apj, 604, 671

\bibitem[{{Felice} \& {Kulsrud}(2001)}]{FeliceKulsrud01}
{Felice}, G.~M. \& {Kulsrud}, R.~M. 2001, \apj, 553, 198

\bibitem[Ferri{\`e}re(2001)]{Ferriere01} Ferri{\`e}re, K.~M.\ 2001, Reviews of Modern Physics, 73, 1031 

\bibitem[Fujita \& Ohira(2011)]{Fujita11} Fujita, Y., \& Ohira, Y.\ 2011, \apj, 738, 182 

\bibitem[{{Gardiner} \& {Stone}(2005)}]{GardinerStone05}
{Gardiner}, T.~A. \& {Stone}, J.~M. 2005, Journal of Computational Physics,
  205, 509

\bibitem[{{Gardiner} \& {Stone}(2008)}]{GardinerStone08}
---. 2008, Journal of Computational Physics, 227, 4123

\bibitem[{{Ginzburg} \& {Syrovatskii}(1964)}]{GinzburgSyrovatskii64}
{Ginzburg}, V.~L. \& {Syrovatskii}, S.~I. 1964, {The Origin of Cosmic Rays}

\bibitem[{{Girichidis} {et~al.}(2016){Girichidis}, {Naab}, {Walch}, {Hanasz},
  {Mac Low}, {Ostriker}, {Gatto}, {Peters}, {W{\"u}nsch}, {Glover}, {Klessen},
  {Clark}, \& {Baczynski}}]{Girichidis_etal16}
{Girichidis}, P., {Naab}, T., {Walch}, S., {Hanasz}, M., {Mac Low}, M.-M.,
  {Ostriker}, J.~P., {Gatto}, A., {Peters}, T., {W{\"u}nsch}, R., {Glover},
  S.~C.~O., {Klessen}, R.~S., {Clark}, P.~C., \& {Baczynski}, C. 2016, \apjl,
  816, L19

\bibitem[Girichidis et al.(2018)]{Girichidis18} Girichidis, P., Naab, T., Hanasz, M., \& Walch, S.\ 2018, \mnras, 479, 3042 


\bibitem[{{Goldreich} \& {Sridhar}(1995)}]{GoldreichSridhar95}
{Goldreich}, P. \& {Sridhar}, S. 1995, \apj, 438, 763

\bibitem[{{Guo} \& {Oh}(2008)}]{GuoOh08}
{Guo}, F. \& {Oh}, S.~P. 2008, \mnras, 384, 251

\bibitem[Grenier et al.(2015)]{Grenier15} Grenier, I.~A., Black, J.~H., \& Strong, A.~W.\ 2015, \araa, 53, 199 


\bibitem[{{Hanasz} {et~al.}(2013){Hanasz}, {Lesch}, {Naab}, {Gawryszczak},
  {Kowalik}, \& {W{\'o}lta{\'n}ski}}]{Hanasz_etal13}
{Hanasz}, M., {Lesch}, H., {Naab}, T., {Gawryszczak}, A., {Kowalik}, K., \&
  {W{\'o}lta{\'n}ski}, D. 2013, \apjl, 777, L38

\bibitem[{{Holcomb} \& {Spitkovsky}(2018)}]{HolcombSpitkovsky19}
{Holcomb}, C. \& {Spitkovsky}, A. 2018, arXiv e-prints

\bibitem[Holguin et al.(2018)]{Holguin18} Holguin, F., Ruszkowski, M., Lazarian, A., Farber, R., \& Yang, H.-Y.~K.\ 2018, arXiv:1807.05494 


\bibitem[{{Hu} \& {Krommes}(1994)}]{HuKrommes94}
{Hu}, G. \& {Krommes}, J.~A. 1994, Physics of Plasmas, 1, 863
\bibitem[Ipavich(1975)]{Ipavich75} Ipavich, F.~M.\ 1975, \apj, 196, 107 

\bibitem[Jacob \& Pfrommer(2017)]{Jacob17} Jacob, S., \& Pfrommer, C.\ 2017, \mnras, 467, 1449 


\bibitem[Jacob et al.(2018)]{Jacob18} Jacob, S., Pakmor, R., Simpson, C.~M., Springel, V., \& Pfrommer, C.\ 2018, \mnras, 475, 570 


\bibitem[{{Jiang} \& {Oh}(2018)}]{JiangOh18}
{Jiang}, Y.-F. \& {Oh}, S.~P. 2018, \apj, 854, 5

\bibitem[{{Jokipii}(1966)}]{Jokipii66}
{Jokipii}, J.~R. 1966, \apj, 146, 480

\bibitem[{{Jubelgas} {et~al.}(2008){Jubelgas}, {Springel}, {En{\ss}lin}, \&
  {Pfrommer}}]{Jubelgas_etal08}
{Jubelgas}, M., {Springel}, V., {En{\ss}lin}, T., \& {Pfrommer}, C. 2008, \aap,
  481, 33

\bibitem[Krymskii(1977)]{Krymsky77} Krymskii, G.~F.\ 1977, Akademiia Nauk SSSR Doklady, 234, 1306

\bibitem[{{Kulsrud} \& {Pearce}(1969)}]{KulsrudPearce69}
{Kulsrud}, R. \& {Pearce}, W.~P. 1969, \apj, 156, 445

\bibitem[{{Kulsrud}(2005)}]{Kulsrud05}
{Kulsrud}, R.~M. 2005, {Plasma physics for astrophysics} (Princeton University
  Press)

\bibitem[{{Kunz} {et~al.}(2014){Kunz}, {Stone}, \& {Bai}}]{Kunz_etal14}
{Kunz}, M.~W., {Stone}, J.~M., \& {Bai}, X.-N. 2014, Journal of Computational
  Physics, 259, 154
  
\bibitem[{{Lazarian}(2016)}]{Lazarian16}
{Lazarian}, A. 2016, \apj, 833, 131

\bibitem[Lazarian \& Beresnyak(2006)]{LazarianBeresnyak06} Lazarian, A., \& Beresnyak, A.\ 2006, \mnras, 373, 1195

\bibitem[Lebiga et al.(2018)]{Lebiga18} Lebiga, O., Santos-Lima, R., \& Yan, H.\ 2018, \mnras, 476, 2779

\bibitem[{{Lee} \& {V{\"o}lk}(1973)}]{LeeVolk73}
{Lee}, M.~A. \& {V{\"o}lk}, H.~J. 1973, \apss, 24, 31

\bibitem[Mao \& Ostriker(2018)]{Mao18} Mao, S.~A., \& Ostriker, E.~C.\ 2018, \apj, 854, 89 

\bibitem[Naab \& Ostriker(2017)]{Naab17} Naab, T., \& Ostriker, J.~P.\ 2017, \araa, 55, 59 

\bibitem[{{Pakmor} {et~al.}(2016){Pakmor}, {Pfrommer}, {Simpson}, \&
  {Springel}}]{Pakmor_etal16}
{Pakmor}, R., {Pfrommer}, C., {Simpson}, C.~M., \& {Springel}, V. 2016, \apjl,
  824, L30

\bibitem[{{Parker}(1966)}]{Parker66}
{Parker}, E.~N. 1966, \apj, 145, 811

\bibitem[{{Parker}(1992)}]{Parker92}
---. 1992, \apj, 401, 137

\bibitem[{{Parker} \& {Lee}(1993)}]{ParkerLee93}
{Parker}, S.~E. \& {Lee}, W.~W. 1993, Physics of Fluids B, 5, 77



\bibitem[{{Pfrommer} {et~al.}(2017){Pfrommer}, {Pakmor}, {Schaal}, {Simpson},
  \& {Springel}}]{Pfrommer_etal17}
{Pfrommer}, C., {Pakmor}, R., {Schaal}, K., {Simpson}, C.~M., \& {Springel}, V.
  2017, \mnras, 465, 4500

\bibitem[{{Recchia} {et~al.}(2016){Recchia}, {Blasi}, \&
  {Morlino}}]{Recchia_etal16}
{Recchia}, S., {Blasi}, P., \& {Morlino}, G. 2016, \mnras, 462, 4227

\bibitem[{{Roe}(1981)}]{Roe81}
{Roe}, P.~L. 1981, Journal of Computational Physics, 43, 357

\bibitem[{{Ruszkowski} {et~al.}(2017){Ruszkowski}, {Yang}, \&
  {Zweibel}}]{Ruszkowski_etal17}
{Ruszkowski}, M., {Yang}, H.-Y.~K., \& {Zweibel}, E. 2017, \apj, 834, 208

\bibitem[{{Salem} \& {Bryan}(2014)}]{SalemBryan14}
{Salem}, M. \& {Bryan}, G.~L. 2014, \mnras, 437, 3312

\bibitem[Salem et al.(2014)]{Salem_etal14} Salem, M., Bryan, G.~L., \& Hummels, C.\ 2014, \apjl, 797, L18 


\bibitem[{{Schlickeiser}(1989)}]{Schlickeiser89}
{Schlickeiser}, R. 1989, \apj, 336, 264

\bibitem[{{Schlickeiser} \& {Miller}(1998)}]{SchlickeiserMiller98}
{Schlickeiser}, R. \& {Miller}, J.~A. 1998, \apj, 492, 352

\bibitem[Sharma et al.(2010)]{Sharma10} Sharma, P., Colella, P., \& Martin, D.~F.\ 2010, SIAM J Sci. Comp. 32, 3564 

\bibitem[Simpson et al.(2016)]{Simpson16} Simpson, C.~M., Pakmor, R., Marinacci, F., et al.\ 2016, \apjl, 827, L29 


\bibitem[{{Skilling}(1975{\natexlab{a}})}]{Skilling75a}
{Skilling}, J. 1975{\natexlab{a}}, \mnras, 172, 557

\bibitem[{{Skilling}(1975{\natexlab{b}})}]{Skilling75b}
---. 1975{\natexlab{b}}, \mnras, 173, 245

\bibitem[{{Skilling}(1975{\natexlab{c}})}]{Skilling75c}
---. 1975{\natexlab{c}}, \mnras, 173, 255


\bibitem[{{Soler} {et~al.}(2016){Soler}, {Terradas}, {Oliver}, \&
  {Ballester}}]{Soler_etal16}
{Soler}, R., {Terradas}, J., {Oliver}, R., \& {Ballester}, J.~L. 2016, \aap,
  592, A28
  
\bibitem[{{Stone} {et~al.}(2008){Stone}, {Gardiner}, {Teuben}, {Hawley}, \&
  {Simon}}]{Stone_etal08}
{Stone}, J.~M., {Gardiner}, T.~A., {Teuben}, P., {Hawley}, J.~F., \& {Simon},
  J.~B. 2008, \apjs, 178, 137

\bibitem[{{Summers} \& {Thorne}(1991)}]{SummersThorne91}
{Summers}, D. \& {Thorne}, R.~M. 1991, Physics of Fluids B, 3, 1835

\bibitem[{{Thomas} \& {Pfrommer}(2018)}]{ThomasPfrommer19}
{Thomas}, T. \& {Pfrommer}, C. 2018, arXiv e-prints

\bibitem[{{Uhlig} {et~al.}(2012){Uhlig}, {Pfrommer}, {Sharma}, {Nath},
  {En{\ss}lin}, \& {Springel}}]{Uhlig_etal12}
{Uhlig}, M., {Pfrommer}, C., {Sharma}, M., {Nath}, B.~B., {En{\ss}lin}, T.~A.,
  \& {Springel}, V. 2012, \mnras, 423, 2374

\bibitem[{{V{\"o}lk}(1973)}]{Volk73}
{V{\"o}lk}, H.~J. 1973, \apss, 25, 471

\bibitem[Wentzel(1974)]{Wentzel74} Wentzel, D.~G.\ 1974, \araa, 12, 71 

\bibitem[{{Wiener} {et~al.}(2013){Wiener}, {Oh}, \& {Guo}}]{Wiener_etal13}
{Wiener}, J., {Oh}, S.~P., \& {Guo}, F. 2013, \mnras, 434, 2209

\bibitem[{{Wiener} {et~al.}(2017){Wiener}, {Pfrommer}, \& {Oh}}]{Wiener_etal17}
{Wiener}, J., {Pfrommer}, C., \& {Oh}, S.~P. 2017, \mnras, 467, 906

\bibitem[Wiener et al.(2018)]{Wiener18a} Wiener, J., Zweibel, E.~G., \& Oh, S.~P.\ 2018, \mnras, 473, 3095 

\bibitem[Wiener \& Zweibel(2018)]{Wiener18b} Wiener, J., \& Zweibel, E.~G.\ 2018, arXiv:1812.02179 


\bibitem[{{Yan} \& {Lazarian}(2002)}]{YanLazarian02}
{Yan}, H. \& {Lazarian}, A. 2002, Physical Review Letters, 89, 281102

\bibitem[{{Yan} \& {Lazarian}(2008)}]{YanLazarian08}
---. 2008, \apj, 673, 942

\bibitem[{{Zirakashvili} {et~al.}(1996){Zirakashvili}, {Breitschwerdt},
  {Ptuskin}, \& {Voelk}}]{Zirakashvili_etal96}
{Zirakashvili}, V.~N., {Breitschwerdt}, D., {Ptuskin}, V.~S., \& {Voelk}, H.~J.
  1996, \aap, 311, 113

\bibitem[{{Zweibel}(2013)}]{Zweibel13}
{Zweibel}, E.~G. 2013, Physics of Plasmas, 20, 055501

\bibitem[Zweibel(2017)]{Zweibel17} Zweibel, E.~G.\ 2017, Physics of Plasmas, 24, 055402 



\end{thebibliography}

\label{lastpage}
\end{document}